\documentclass[reqno,11pt]{amsart}
\usepackage[dvipsnames]{xcolor}
\usepackage{amssymb}
\usepackage{amsmath}
\usepackage{amsthm}
\usepackage{amsfonts}
\usepackage{amsxtra}
\usepackage{mathtools}

\usepackage{graphicx}   
\usepackage{hyperref}
\usepackage{tikz}
\usepackage{pgf}

\newtheorem{theorem}{Theorem}[section]

\newtheorem{lemma}[theorem]{Lemma}

\newtheorem{proposition}[theorem]{Proposition}
\newtheorem{corollary}[theorem]{Corollary}

\newcommand{\Rz}{\mathbb{R}}

\newcommand{\Zz}{\mathbb{Z}}
 
\newcommand{\epsi}{\varepsilon}
\newcommand{\eps}{\varepsilon}

\newcommand{\D}{\nabla}

\newcommand{\R}{\ensuremath{\mathbb{R}}}				
\newcommand{\Z}{\ensuremath{\mathbb{Z}}}				
\newcommand{\ra}{\ensuremath{\rightarrow}}				
\newcommand{\ori}{\ensuremath{\mathcal{O}}}				

\usepackage{bm}  				
\usepackage{aligned-overset}	

\usepackage{caption}
\usepackage{subcaption}

\makeatletter
\def\p@subfigure{\thefigure\,}
\def\p@subtable{\thetable\,}
\makeatother

\usepackage{import}
\usepackage{pdfpages}
\usepackage{transparent}

\usepackage{multirow}	
\usepackage{makecell}	

\usepackage{relsize}
\usepackage{stackengine}

\def\INNERSEP{2pt}
\def\SHIFT{.2em}

\newcommand{\ph}{\ensuremath{\phantom{\circledcirc}}} 

\usepackage{tikz}
\usetikzlibrary{shapes, arrows.meta}

\newcommand\mborder[2][]{%
	\tikz[anchor=base,baseline]{
		\node[inner sep=1pt,#1](h){$\displaystyle#2\mathstrut$};
		\draw(h.south west)--(h.north west)
		--(h.north east)--(h.south east) -- cycle;
	}%
}

\newcommand\borderIIII[2][]{%
	\tikz[anchor=base,baseline]{
		\node[inner sep=\INNERSEP,#1](h){$\displaystyle#2\mathstrut$};
		\draw(h.south west)-- ([xshift=\SHIFT]h.west) -- (h.north west)-- ([yshift=-.2em] h.north)
		--(h.north east)--([xshift=-.2em]h.east) -- (h.south east) -- ([yshift = .2em] h.south)  -- cycle;
	}%
}

\newcommand\borderIIOO[2][]{%
	\tikz[anchor=base,baseline]{
		\node[inner sep=2pt,#1](h){$\displaystyle#2\mathstrut$};
		\draw(h.south west)-- ([xshift=-.2em]h.west) -- (h.north west)-- ([yshift=.2em] h.north)
		--(h.north east)--([xshift=-.2em]h.east) -- (h.south east) -- ([yshift = .2em] h.south)  -- cycle;
	}%
}

\newcommand\borderOOII[2][]{%
	\tikz[anchor=base,baseline]{
		\node[inner sep=2pt,#1](h){$\displaystyle#2\mathstrut$};
		\draw(h.south west)-- ([xshift=.2em]h.west) -- (h.north west)-- ([yshift=-.2em] h.north)
		--(h.north east)--([xshift=.2em]h.east) -- (h.south east) -- ([yshift =-.2em] h.south)  -- cycle;
	}%
}

\newcommand\borderOOOO[2][]{%
	\tikz[anchor=base,baseline]{
		\node[inner sep=2pt,#1](h){$\displaystyle#2\mathstrut$};
		\draw(h.south west)-- ([xshift=-.2em]h.west) -- (h.north west)-- ([yshift=.2em] h.north)
		--(h.north east)--([xshift=.2em]h.east) -- (h.south east) -- ([yshift =-.2em] h.south)  -- cycle;
	}%
}

\newcommand\borderOOOOnwd[2][]{%
	\tikz[anchor=base,baseline]{
		\node[inner sep=2pt,#1](h){$\displaystyle#2\mathstrut$};
		\draw(h.south west)-- ([xshift=-.2em]h.west) -- (h.north west)-- ([yshift=.2em] h.north)
		--(h.north east)--([xshift=.2em]h.east) -- (h.south east) -- ([yshift =-.2em] h.south)  -- cycle;
		\draw[fill=white] (h.north west) circle (0.05);
	}%
}

\newcommand\borderOOOOseu[2][]{%
	\tikz[anchor=base,baseline]{
		\node[inner sep=2pt,#1](h){$\displaystyle#2\mathstrut$};
		\draw(h.south west)-- ([xshift=-.2em]h.west) -- (h.north west)-- ([yshift=.2em] h.north)
		--(h.north east)--([xshift=.2em]h.east) -- (h.south east) -- ([yshift =-.2em] h.south)  -- cycle;
		\draw[fill=black] (h.south east) circle (0.05);
	}%
}

\newcommand\borderIIIInwu[2][]{%
	\tikz[anchor=base,baseline]{
		\node[inner sep=\INNERSEP,#1](h){$\displaystyle#2\mathstrut$};
		\draw(h.south west)-- ([xshift=\SHIFT]h.west) -- (h.north west)-- ([yshift=-.2em] h.north)
		--(h.north east)--([xshift=-.2em]h.east) -- (h.south east) -- ([yshift = .2em] h.south)  -- cycle;
		\draw[fill=black] (h.north west) circle (0.05);
	}%
}

\newcommand\borderIIIIsed[2][]{%
	\tikz[anchor=base,baseline]{
		\node[inner sep=\INNERSEP,#1](h){$\displaystyle#2\mathstrut$};
		\draw(h.south west)-- ([xshift=\SHIFT]h.west) -- (h.north west)-- ([yshift=-.2em] h.north)
		--(h.north east)--([xshift=-.2em]h.east) -- (h.south east) -- ([yshift = .2em] h.south)  -- cycle;
		\draw[fill=white] (h.south east) circle (0.05);
	}%
}

\newcommand\borderOOIInwseu[2][]{%
	\tikz[anchor=base,baseline]{
		\node[inner sep=2pt,#1](h){$\displaystyle#2\mathstrut$};
		\draw(h.south west)-- ([xshift=.2em]h.west) -- (h.north west)-- ([yshift=-.2em] h.north)
		--(h.north east)--([xshift=.2em]h.east) -- (h.south east) -- ([yshift =-.2em] h.south)  -- cycle;
		\draw[fill=black] (h.north west) circle (0.05);
		\draw[fill=black] (h.south east) circle (0.05);
	}%
}

\newcommand\borderOOIIneswu[2][]{%
	\tikz[anchor=base,baseline]{
		\node[inner sep=2pt,#1](h){$\displaystyle#2\mathstrut$};
		\draw(h.south west)-- ([xshift=.2em]h.west) -- (h.north west)-- ([yshift=-.2em] h.north)
		--(h.north east)--([xshift=.2em]h.east) -- (h.south east) -- ([yshift =-.2em] h.south)  -- cycle;
		\draw[fill=black] (h.north east) circle (0.05);
		\draw[fill=black] (h.south west) circle (0.05);
	}%
}

\newcommand\borderIIOOnwsed[2][]{%
	\tikz[anchor=base,baseline]{
		\node[inner sep=2pt,#1](h){$\displaystyle#2\mathstrut$};
		\draw(h.south west)-- ([xshift=-.2em]h.west) -- (h.north west)-- ([yshift=.2em] h.north)
		--(h.north east)--([xshift=-.2em]h.east) -- (h.south east) -- ([yshift = .2em] h.south)  -- cycle;
		\draw[fill=white] (h.north west) circle (0.05);
		\draw[fill=white] (h.south east) circle (0.05);
	}%
}

\newcommand\borderIIOOneswd[2][]{%
	\tikz[anchor=base,baseline]{
		\node[inner sep=2pt,#1](h){$\displaystyle#2\mathstrut$};
		\draw(h.south west)-- ([xshift=-.2em]h.west) -- (h.north west)-- ([yshift=.2em] h.north)
		--(h.north east)--([xshift=-.2em]h.east) -- (h.south east) -- ([yshift = .2em] h.south)  -- cycle;
		\draw[fill=white] (h.north east) circle (0.05);
		\draw[fill=white] (h.south west) circle (0.05);
	}%
}

\newcommand\borderIOIOned[2][]{%
	\tikz[anchor=base,baseline]{
		\node[inner sep=2pt,#1](h){$\displaystyle#2\mathstrut$};
		\draw(h.south west)-- ([xshift=.2em]h.west) -- (h.north west)-- ([yshift=.2em] h.north)
		--(h.north east)--([xshift=-.2em]h.east) -- (h.south east) -- ([yshift = -.2em] h.south)  -- cycle;
		\draw[fill=white] (h.north east) circle (0.05);
	}%
}

\newcommand\borderIOIOswu[2][]{%
	\tikz[anchor=base,baseline]{
		\node[inner sep=2pt,#1](h){$\displaystyle#2\mathstrut$};
		\draw(h.south west)-- ([xshift=.2em]h.west) -- (h.north west)-- ([yshift=.2em] h.north)
		--(h.north east)--([xshift=-.2em]h.east) -- (h.south east) -- ([yshift = -.2em] h.south)  -- cycle;
		\draw[fill=black] (h.south west) circle (0.05);
	}%
}

\newcommand\borderOIOIswd[2][]{%
	\tikz[anchor=base,baseline]{
		\node[inner sep=2pt,#1](h){$\displaystyle#2\mathstrut$};
		\draw(h.south west)-- ([xshift=-.2em]h.west) -- (h.north west)-- ([yshift=-.2em] h.north)
		--(h.north east)--([xshift=.2em]h.east) -- (h.south east) -- ([yshift = .2em] h.south)  -- cycle;
		\draw[fill=white] (h.south west) circle (0.05);
	}%
}

\newcommand\borderOIOIneu[2][]{%
	\tikz[anchor=base,baseline]{
		\node[inner sep=2pt,#1](h){$\displaystyle#2\mathstrut$};
		\draw(h.south west)-- ([xshift=-.2em]h.west) -- (h.north west)-- ([yshift=-.2em] h.north)
		--(h.north east)--([xshift=.2em]h.east) -- (h.south east) -- ([yshift = .2em] h.south)  -- cycle;
		\draw[fill=black] (h.north east) circle (0.05);
	}%
}

\newcommand\nborder[1]{%
	\tikz[anchor = base, baseline]{
		\node[inner sep = 1pt]{$\displaystyle#1\mathstrut$};
		\draw (-0.25,-0.15) rectangle (0.25,0.4);
		\path (-0.35,-0.2) -- (0.3,0.45);
	}%
}

\newcommand\nbordernwu[1]{%
	\tikz[anchor = base, baseline]{
		\node[inner sep = 1pt]{$\displaystyle#1\mathstrut$};
		\draw (-0.25,-0.15) rectangle (0.25,0.4);
		\draw[fill=black] (-0.25,0.4) circle (0.05);
		\path (-0.35,-0.2) -- (0.3,0.45);
	}%
}

\newcommand\nborderneu[1]{%
	\tikz[anchor = base, baseline]{
		\node[inner sep = 1pt]{$\displaystyle#1\mathstrut$};
		\draw (-0.25,-0.15) rectangle (0.25,0.4);
		\draw[fill=black] (0.25,0.4) circle (0.05);
		\path (-0.35,-0.2) -- (0.3,0.45);
	}%
}

\newcommand\nbordernwseu[1]{%
	\tikz[anchor = base, baseline]{
		\node[inner sep = 1pt]{$\displaystyle#1\mathstrut$};
		\draw (-0.25,-0.15) rectangle (0.25,0.4);
		
		\draw[fill=black] (0.25,-0.15) circle (0.05);
		\draw[fill=black] (-0.25,0.4) circle (0.05);

		\path (-0.35,-0.2) -- (0.3,0.45);
	}%
}

\newcommand\nborderned[1]{%
	\tikz[anchor = base, baseline]{
		\node[inner sep = 1pt]{$\displaystyle#1\mathstrut$};
		\draw (-0.25,-0.15) rectangle (0.25,0.4);
		\draw[fill=white] (0.25,0.4) circle (0.05);
		\path (-0.35,-0.2) -- (0.3,0.45);
	}%
}

\newcommand\nborderne[1]{%
	\tikz[anchor = base, baseline]{
		\node[inner sep = 1pt]{$\displaystyle#1\mathstrut$};
		\draw (-0.25,-0.15) rectangle (0.25,0.4);

		\path (-0.35,-0.2) -- (0.3,0.45);
		\node (h) at (0.25,0.4) {};
		\draw[fill = white] ([xshift = -0.2em,yshift = -.2em] h) rectangle ([xshift = 0.2em,yshift = .2em] h) ;
		
	}%
}

\newcommand\nborderneStar[1]{%
	\tikz[anchor = base, baseline]{
		\node[inner sep = 1pt]{$\displaystyle#1\mathstrut$};
		\draw (-0.25,-0.15) rectangle (0.25,0.4);
		
		\path (-0.35,-0.2) -- (0.3,0.45);
		\node[star,fill=black,minimum width=0.3em, star point height= 0.2em, inner sep = 1pt] at (0.25,0.4) {};
	}%
}

\newcommand\nbordernwseune[1]{%
	\tikz[anchor = base, baseline]{
		\node[inner sep = 1pt]{$\displaystyle#1\mathstrut$};
		\draw (-0.25,-0.15) rectangle (0.25,0.4);
		
		\draw[fill=black] (0.25,-0.15) circle (0.05);
		\draw[fill=black] (-0.25,0.4) circle (0.05);
		
		\path (-0.35,-0.2) -- (0.3,0.45);
		
		\node (h) at (0.25,0.4) {};
		\draw[fill = white] ([xshift = -0.2em,yshift = -.2em] h) rectangle ([xshift = 0.2em,yshift = .2em] h) ;
	}%
}

\newcommand\nbordernwseuneStar[1]{%
	\tikz[anchor = base, baseline]{
		\node[inner sep = 1pt]{$\displaystyle#1\mathstrut$};
		\draw (-0.25,-0.15) rectangle (0.25,0.4);
		
		\draw[fill=black] (0.25,-0.15) circle (0.05);
		\draw[fill=black] (-0.25,0.4) circle (0.05);
		
		\path (-0.35,-0.2) -- (0.3,0.45);
		\node[star,fill=black,minimum width=0.3em, star point height= 0.2em, inner sep = 1pt] at (0.25,0.4) {};
	}%
}

\newcommand\sborder{%
	\tikz[anchor = base, baseline]{
		\draw (0.0,0.0) rectangle (0.3,0.3);
		\path (-0.05,-0.05) -- (0.35,0.35);
	}%
}

\newcommand\sborderswu{%
	\tikz[anchor = base, baseline]{
		\draw (0.0,0.0) rectangle (0.3,0.3);
		\path (-0.05,-0.05) -- (0.35,0.35);
		\draw[fill=black] (0.0,0.0) circle (0.05);
		
	}%
}
\newcommand\sbordernwu{%
		\tikz[anchor = base, baseline]{
			\draw (0.0,0.0) rectangle (0.3,0.3);
			\path (-0.05,-0.05) -- (0.35,0.35);
			\draw[fill=black] (0.0,0.3) circle (0.05);	
		}%
}

\newcommand\sborderneu{%
		\tikz[anchor = base, baseline]{
			\draw (0.0,0.0) rectangle (0.3,0.3);
			\path (-0.05,-0.05) -- (0.35,0.35);
			\draw[fill=black] (0.3,0.3) circle (0.05);
		}%
}

\newcommand\sborderseu{%
	\tikz[anchor = base, baseline]{
		\draw (0.0,0.0) rectangle (0.3,0.3);
		\path (-0.05,-0.05) -- (0.35,0.35);
		\draw[fill=black] (0.3,0.0) circle (0.05);
	}%
}

\newcommand\sbordernwseu{%
	\tikz[anchor = base, baseline]{
	\draw (0.0,0.0) rectangle (0.3,0.3);
	\path (-0.05,-0.05) -- (0.35,0.35);
	\draw[fill=black] (0.0,0.3) circle (0.05);
	\draw[fill=black] (0.3,0.0) circle (0.05);
}%
}

\newcommand\sborderswd{%
	\tikz[anchor = base, baseline]{
		\draw (0.0,0.0) rectangle (0.3,0.3);
		\path (-0.05,-0.05) -- (0.35,0.35);
		\draw[fill=white] (0.0,0.0) circle (0.05);
		
	}%
}
\newcommand\sbordernwd{%
	\tikz[anchor = base, baseline]{
		\draw (0.0,0.0) rectangle (0.3,0.3);
		\path (-0.05,-0.05) -- (0.35,0.35);
		\draw[fill=white] (00.0,0.3) circle (0.05);	
	}%
}

\newcommand\sborderned{%
	\tikz[anchor = base, baseline]{
		\draw (0.0,0.0) rectangle (0.3,0.3);
		\path (-0.05,-0.05) -- (0.35,0.35);
		\draw[fill=white] (0.3,0.3) circle (0.05);
	}%
}

\newcommand\sbordersed{%
	\tikz[anchor = base, baseline]{
		\draw (0.0,0.0) rectangle (0.3,0.3);
		\path (-0.05,-0.05) -- (0.35,0.35);
		\draw[fill=white] (0.3,0.0) circle (0.05);
	}%
}

\newcommand\borderDashedSSOO[2][]{%
	\tikz[anchor=base,baseline]{
		\node[inner sep=2pt,#1](h){$\displaystyle#2\mathstrut$};
		\draw[dotted](h.south west)-- (h.west) -- (h.north west)-- (h.north)
		--(h.north east)--([xshift=.2em]h.east) -- (h.south east) -- ([yshift = -.2em] h.south)  -- cycle;
	}%
}

\newcommand\borderDashedIISS[2][]{%
	\tikz[anchor=base,baseline]{
		\node[inner sep=2pt,#1](h){$\displaystyle#2\mathstrut$};
		\draw[dotted](h.south west)-- ([xshift=.2em]h.west) -- (h.north west)-- ([yshift = -.2em]h.north)
		--(h.north east)--(h.east) -- (h.south east) -- (h.south)  -- cycle;
	}%
}

\newcommand\borderDashedSSOS[2][]{%
	\tikz[anchor=base,baseline]{
		\node[inner sep=2pt,#1](h){$\displaystyle#2\mathstrut$};
		\draw[dotted](h.south west)-- (h.west) -- (h.north west)-- (h.north)
		--(h.north east)--(h.east) -- (h.south east) -- ([yshift = -.2em] h.south)  -- cycle;
	}%
}

\newcommand\borderDashedSSIS[2][]{%
	\tikz[anchor=base,baseline]{
		\node[inner sep=2pt,#1](h){$\displaystyle#2\mathstrut$};
		\draw[dotted](h.south west)-- (h.west) -- (h.north west)-- (h.north)
		--(h.north east)--(h.east) -- (h.south east) -- ([yshift = .2em] h.south)  -- cycle;
	}%
}

\newcommand\TileA{%
	\ensuremath{\stackinset{c}{0pt}{c}{0pt}{\mathfrak{A}}{\borderDashedSSOS{\textcolor{white}{\tileD{\diagdown}{\diagdown}{\diagdown}{\diagdown}}}}}
}

\newcommand\TileB{%
	\ensuremath{\stackinset{c}{0pt}{c}{0pt}{\mathfrak{B}}{\borderDashedSSIS{\textcolor{white}{\tileD{\diagdown}{\diagdown}{\diagdown}{\diagdown}}}}}
}

\newcommand{\tile}[4]{\ensuremath{\begin{smallmatrix}	#1 & #2 \\ #3 & #4 \end{smallmatrix}}}

\stackMath
\newcommand{\tileU}[4]
{\stackinset{c}{}{c}{0pt}{+}{\tile{#1}{#2}{#3}{#4}}
}
\newcommand{\tileD}[4]
{\stackinset{c}{}{c}{0pt}{-}{\tile{#1}{#2}{#3}{#4}}
}

\newcommand{\DU}{\diagup}
\newcommand{\DD}{\diagdown}

\newcommand{\df}[1]{\textit{#1}}					
\newcommand{\cell}{\textrm{cell}}					

\newcommand{\bz}{{\boldsymbol z}}

\theoremstyle{definition}
\begingroup

\endgroup

\numberwithin{equation}{section}

\begin{document}

\title[Tilings with nonflat squares: a characterization]{Tilings with nonflat squares: a characterization}


\author[M. Friedrich]{Manuel Friedrich} 
\address[Manuel Friedrich]{Applied Mathematics,  
University of M\"{u}nster, Einsteinstr. 62, D-48149 M\"{u}nster, Germany
}
\email{manuel.friedrich@uni-muenster.de}
\urladdr{https://www.uni-muenster.de/AMM/en/Friedrich/}

\author[M. Seitz]{Manuel Seitz} 
\address[Manuel Seitz]{Faculty of Mathematics, University of
  Vienna, Oskar-Morgenstern-Platz~1, A-1090 Vienna, Austria 
}
\email{manuel.seitz@univie.ac.at}

\author[U. Stefanelli]{Ulisse Stefanelli} 
\address[Ulisse Stefanelli]{Faculty of Mathematics, University of
  Vienna, Oskar-Morgenstern-Platz 1, A-1090 Vienna, Austria,
Vienna Research Platform on Accelerating
  Photoreaction Discovery, University of Vienna, W\"ahringerstra\ss e 17, 1090 Wien, Austria,
 \& Istituto di
  Matematica Applicata e Tecnologie Informatiche {\it E. Magenes}, via
  Ferrata 1, I-27100 Pavia, Italy
}
\email{ulisse.stefanelli@univie.ac.at}
\urladdr{http://www.mat.univie.ac.at/$\sim$stefanelli}



\begin{abstract} 
	Inspired by the modelization of 2D materials systems, we characterize
	arrangements of identical nonflat squares in 3D. We prove that
	the fine geometry of such arrangements is completely characterized in
	terms of patterns of mutual orientations of the squares and that these
	patterns are periodic and one-dimensional. In contrast to the flat
	case, the nonflatness of the tiles gives rise to nontrivial
	geometries, with configurations bending, wrinkling, or even rolling up
	in one direction.
\end{abstract}

%
\maketitle

\pagestyle{myheadings}

\section{Introduction}

The serendipitous isolation of graphene in 2004 \cite{Novoselov}
attracted enormous interest on the physics of 2D  materials  systems. Driven by their fascinating electronic and
mechanical properties \cite{Zhang}, research on 2D systems is currently 
witnessing an exponential growth. Beyond graphene \cite{Allen,Geim}, 2D material systems are continuously  syntetized
and investigated \cite{Cahangirov,Davila,Liu,Zhu} and
findings are emerging at an always
increasing pace, ranging from fundamental understanding to applications \cite{Akinwande}.

Free standing 2D material samples are often not flat, but rather present
rippling patterns at specific   length scales  \cite{Li}. The origin of such
nonflatness is currently debated, one possible explanation being the
instability of perfectly flat arrangements at finite temperatures, as
predicted by the classical
Mermin-Wagner  theory  \cite{Mermin,Mermin2}.
In the case of graphene, ripples have been experimentally
observed \cite{Jani,Meyer}, computationally investigated 
\cite{Fasolino}, and analytically assessed 
\cite{emergence,ripples}. The phenomenon is however not restricted to
graphene, and surface rippling has been detected in other 2D
systems as well \cite{Boustani,Wang}. 
Understanding the global geometry of 2D materials  is of the
greatest importance,  as 
flatness is known to  influence crucially the
electronic, thermal, and mechanical behavior of these systems
\cite{Chu,Deng,Xu,Zhang2}.

\begin{figure}[h]
	\pgfdeclareimage[width=85mm]{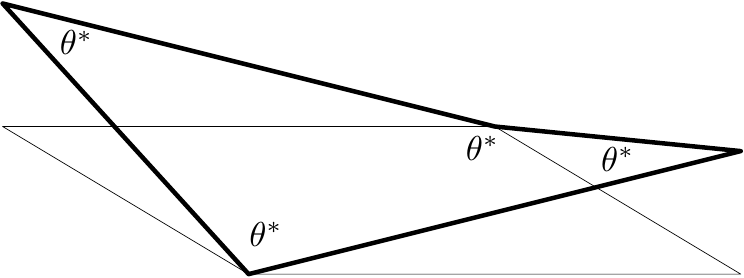}{Images/one}
	
	\centering
	\pgfuseimage{one}
	\caption{The regular nonflat square}\label{one}
\end{figure}

In this paper, we tackle the question of flatness of 2D systems with square symmetry. Our
interest is theoretical and our arguments are not tailored to a
specific material system. Still, we remark that square-like 2D crystals have been
predicted in selenene and tellurene \cite{Xian17}. We formulate the problem in the setting of molecular
mechanics \cite{Allinger,Lewars,Rappe} by associating to each point
configuration a scalar {\it configurational energy} and focusing on
its ground states in the quest for optimal geometries
\cite{Blanc,Friesecke-Theil15}. In the square-symmetric case, each
atom has  four first
neighbors and the topology of the configuration is that of the square
lattice $\Zz^2$ \cite{Mainini14}. The configurational energy is assumed to
feature both two- and three-body effects
\cite{Brenner,Stillinger-Weber85,Tersoff}, depending  on bond
lengths (distances between atoms) and angles between bonds, respectively. 
We present conditions ensuring
that global minimizers of the configurational energy have all
bonds of equal length, all angles formed by
bonds to first neighbors of equal amplitude $\theta^*$, and the four
first neighbors of each atom  are  coplanar. 
As a result, minimal cycles of four atoms form {\it regular} squares featuring
equal sides and equal
angles $\theta^*$, see Figure~\ref{one}.
Such
identical squares arrange then in an infinite 3D configuration, which
under the above provisions we call {\it admissible} and 
which we interpret as the actual geometry of the crystal. 

The goal of
this paper is to classify all admissible configurations, namely
all possible 
3D arrangements of identical regular squares. In case the squares are flat,
namely if $\theta^*=\pi/2$,
the result is straightforward: the only configuration of flat
squares where all first neighbors of each atom are coplanar is
 the plane.  In order to tackle  genuinely 3D  geometries, we hence need to focus on
the case $\theta^*<\pi/2$ instead, which induces nonflatness, as per Figure~\ref{one}.

Our main result is a complete characterization of admissible arrangements of identical
regular nonflat squares in 3D,  see   Theorem \ref{thm: main thm 4-tiles}. We prove in particular
that admissible configurations can bend, wrinkle, and roll in one direction and
that such flexural behavior is completely characterized by specifying
a suitably defined  
section of the configuration in the bending direction, see Figure \ref{fig:gs are one-dim}
below. 
More precisely, one classifies
patches of four squares sharing an atom ({\it 4-tiles}) in six
different {\it classes}, in terms of their mutual orientation, see Figure \ref{fig:4-tiles}. We prove that just {\it three} of these
classes actually give rise to admissible configurations, that the whole
geometry is specified by knowing the pattern of such classes, and that
such pattern is periodic. 

One can visualize the square in Figure \ref{one} as (the boundary of)
a nonflat tile. Our result can hence be interpreted as a  classification
of all possible {\it tilings} with such nonflat tiles under the
condition that the four neighbors of each atom are coplanar. The
relevance of this coplanarity condition is revealed by
considering the limiting flat case. In case tiles are flat and the four neighbors of each
atom are coplanar, the only possible tiling is the plane. By
dropping the
coplanarity requirement, we however allow for tilings ensuing from
foldings of the reference square lattice $\Zz^2$ along a set of
parallel coordinate
directions.  Thus,  the coplanarity requirement serves  the purpose of
excluding the effect of the symmetry
of the reference lattice on the onset of nontrivial geometries.

In the case of hexagonal symmetry, the characterization of global
arrangements of regular nonflat hexagons has been obtained in
\cite{emergence,ripples}. To some extent, the results in this paper for
squares are akin to the
hexagonal case, for in both cases the arrangement shows
some distinguished one-dimensional patterning. Compared with the
hexagonal setting, the  present  square-symmetric case
is however much more involved. This is an effect of  the  different symmetry
of the underlying reference lattices.
In the square case, arguments require to consider the detailed geometry of
patches of up to sixteen neighboring squares, which makes the combinatorial
picture much richer. 

The paper is organized as follows. Section 2 is devoted to the
statement of our main results. The
molecular-mechanical model is discussed first and the detailed
geometry of ground states is assessed. A first description of
admissible configurations is presented  in
Theorem \ref{thm: main optimal cells}. We then introduce the
concept of {\it $4$-tile} and of its {\it type},  collect  all possible types  and classes,  and
discuss the possibility of attaching two $4$-tiles by analyzing the
corresponding boundary, see Lemma \ref{prop: attach}.
This eventually paves the way to the statement of our main result, namely
the characterization of
Theorem \ref{thm: main thm 4-tiles}. Section~\ref{sec:coplanar} is
entirely devoted to the proof of the main result, hinging both on
combinatorial and geometrical arguments.  Some proofs are  postponed to
the Appendix in order to enhance the readability of the arguments.


\section{The setting and main results}

 \subsection{Ground states of configurational energies}
 
We focus on three-dimensional deformations  $y\colon \Z^2 \ra \R^3$, defined on  the two-dimensional \df{reference lattice} $\Z^2$. For any open, bounded subset $\Omega \subset \R^2$ we define the \emph{configurational energy}  of a deformation on $\Omega$ by
\begin{align}\label{eq: restrict energy}
E(y,\Omega) & := \frac{1}{2}\sum_{(x,x') \in N_1(\Omega)} v_2\big(|y(x) - y(x')|\big) +  \frac{1}{2}  \sum_{(x,x') \in N_2(\Omega)}  v_2\big(|y(x) - y(x')|\big)\notag  \\ & \ \ \ \  + \frac{1}{2}\sum_{(x,x',x'') \in T(\Omega)} v_3\big(\measuredangle y(x)\, y(x') \, y(x'')\big), 
\end{align}
where 
$$N_1(\Omega):=\big\{ (x,x')\colon x,x' \in  \Z^2, \, x \in \Omega, \,  x' \in \overline{\Omega}, \,   |x-x'|=1\big\}$$
 denotes the set of  {\it nearest-neighbors} and  
 $${N_2(\Omega):=\lbrace (x,x') \colon x,x' \in  \mathbb{Z}^2 \cap   \overline{\Omega}, \,     |x-x'|=\sqrt{2};   \text{
   $(x-x')\cdot e_1>0$ if $x\in \partial \Omega$ or $x'\in \partial \Omega$}\rbrace}$$
     is the set of  {\it closest  next-to-nearest-neighbors}.  Moreover, by $\measuredangle y(x)\, y(x') \, y(x'')$ we denote the {\it bond angle}  in $[0,\pi]$  at  $y(x')$ formed by the   
  the vectors $y(x)-y(x')$ and $y(x'')- y(x')$, where the set of {\it triplets} $T(\Omega)$ is defined by  
  $$T(\Omega):=\lbrace (x,x',x'')\colon (x',x) \in N_1(\Omega), (x',x'') \in N_1(\Omega), x \neq x'' \rbrace.$$  
  The factor $1/2$ reflects the fact that {\it bonds} $\lbrace y(x),
  y(x') \rbrace$, $(x,x') \in N_1(\Omega)  \cup N_2(\Omega)$, and bond angles
  $\measuredangle y(x)\, y(x') \, y(x'')$ appear twice in the 
  corresponding  sums.  Let us point out that in order to take surface effects at $\partial \Omega$ properly into account, the bonds $\lbrace y(x), y(x') \rbrace$ in $N_1(\Omega) \cup  N_2(\Omega)$  touching $\partial \Omega$   are only counted once. 
  
 We assume  the    {\it two-body}  interaction potential
$v_2\colon\Rz^+\to[-1,\infty)$  to be of short-range repulsive and
long-range attractive type. In particular, we assume that $v_2$ 
is continuous and attains its minimum value  only   at
$1$ with $v_2(1) = -1$. Moreover, we suppose that $v_2$ is decreasing on $(0,1)$,  increasing  on $[1,\infty)$,  and that  $v_2$ is continuously differentiable  with $v_2' >0$ on  $(1,2]$. The  {\it three-body} interaction density
$v_3\colon  [0,\pi] \to[0,\infty)$
is assumed to be strictly convex and smooth, with   ${v}_3(\pi) = 0$.

In the following, we will be interested in minimizing the energy of a configuration on the \emph{whole} reference lattice. To this end, we define the \emph{normalized energy} of $y\colon \Z^2 \to \R^3$ by
\begin{align}\label{eq: general energy}
E(y) =  \sup_{m \in \mathbb{N}}  \frac{1}{(2m-1)^2} E(y,Q_m), 
\end{align}
where $Q_m \subset \Rz^2$ is the open square centered at $0$ with sidelength $2m$.  A deformation is called a {\it ground state} if it minimizes  the energy $E$.

For a fine
characterization of the minimizers,   some additional qualification on
$v_2$ and $v_3$ will be needed. More precisely, we suppose that there exist small parameters    $\eta, \eps >0$  such that 
\begin{align}
	&v_2(1-\eta) > 3 + 4 v_2(\sqrt{2}) +8v_3(\pi/2), \label{eq:energy a1} \\
	&v_2(1+\eta) > -1 + 4 v_2(\sqrt{2}) -  4v_2(\sqrt{2}(1-\eta)^2) +8v_3(\pi/2), \label{eq:energy a2} \\ 
	&v_3(\theta) > 2 + 2 v_2(\sqrt{2}) +4v_3(\pi/2)  \text{  if   }   \theta  \le \pi/2 - \eta,   \label{eq:energy a3} \\
	&( \ell_1,\ell_{2} ,\theta) \mapsto \dfrac{1}{4}
   v_2( \ell_1) + \dfrac{1}{4} v_2( \ell_2) +  v_2  \left( \left(  \ell_1^2+  \ell_2^2 - 2  \ell_1  \ell_2 \cos\theta \right)^{1/2} \right) + v_3(\theta) \nonumber \\
	&\quad \quad  \text{strictly convex on $[1-\eta,1+\eta]^2\times [\pi/2- \eta,\pi]$ and }\nonumber \\ & \quad \quad \text{strongly  convex for $\theta \in [\pi/2- \eta,\pi/2+3\eta]$}, \label{eq:energy a4} \\
	&  |v_3|,|v_3'| \le \eps \text{ in a neighborhood of $\pi$}, \label{eq:energy a5}   \\
	&   0 < -2\sqrt{2} \sqrt{1-\cos\theta} v_3'(\theta) <  \ell \sin\theta \, v_2'\big(\sqrt{2}\ell\sqrt{1-\cos\theta}\big)  \nonumber \\ &\quad\quad \text{ for   $\ell \in [1-\eta, 1]$  and $\theta \in [\pi/2-\eta,\pi]$}.\label{eq:energy a6} 
\end{align} 
Properties  \eqref{eq:energy a1}--\eqref{eq:energy a2} entail that
first-neighbor bond lengths range between $1-\eta$ and $1+\eta$,
whereas \eqref{eq:energy a3} ensures that bond angles are not significantly smaller than $\pi/2$.   Eventually, assumptions
\eqref{eq:energy a4}--\eqref{eq:energy a6}  yield  that the contributions of first and second neighbors  are
strong enough to induce  local geometric symmetry of ground states, i.e., bonds and bond angles  will be constant, see \eqref{eq: dist}--\eqref{eq: angles2} below.

Note that the assumptions  \eqref{eq:energy a1}--\eqref{eq:energy a6}  are compatible with a choice of a density
$v_2$  growing sufficiently fast out of its minimum.  In
particular, the quantitative Lennard-Jones-like case of {\sc Theil}
\cite{Theil06} (see also \cite{E-Li09,Smereka15}) can be reconciled with assumptions
\eqref{eq:energy a1}-\eqref{eq:energy a2}, upon suitably choosing densities and parameters. 
Let us however remark that the specific form
of  \eqref{eq:energy a1}--\eqref{eq:energy a6}  is here chosen for the sake of
definiteness and simplicity. Indeed, these assumptions could be weakened, 
at the expense of additional notational intricacies. Under the above assumptions we have the following result, where we define  $N_i = \bigcup_{m \in \mathbb{N}} N_i(Q_m)$ for $i=1,2$ and $T = \bigcup_{m \in \mathbb{N}} T(Q_m)$.

\begin{proposition}[Ground states]\label{prop:minimizers of E}
 For $\eta$ small enough and $\eps = \eps(\eta)$ small enough   there exist $\ell \le 1$,  $\theta <\pi/2$, and  $\delta_\theta < \pi$  only depending on $v_2$ and $v_3$ such that a deformation  $y\colon \Z^2 \ra \R^3$ is a ground state of the energy $E$ if and only if $y$ satisfies 
\begin{align}\label{eq: dist}
\text{$|y(x) - y(x')| =\ell$ for all $(x,x') \in N_1$,}
\end{align}
and 
\begin{align}\label{eq: angles}
\text{$\measuredangle y(x)\, y(x') \, y(x'') =  \theta$ for all $(x,x',x'') \in T$ with $(x,x'') \in N_2$,  } 
\end{align}
as well as
\begin{align}\label{eq: angles2}
\text{$\measuredangle y(x)\, y(x') \, y(x'')  =\delta_\theta$ for all $(x,x',x'') \in T$ with $(x,x'') \notin N_2$.  } 
\end{align}
\end{proposition} 
Here, the conditions $(x,x'') \in N_2$ and $(x,x'') \notin N_2$ correspond to the  case  that the vectors $x-x'$ and $x''-x'$ form an angle  $\pi/2$ or $\pi$, respectively, in the reference lattice. We will see later that $\delta_\theta$ is uniquely determined by $\theta$ due to a geometric compatibility condition, see Lemma \ref{prop:delta bounded by 2 theta} below. 

The proof of Proposition~\ref{prop:minimizers of E} is similar to the one in \cite[Proposition 3.1]{emergence} and is postponed to Appendix \ref{sec:energy}. At this stage, let us just comment on the effect of condition $v_2' > 0$ in a   neighborhood  of $\sqrt{2}$, see \eqref{eq:energy a6},  which guarantees that $\theta$ is strictly smaller than $\pi/2$. Indeed if $v_2' = 0$ in a neighborhood of $\sqrt{2}$,  we would obtain $\ell = 1$ and  $\theta = \pi/2$, i.e., $y(\Z^2)$ would coincide with $\Z^2$ up to isometries. For $\theta <\pi/2$ instead, ground states exhibit interesting nontrivial geometries. The aim of this paper is precisely that of characterizing these nontrivial geometries.

\subsection{Necessary conditions for admissibility}\label{sec: opti-thm}
 
Deformations  $y\colon \Z^2 \ra \R^3$ satisfying the conditions \eqref{eq: dist}--\eqref{eq: angles2} are called \emph{admissible}.  Without restriction, we suppose for notational convenience that $\ell=1$. Indeed, this can be achieved by replacing $y$ by $\frac{1}{\ell}y$ without effecting the geometry of admissible configurations.

\begin{figure}[h]
	\centering
	\begin{subfigure}[b]{0.45\textwidth}
		\centering
	\def\svgwidth{ \columnwidth}
	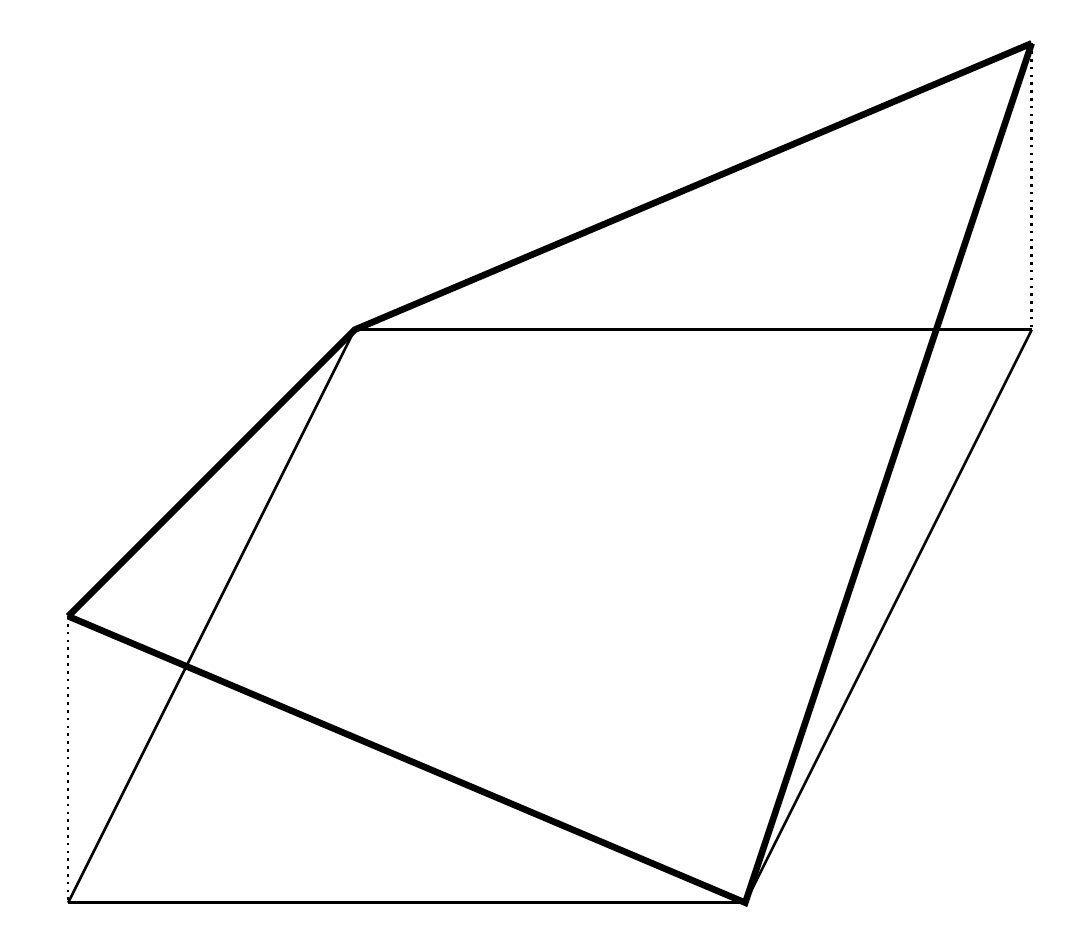

		\caption{Optimal cell of  form  $\diagdown$}
		\label{fig:optcell1}
	\end{subfigure}
	\hfill
	\begin{subfigure}[b]{0.45\textwidth}
		\centering
	\def\svgwidth{ \columnwidth}
	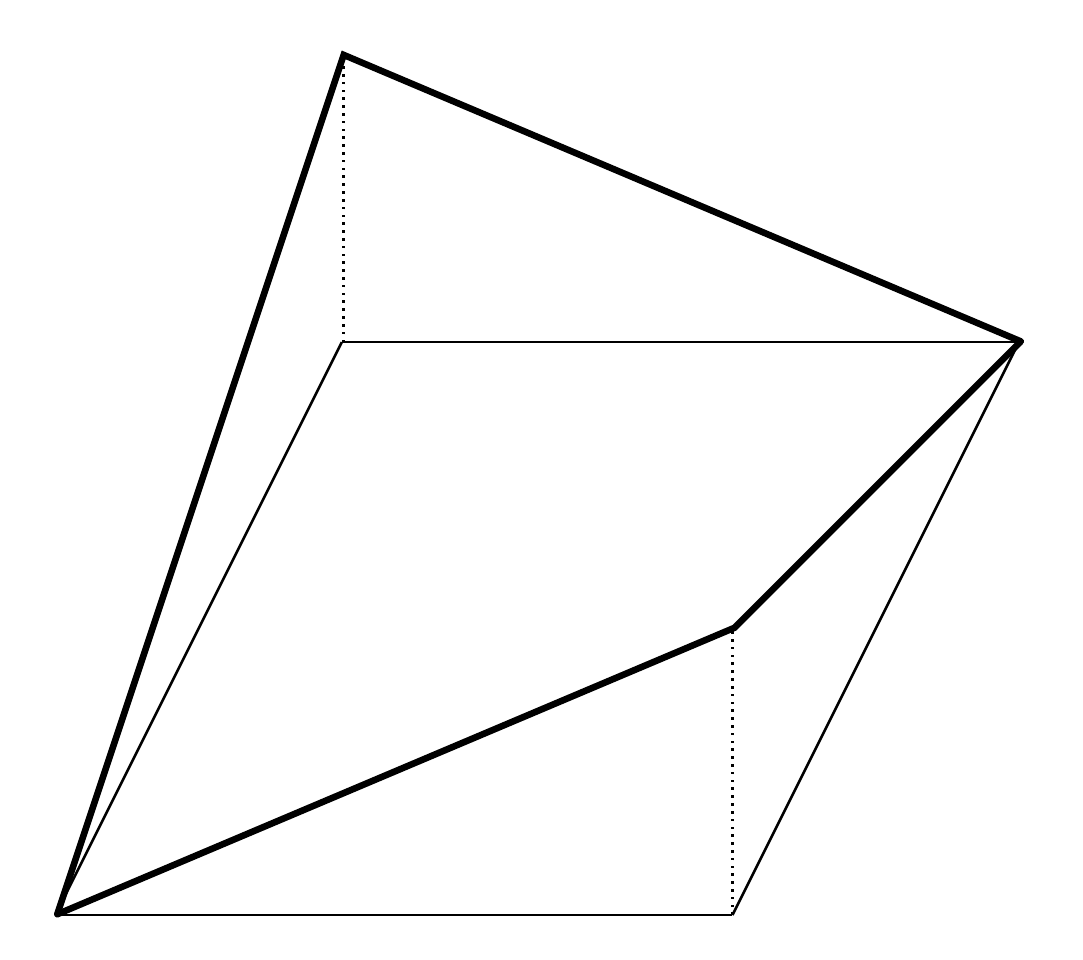

		\caption{Optimal cell of  form  $\diagup$}
		\label{fig:optcell0}
	\end{subfigure}
	\caption{The two optimal cells, defined via the
          vector $p$ and the normal vector of one face. In
          symbols, we indicate optimal cells with $\diagup$ or
          $\diagdown$, 
          according to the direction of the lower diagonal.}
	\label{fig:types optimal cells}
\end{figure}

Obviously, conditions \eqref{eq: dist}--\eqref{eq: angles} constrain the local geometry of configurations: let $\{x_1, x_2, x_3, x_4\}$ be a simple cycle in $\Z^2$, called a \df{reference cell}, where here and in the following the labeling is counterclockwise and counted modulo $4$.  The image via $y$ is the simple cycle $\{y_1, y_2, y_3, y_4\}$, where $y_i = y(x_i)$, called an \df{optimal cell}.  Since $\theta < \pi/2$ from \eqref{eq: angles}, optimal cells are not flat. Indeed, the sum of interior angles is strictly less then $2\pi$, i.e., $\sum_{i = 1}^{4}  \measuredangle y_{i-1} \, y_i \, y_{i+1} = 4 \theta < 2 \pi$, see also Figure~\ref{fig:types optimal cells}.

The kink of an optimal cell can equivalently be visualized as occurring along the diagonal $x_3 - x_1$ or along the diagonal $x_4 - x_2$ of the corresponding reference cell. We set $m_1 := (y_1 + y_3)/2$ and $m_2 := (y_2 + y_4)/2$ and define $p := m_1 - m_2$.  Let  $n$ be the normal vector of the triangle formed by $ y_1$, $y_2$, and $y_4$, in direction $(y_2 - y_1) \times (y_4-y_2)$. Then, we  say that the optimal cell is of  \df{ form  $\diagdown$}  if $p \cdot n > 0$ and of  \df{ form  $\diagup$}  if $p \cdot n < 0$, see Figure~\ref{fig:types optimal cells}. An optimal cell of any  form  can be  transformed  into a cell of the other form simply via a rotation by $\pi/2$ along the vector $p$ or via a reflection with respect to the plane with normal $p$.

Our goal is to provide a complete characterization of admissible configurations. In a first step, we will present  necessary conditions for admissibility   in terms of \emph{optimal cells}.  To obtain a complete  characterization,   we will subsequently present a refined formulation  in  terms of so-called \emph{$4$-tiles}, namely, $2\times 2$ groups of optimal cells, see Subsection~\ref{sec: thm-4-ti}.  To state our first main result,  we need to introduce some further notation. 

\emph{Form  function.} Given a reference cell $\{x_1, x_2, x_3, x_4\}$ labeled in such a way that for the lower-left corner $x_1$ we have $x_1 = (s,t)$, we define the \df{barycenter} $z$ of the reference cell via $z(s,t) := (1/2 + s, 1/2 + t)$. Thus, $z(\Z^2)= {\Z^2}^*$, where ${\Z^{2}}^*$ denotes the dual lattice of $\Z^2$. We define the \df{form  function} on the dual lattice $\tau\colon {\Z^{2}}^* \ra \{\diagdown, \diagup \}$  as the map assigning to each reference cell the  form  of the optimal cell in the deformed configuration. In other words, the deformation $y$ maps a reference cell with barycenter $z(s,t)$ to an optimal cell of  form  $\tau(z(s,t))$.

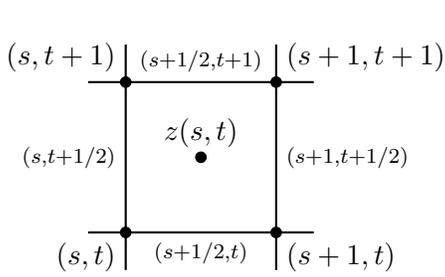
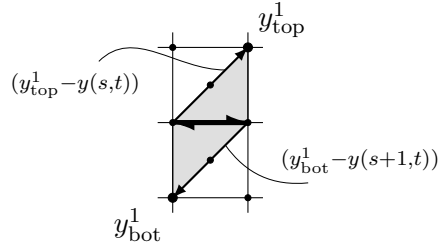
\begin{figure}[h]
	\begin{subfigure}[b]{0.43\textwidth}
		\centering
		\begin{tikzpicture}		
			\draw[thick] (-0.5,0) -- (2.5,0); 		
			\draw[thick] (-0.5,2) -- (2.5,2);		
			\draw[thick] (0,-0.5) -- (0,2.5);		
			\draw[thick] (2,-0.5) -- (2,2.5);		
			\draw[fill=black] (0,0) circle (0.07);		
			\draw[fill=black] (2,0) circle (0.07);		
			\draw[fill=black] (2,2) circle (0.07);		
			\draw[fill=black] (0,2) circle (0.07);		
			\draw[fill=black] (1,1) circle (0.07);		

			\node[below left] at (0,0) {$(s,t)$};
			\node[above] at (1,1) {$z(s,t)$};
			\node[below right] at (2,0) {$(s+1, t)$};
			\node[above right] at (2,2) {$(s+1, t+1)$};
			\node[above left] at (0,2) {$(s, t+1)$};

			\node[below] at (1,0) {${\scriptstyle (s+1/2,t)}$}; 	
			\node[above] at (1,2) {${\scriptstyle (s+1/2,t+1)}$};	
			\node[left] at (0,1) {${\scriptstyle (s,t+1/2)}$};	
			\node[right] at (2,1) {${\scriptstyle (s+1,t+1/2)}$};	
		\end{tikzpicture}
		\caption{Parametrization in the reference lattice.}
		\label{fig:param ref}
	\end{subfigure}
	\hfill
	\begin{subfigure}[b]{0.53\textwidth}
		\centering
		\begin{tikzpicture}
			\draw[fill=gray!25]    (0,1) -- (1,1) -- (1,2) -- cycle;
			\draw[fill=gray!25]    (0,1) -- (1,1) -- (0,0) -- cycle;
			
			\draw[] (-0.2,0) -- (1.2,0);
			\draw[] (-0.2,1) -- (1.2,1);
			\draw[] (-0.2,2) -- (1.2,2);
			\draw[] (0,-0.2) -- (0,2.2);
			\draw[] (1,-0.2) -- (1,2.2);
			
			\draw[fill=black] (0,0) circle (0.06);		
			\draw[fill=black] (0,1) circle (0.04);	
			\draw[fill=black] (0,2) circle (0.04);	
			\draw[fill=black] (1,0) circle (0.04);		
			\draw[fill=black] (1,1) circle (0.04);	
			\draw[fill=black] (1,2) circle (0.06);		
			
			\draw[fill=black] (0.5,0.5) circle (0.04);		
			\draw[fill=black] (0.5,1.5) circle (0.04);		
			
			\draw[ultra thick] (0,1) -- (1,1);
			
			\draw[thick,->,> = latex] (0,1) -- (1,2);
			\draw[thick,->,-{Latex[left]}] (0,1.02)--(1,1.02);
			
			\draw[thick,<-,-{Latex[left]}] (1,0.98) --(0,0.98);
			\draw[thick, ->,> = latex] (1,1) -- (0,0);		
			
			\node[below left] at (0,0) {$y^1_{\rm bot}$};
			\node[above right] at (1,2) {$y^1_{\rm top}$};
			\node[left] (bes1) at (-0.3,1.5) {${\scriptstyle (y_{\rm top}^1 - y(s,t))}$};
			\node[right] (bes2) at (1.3,0.5) {${\scriptstyle (y_{\rm bot}^1 - y(s+1,t))}$};
			
			\draw[very thin] (bes1) to[out = 45, in = 180] (0.7,1.7);
			\draw[very thin] (bes2) to[bend left] (0.7,0.7); 
			
		\end{tikzpicture}
		\caption{Definition of the incidence angle along $d_1$ for a horizontal bond. The shaded areas correspond to the bond planes.}
		\label{fig:inc angle}
	\end{subfigure}
	\caption{Notions for Theorem~\ref{thm: main optimal cells}.}  
\end{figure}

\emph{Incidence angles.} We define the diagonals $d_1 = (1,1)$ and
$d_2 = (-1,1)$. For $i=1,2$, we indicate  \emph{signed incidence
  angles along the diagonal $d_i$} for each bond of the configuration
 via the   mappings  $\gamma_i\colon ((\Z +1/2) \times \Z) \cup (\Z\times (\Z +1/2)) \to [-\pi,\pi]$ defined as follows: first, for $s, t \in \Z$, $(s+1/2,t)$ parametrizes the horizontal bond in the reference lattice connecting $(s,t)$ and $(s+1,t)$, and  $(s,t+1/2)$   parametrizes the vertical bond in the reference lattice connecting $(s,t)$ and $(s,t+1)$, see Figure~\ref{fig:param ref}.  In the following,  we explicitly give the definition of the incidence angle $\gamma_i(s+1/2,t)$, $i=1,2$, for horizontal bonds. The definition associated to vertical bonds follows analogously, up to a rotation of the reference lattice by $\pi /2$. 

Consider a horizontal bond parametrized by $(s+1/2,t)$, which is shared by the two cells with barycenters $z(s,t-1) = (s+1/2,t-1/2)$ and $z(s,t) = (s+1/2, t+1/2)$.  By $n^i_{\rm top}$ we denote the unit normal vector to the plane spanned by the points  $y(s,t)$, $y(s+1,t)$, and $y_{\rm top}^i := y((s,t)+v_i),$  with direction  $(y(s+1,t) - y(s,t)) \times (y_{\rm top}^i - y(s,t))$, where for convenience we set  $v_1 := d_1 = (1,1)$ and $v_2 := (0,1)$. Analogously, we let $n^i_{\rm bot}$  be the unit normal vector to the plane spanned by  $ y(s,t)$, $y(s+1,t)$, and $y_{\rm bot}^i:= y((s+1,t)-v_i)$ with direction  $(y(s,t) - y(s+1,t)) \times (y_{\rm bot}^i - y(s+1,t))$,  see Figure~\ref{fig:inc angle}. 
 
Then, for all $s,t \in \Z$,   the \emph{signed incidence angles  along the diagonal $d_i$} of horizontal bonds are  given by
\begin{equation}\label{eq:def of gamma hor}
	\gamma_i(s+1/2, t) = \left\lbrace \begin{array}{rcl}
		\arccos(n^i_{\mathrm{top}}\cdot n^i_{\mathrm{bot}})  &  \text{if} & (y^i_{\rm top}- y_{\rm bot}^i) \cdot (n^i_{\mathrm{top}} - n^i_{\rm bot}) \geq 0\\ 
		-\arccos(n^i_{\mathrm{top}}\cdot n^i_{\mathrm{bot}})  &  \text{if} & (y^i_{\rm top}- y_{\rm bot}^i) \cdot (n^i_{\mathrm{top}} - n^i_{\rm bot})  < 0.
		\end{array} \right. 
\end{equation} 

Making use of the introduced notation, we are now in the position of
formulating our first result.  This is  a simplified version
of  the later  Theorem~\ref{thm: main thm 4-tiles}  and
provides {\it necessary}  conditions on the existence of admissible configurations.

\begin{theorem}[Basic structure of admissible configurations]\label{thm: main optimal cells}
  There exists $\gamma^* \in (0,\pi)$, depending only on $\theta$, such that  for every admissible configuration  $y\colon \Z^2\to \R^3$,  possibly up to reorientation of the reference lattice, the following holds true: 
	\begin{itemize}
	\item (Constant form function along $d_1$) We have $\tau(s,t) = \tau(s+1, t+1)$ for all $s,t \in \Z$.
	\item (Vanishing incidence angle along $d_1$) We have $\gamma_1(s+1/2,t) = 0 = \gamma_1(s,t+1/2)$ for all $s,t\in\Z$.
	\item (Incidence angle along $d_2$)  It holds that  $\gamma_2(s,t) = \gamma_2( s+1/2,t+1/2) \in \{ \pm \gamma^*, 0\}$ for all $s,t\in \frac{1}{2}\Z$ with $s+t \in \Z+1/2$.  
	\end{itemize} 

\end{theorem}

\begin{figure}[h]
	\centering
	\begin{subfigure}[t]{0.45\textwidth}
		\centering
		\includegraphics[width=\linewidth]{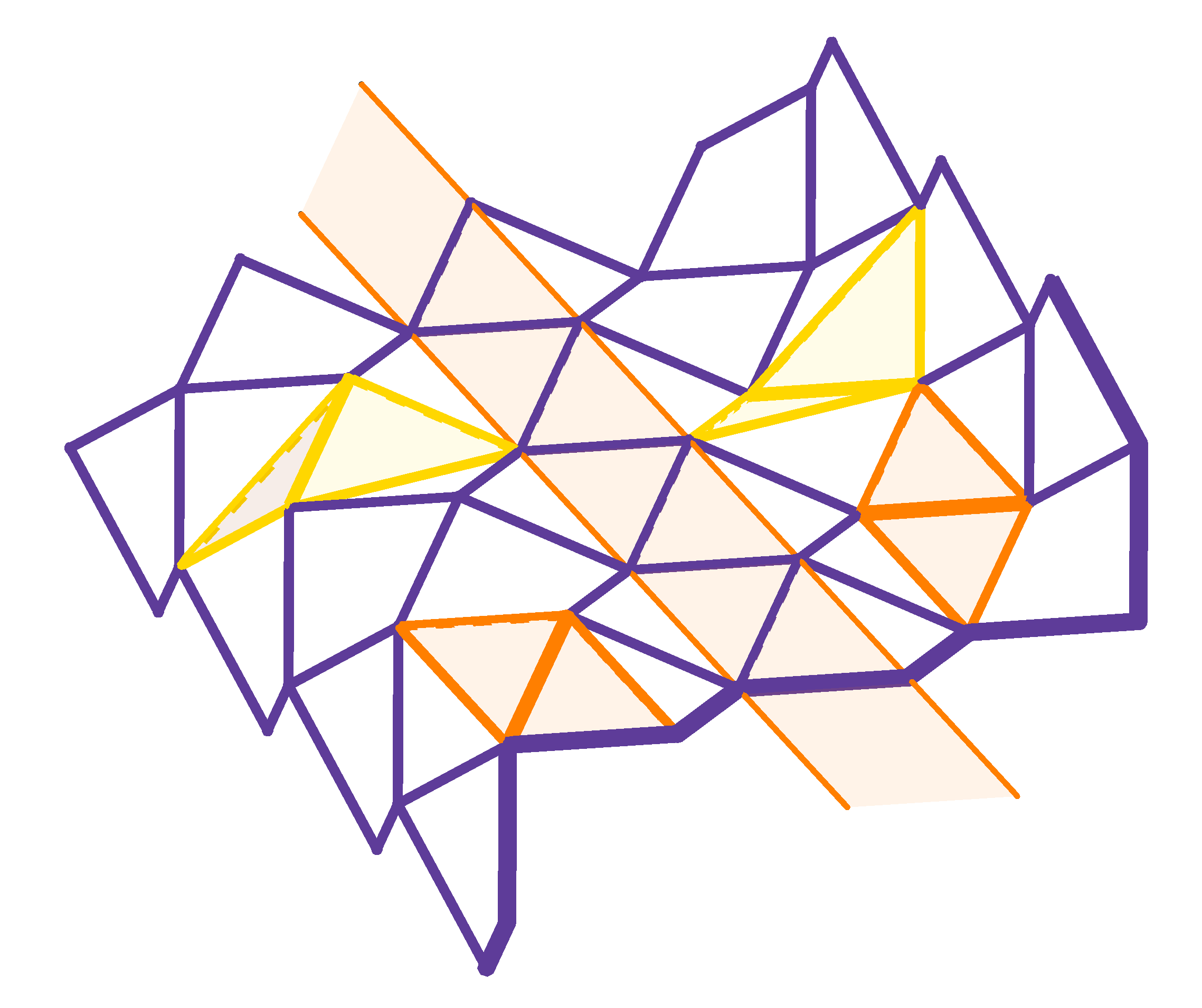}
		 
		\label{fig:ex gs}
	\end{subfigure}
	\hfill
	\begin{subfigure}[t]{0.45\textwidth}
		\centering
	\def\svgwidth{0.8 \columnwidth}
\begingroup%
  \makeatletter%
  \providecommand\color[2][]{%
    \errmessage{(Inkscape) Color is used for the text in Inkscape, but the package 'color.sty' is not loaded}%
    \renewcommand\color[2][]{}%
  }%
  \providecommand\transparent[1]{%
    \errmessage{(Inkscape) Transparency is used (non-zero) for the text in Inkscape, but the package 'transparent.sty' is not loaded}%
    \renewcommand\transparent[1]{}%
  }%
  \providecommand\rotatebox[2]{#2}%
  \newcommand*\fsize{\dimexpr\f@size pt\relax}%
  \newcommand*\lineheight[1]{\fontsize{\fsize}{#1\fsize}\selectfont}%
  \ifx\svgwidth\undefined%
    \setlength{\unitlength}{264.0566424bp}%
    \ifx\svgscale\undefined%
      \relax%
    \else%
      \setlength{\unitlength}{\unitlength * \real{\svgscale}}%
    \fi%
  \else%
    \setlength{\unitlength}{\svgwidth}%
  \fi%
  \global\let\svgwidth\undefined%
  \global\let\svgscale\undefined%
  \makeatother%
  \begin{picture}(1,0.80450403)%
    \lineheight{1}%
    \setlength\tabcolsep{0pt}%
    \put(0,0){\includegraphics[width=\unitlength,page=1]{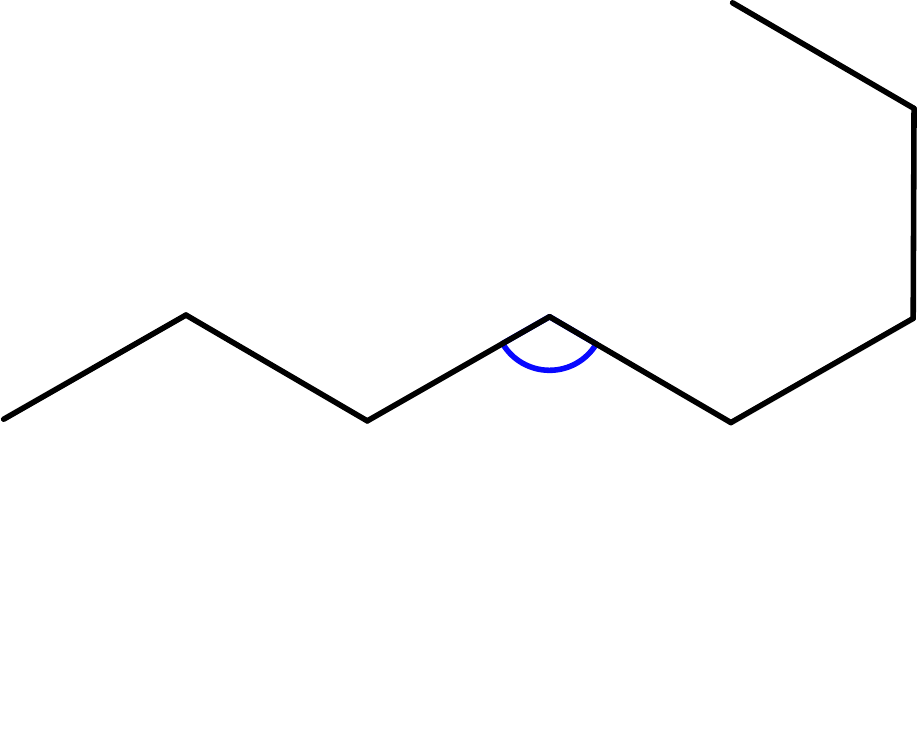}}%
    \put(0.52669954,0.34148694){\color[rgb]{0,0,1}\makebox(0,0)[lt]{\lineheight{1.25}\smash{\begin{tabular}[t]{l}$\kappa^*$\end{tabular}}}}%
    \put(0,0){\includegraphics[width=\unitlength,page=2]{1-dim_2.pdf}}%
  \end{picture}%
\endgroup%

		\label{fig: 2-dim def}
	\end{subfigure}
	\caption{ An  admissible configuration 
	(left).  Since the form function is constant along the
	diagonal $d_1$,  as indicated by the orange area, the same
	pattern repeats periodically and all necessary information is
	contained in one cross section as shown on the right. The
	angle $\kappa^*$ is defined in \eqref{eq:alpha}. The defining
	bond planes  for vertical (on the right) and horizontal
	(on the left) bonds  of  the  incidence angles $\gamma_1$ (orange) and $\gamma_2$ (yellow) are marked, indicating that $\gamma_1= 0 \neq \gamma_2$.}
	\label{fig:gs are one-dim}
\end{figure}

This theorem implies that ground states are essentially one-dimensional, in the sense that they can be characterized as two-dimensional deformations of one-dimensional chains, see Figure~\ref{fig:gs are one-dim}. Indeed, due to $\tau$ being constant along $d_1$, any cross section along $d_2$ contains the same information. In particular, admissible configurations  can be any combination of flat, rolled-up/down areas in relation to the fact that the incidence angle along $d_2$ can be $0$ (flat areas),  $-\gamma^*$ (rolled-up areas) or $+\gamma^*$ (rolled-down areas).

In the next subsections, we will present a refined version  of  Theorem~\ref{thm: main optimal cells}, namely Theorem~\ref{thm: main thm 4-tiles}. We will show that Theorem~\ref{thm: main thm 4-tiles}  below implies Theorem~\ref{thm: main optimal cells}. In Section~\ref{sec:coplanar} we then prove Theorem~\ref{thm: main thm 4-tiles}, which then also implies Theorem~\ref{thm: main optimal cells}.

\subsection{Geometry of optimal cells and construction of $4$-tiles}\label{sec: 4-tiles}

\begin{figure}[h]
	\centering
	\def\svgwidth{1 \columnwidth}
	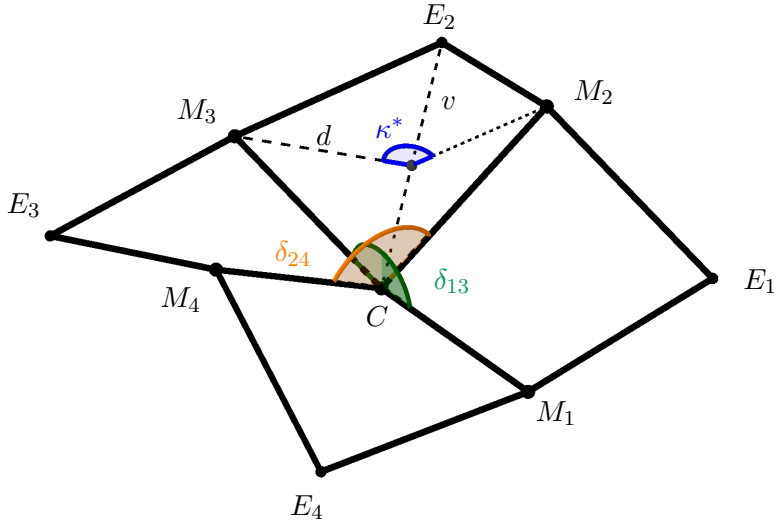

	\caption{Example of a $4$-tile with center $C$, middle points $M_1, \dots, M_4$ and corner points $E_1, \dots, E_4$. We have also indicated $v$, $d$, and $\kappa^*$ of the optimal cell $\{C,M_2, E_2, M_3\}$.}
	\label{fig:ex 4-tile}
\end{figure}

We aim at obtaining a complete  characterization of admissible configurations, by resorting to so-called \emph{$4$-tiles}. To introduce this concept, we first need to investigate the geometry of optimal cells in more detail. First, we consider an admissible  deformation $y$  and an optimal cell of the configuration,  consisting of the points $y_1,\ldots,y_4$ and the corresponding midpoints   $m_1 = (y_1 + y_3)/2$ and $m_2 = (y_2 + y_4)/2$,  as indicated in Figure~\ref{fig:types optimal cells}. We denote the length of the diagonal by $2v:= \vert y_1 - y_3 \vert = \vert y_2 - y_4 \vert$. By the cosine rule  we have  
\begin{equation}\label{eq: def of v}
	v = \sqrt{ (1-\cos\theta)/2 }.
\end{equation} 
Setting  $d := \vert y_1 - m_2 \vert = \vert y_3 - m_2 \vert = \vert y_2 - m_1 \vert = \vert y_4 - m_1 \vert$, we obtain by   Pythagoras' theorem $d  =  \sqrt{1-v^2}=   \sqrt{(1+\cos\theta)/2}$. This allows us to calculate the \emph{kink angle} $\kappa^*$ of an optimal cell by
\begin{equation}\label{eq:alpha}
	\kappa^* = \pi - 2 \kappa, \quad \quad  \text{where   $\kappa:=\arccos(v/d)  = \arctan(h/v)$,}
\end{equation} 
 with $h = \sqrt{1-2v^2}$, see also Lemma~\ref{prop: ref}. 
We refer to Figure~\ref{fig:ex 4-tile} with the optimal cell formed by $\lbrace C,M_2,E_2,M_3 \rbrace$ for an illustration.  For $\theta = \pi/2$ we have $v/d = 1$, and thus $\kappa^* = \pi$. In this case, as expected, optimal cells are flat. Let us firstly observe that an optimal cell is uniquely determined by the coordinates of three points and the choice of the cell form.  

\begin{lemma}[Optimal cell]\label{lem:fourthpoint}
	Given any three points  $y_1, y_2, y_4 \in \R^3$  of an optimal cell, i.e., points satisfying $\vert y_1 - y_4 \vert = \vert y_1 - y_2 \vert = 1$ and $\measuredangle y_4 y_1 y_2 = \theta$, there exists a unique fourth point  $y^\diagdown_3$ and $y_3^\diagup$,  respectively, such that $\{y_1, y_2, y_3^\diagdown, y_4\}$ is optimal of form~$\diagdown$ and $\{y_1, y_2, y_3^\diagup, y_4\}$ is optimal of form~$\diagup$.  
\end{lemma}

For the proof, we refer to Subsection~\ref{subsec:proof of lemma fourthpoint}. A priori, by prescribing only the common angle $\theta$, many configurations are conceivable as each optimal cell can be of  form $\diagdown$ or form $\diagup$, and neighboring cells can in principle be attached to each other with an arbitrary incidence angle.  Condition \eqref{eq: angles2} is therefore essential to reduce the number of admissible deformations. To take \eqref{eq: angles2} into account, we now consider  sub-configurations consisting of four optimal cells which are arranged in a square sharing one common point. Such structures are called \df{$4$-tiles}, and we refer to Figure~\ref{fig:ex 4-tile} for an illustration. 

 The  point shared by all four optimal cells  is called
 \df{center} and  is denoted  by $C$. The additional four points shared by two optimal cells are called \df{middle points} (as they are in the middle of the boundary of the $4$-tile), are denoted by $M_i$ for $i = 1,\dots,4$, and are labeled counter-clockwise such that 
\begin{align*}
	y^{-1}(M_1) -  y^{-1}(M_3) = 2e_1\quad \quad \text{ and }  \quad \quad \quad y^{-1}(M_2) -  y^{-1}(M_4) = 2e_2. 
\end{align*}
By construction, we have $\measuredangle M_i \, C \, M_{i+1} = \theta < \pi/2$ which implies that the five points $C$ and $(M_i)_{i=1}^4$ cannot be coplanar.  We introduce the \df{nonplanarity angles} $\delta_{13}$ and $\delta_{24}$  by 
\begin{align}\label{eq: delta1234}
	\delta_{13} := \measuredangle M_1 \, C \, M_3 \quad \text{  and } \quad \delta_{24}:=   \measuredangle M_2 \, C \, M_4. 
\end{align} 
Note that $|\delta_{13}- \pi|$ and $|\delta_{24}- \pi|$ indicate how far the five  points $C$ and  $(M_i)_{i=1}^4$ are from being coplanar, and again refer to Figure~\ref{fig:ex 4-tile} for an illustration.  The nonplanarity angles $\delta_{13}$ and $\delta_{24}$ are related by the following lemma.  

\begin{lemma}[Nonplanarity angles]\label{prop:delta bounded by 2 theta}
	The nonplanarity angles $\delta_{13}$ and $\delta_{24}$ satisfy 
\begin{equation}\label{eq:delta eta}
	\cos\big( \delta_{13} / 2\big)	\cos\big( \delta_{24}/2\big) = \cos\theta.
\end{equation}	
In particular, $\delta_{13}$ and $\delta_{24}$ coincide if and only if $\delta_{13} = \delta_{24} = \delta_\theta = 2 \arccos(\sqrt{\cos\theta})$. 
\end{lemma}

Indeed, by \eqref{eq: angles2} we always have $\delta_{13} = \delta_{24}$ for every $4$-tile of an  admissible configurations since $M_1,\, C,\, M_3$ and $M_2, \, C,\, M_4$ fulfill the condition in \eqref{eq: angles2}. This yields that $\delta_\theta =  2\arccos(\sqrt{\cos\theta})$ is solely determined by $\theta$.  The proof relies on the geometry of optimal cells, i.e.,  on  assumptions \eqref{eq: dist} and \eqref{eq: angles},  and  will be given in Subsection~\ref{subsec:proof of lemma fourthpoint}.  

We denote the four corner points of the $4$-tile by $E_i$, $i = 1,\dots,4$, as indicated in Figure~\ref{fig:ex 4-tile}. For the classification of all different $4$-tiles, it is convenient to frame $4$-tiles in a \df{reference position}, as given in the following proposition.

\begin{lemma}[Reference position]\label{prop: ref} 
{\rm (i)} By applying a suitable isometry, every $4$-tile can be positioned in such a way that the center $C$ coincides with the origin,  and  we have
$$M_1 = (s,0,\varsigma h), \quad M_2 = (0,s,\varsigma h), \quad M_3 = (-s,0,\varsigma h), \quad M_4 = (0,-s,\varsigma h),$$
where $s = \sqrt{2}v$ (see \eqref{eq: def of v}),  $h = \sqrt{1-2v^2}$, and $\varsigma  \in  \lbrace -1,1 \rbrace$.

{\rm (ii)} Fixing  $\varsigma \in \lbrace -1,1 \rbrace$, and the form of each of the four optimal cells, the positions of $(M_i)_{i=1}^4$ and $(E_i)_{i=1}^4$ are uniquely  determined, up to isometry.  
\end{lemma}


For the proof, we again refer to Subsection~\ref{subsec:proof of lemma fourthpoint}.  Lemma \ref{prop: ref} entails that the middle points $(M_i)_{i = 1}^4$ are coplanar.  For this reason, we call $4$-tiles \emph{coplanar} in the following.  By \eqref{eq: angles2} coplanarity is a necessary condition for the admissibility of $4$-tiles. 
 
\begin{figure}[!h]
	\centering
	\begin{subfigure}[b]{0.3\textwidth}
		\centering
		\includegraphics[width=\textwidth]{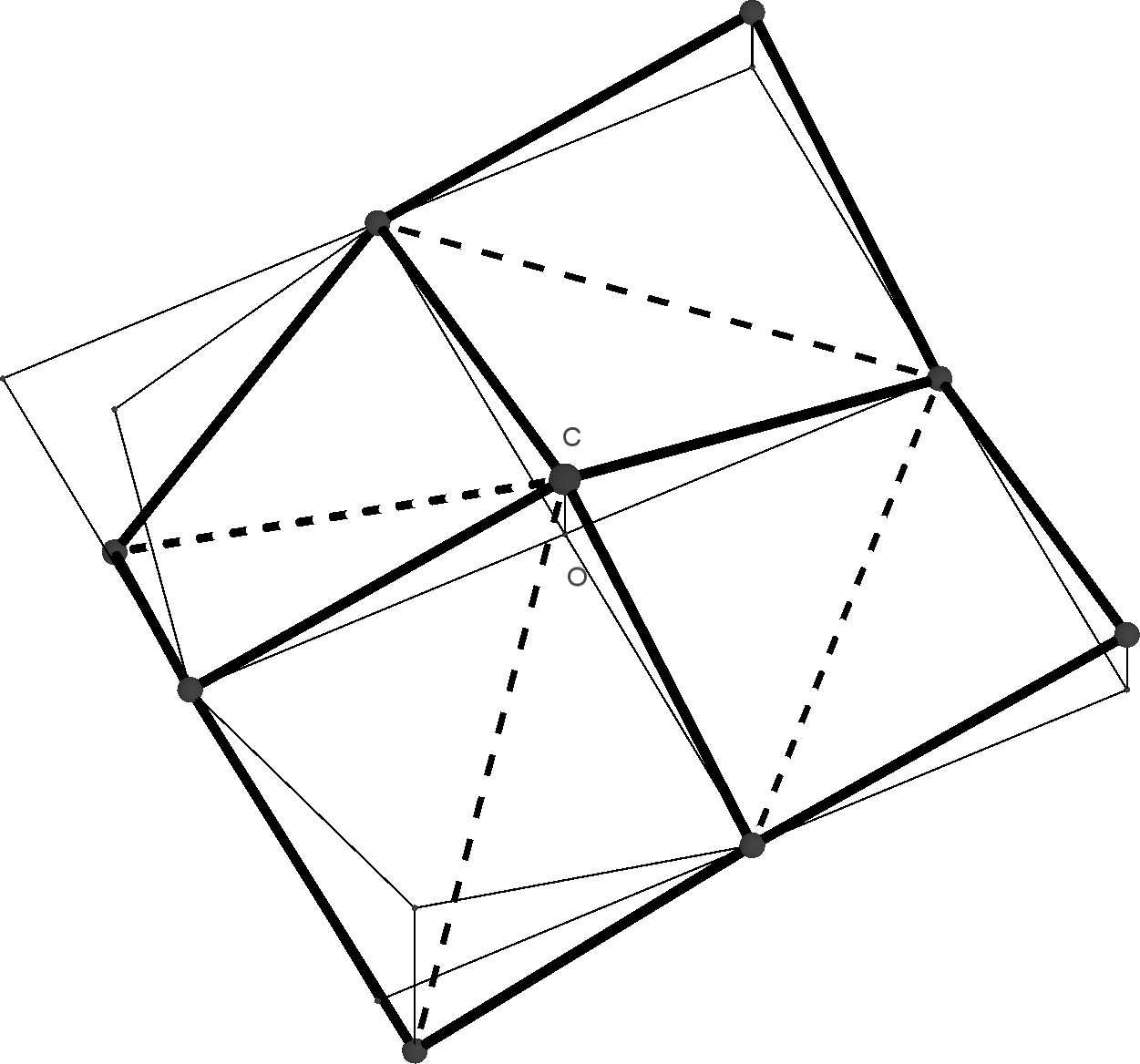}
		\caption{A-tile, $\mborder{\tileU{\diagdown}{\diagdown}{\diagup}{\diagup}}$}
		\label{fig:A1}
	\end{subfigure}
	\hfill
	\begin{subfigure}[b]{0.3\textwidth}
		\centering
		\includegraphics[width=\textwidth]{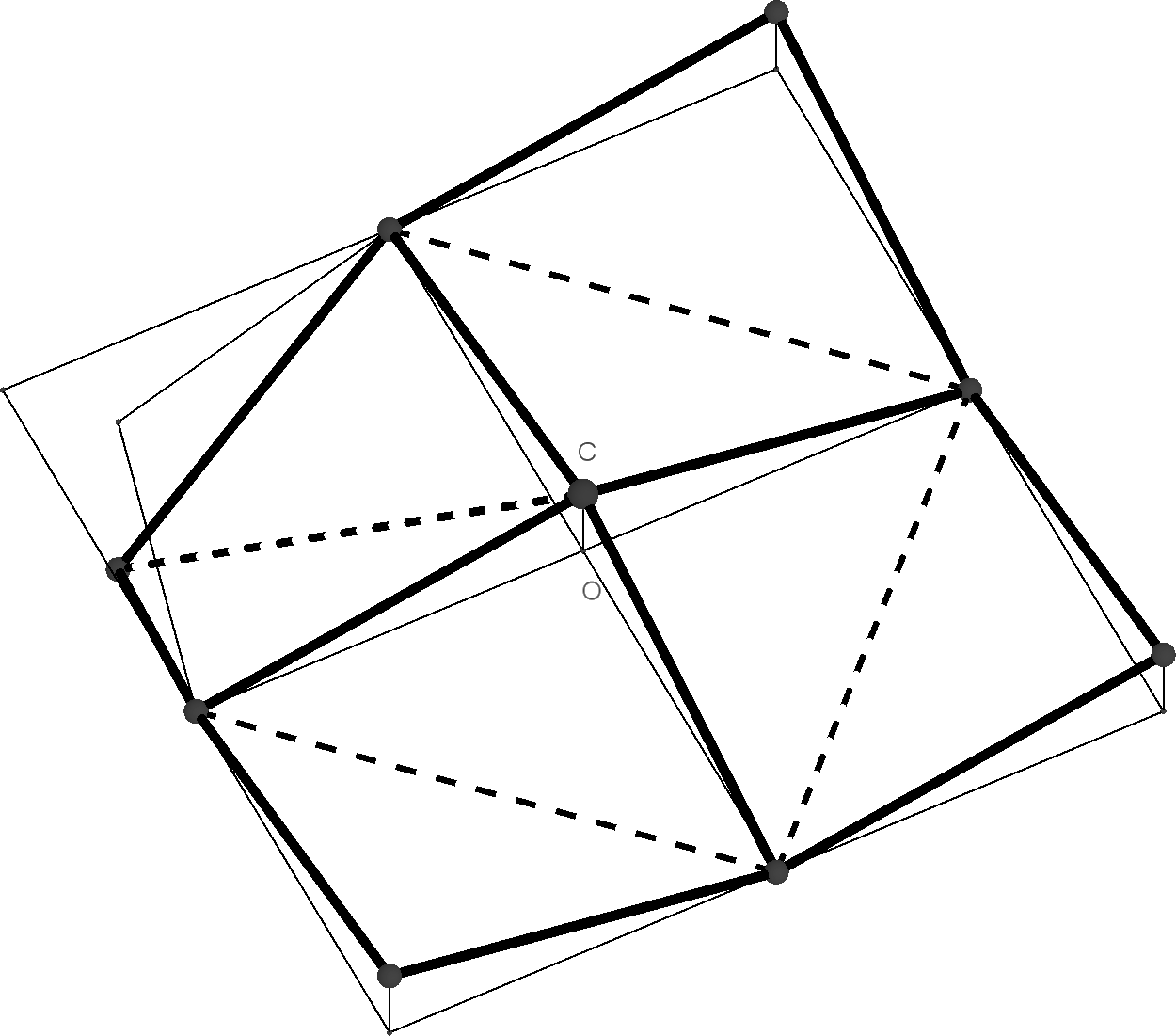}
		\caption{I-tile, $\mborder{\tileU{\diagdown}{\diagdown}{\diagdown}{\diagup}}$}
		\label{fig:I1}
	\end{subfigure}
	\hfill
	\begin{subfigure}[b]{0.3\textwidth}
		\centering
		\includegraphics[width=\textwidth]{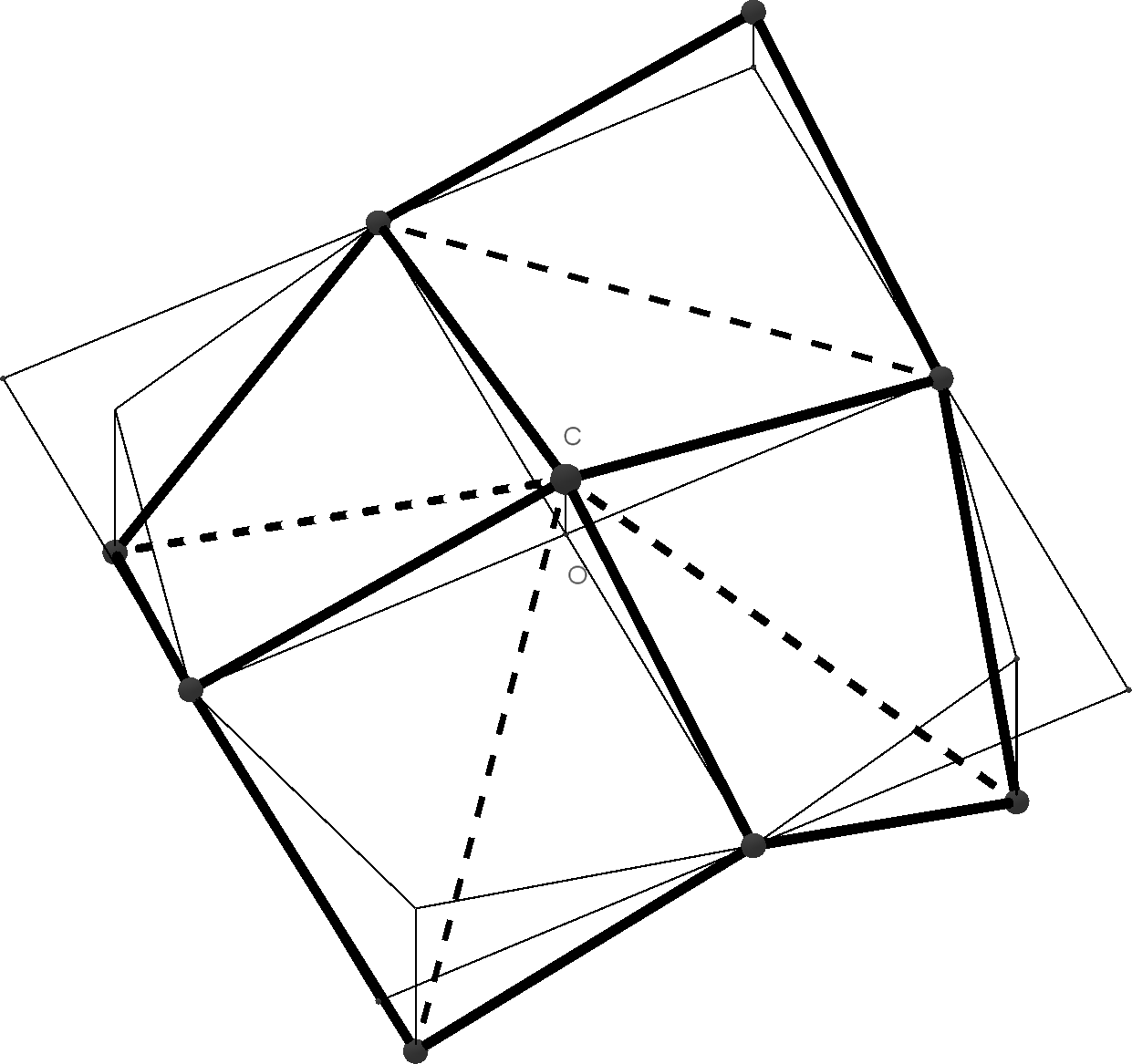}
		\caption{J-tile, $\mborder{\tileU{\diagdown}{\diagdown}{\diagup}{\diagdown}}$}
		\label{fig:J1}
	\end{subfigure}
	
	\begin{subfigure}[b]{0.3\textwidth}
		\centering
		\includegraphics[width=\textwidth]{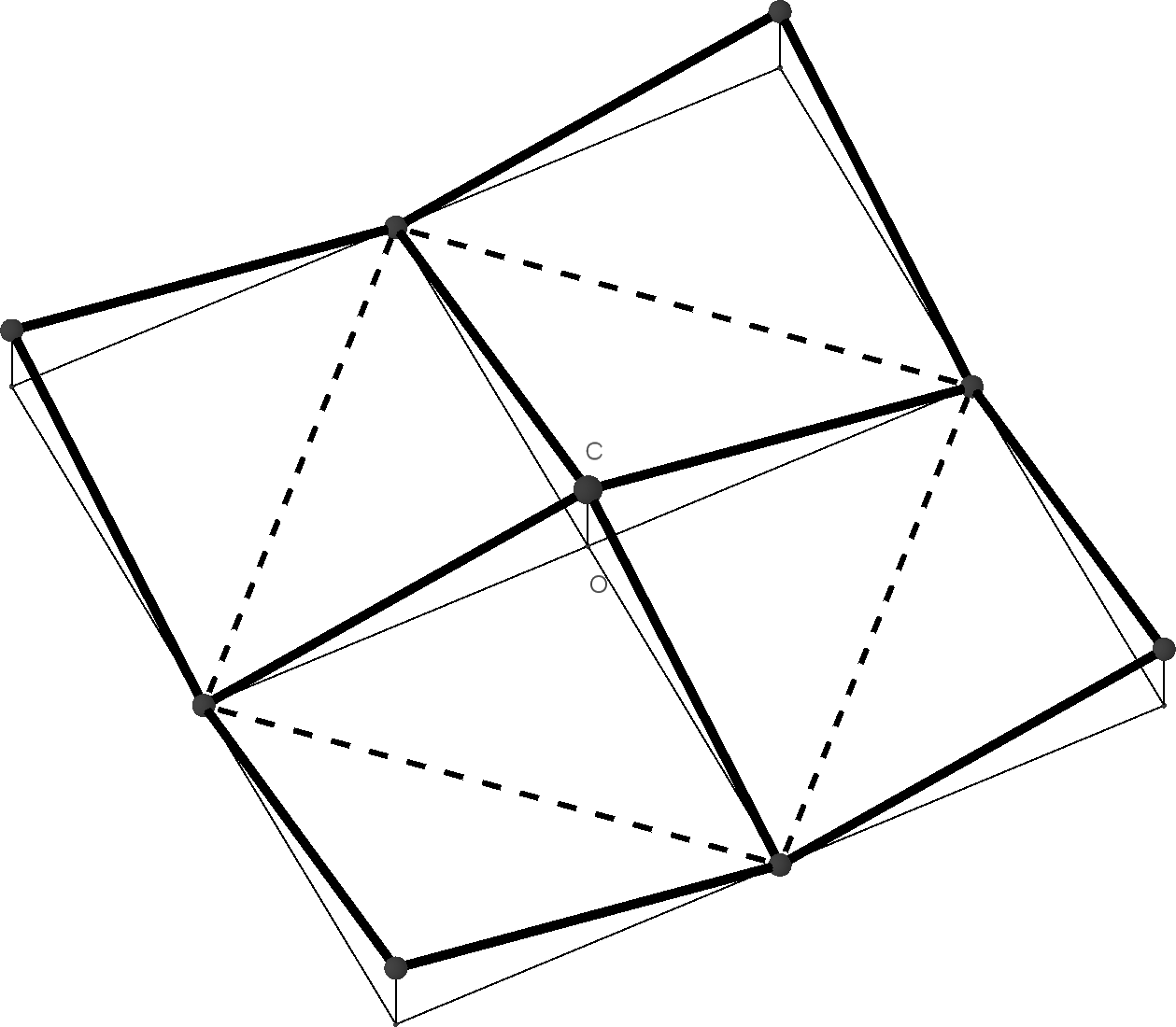}
		\caption{Z-tile, $\mborder{\tileU{\diagup}{\diagdown}{\diagdown}{\diagup}}$}
		\label{fig:Z}
	\end{subfigure}
	\hfill
	\begin{subfigure}[b]{0.3\textwidth}
		\centering
		\includegraphics[width=\textwidth]{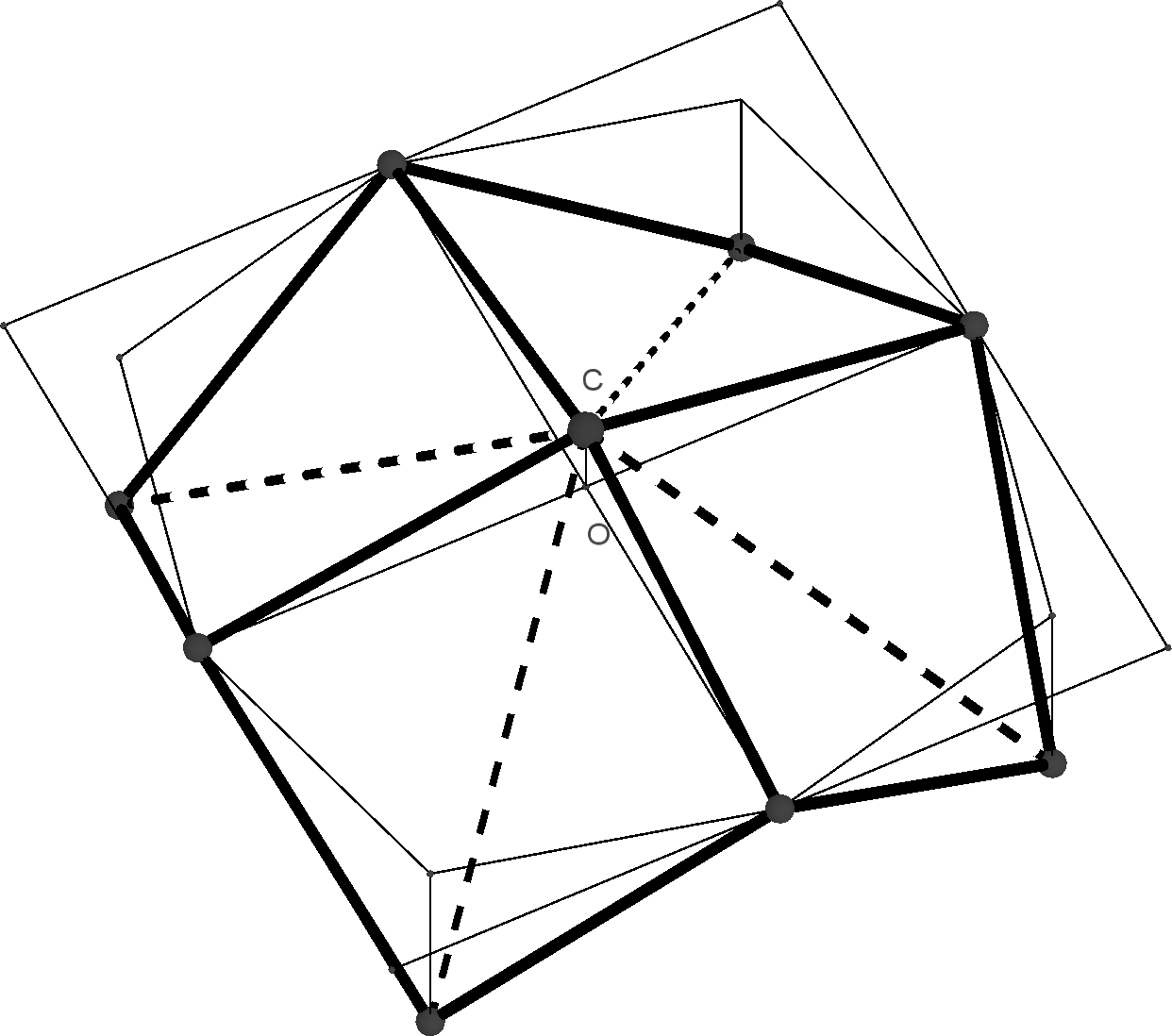}
		\caption{E-tile, $\mborder{\tileU{\diagdown}{\diagup}{\diagup}{\diagdown}}$}
		\label{fig:E}
	\end{subfigure}
	\hfill
	\begin{subfigure}[b]{0.3\textwidth}
		\centering
		\includegraphics[width=\textwidth]{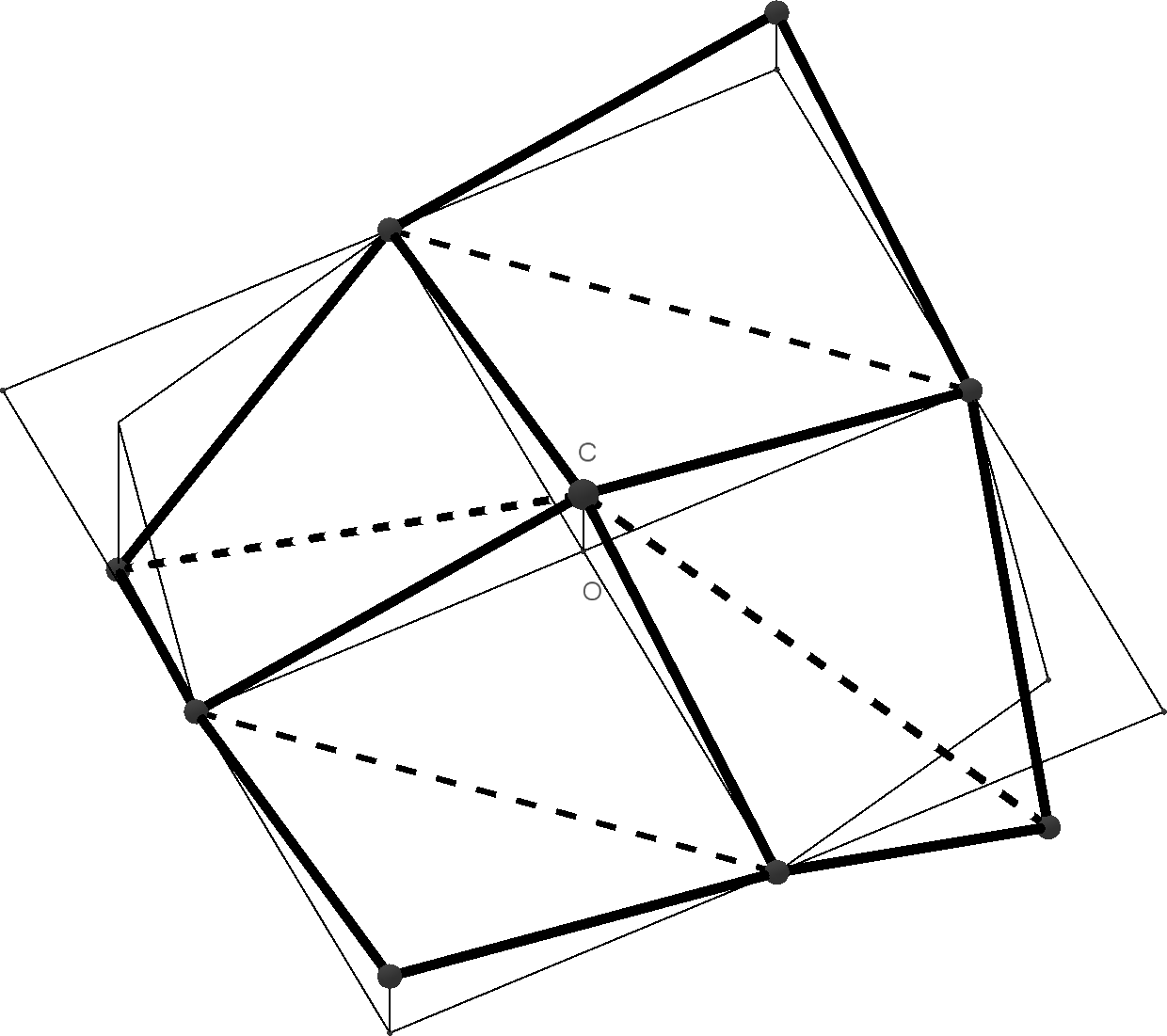}
		\caption{D-tile, $\mborder{\tileU{\diagdown}{\diagdown}{\diagdown}{\diagdown}}$}
		\label{fig:D1}
	\end{subfigure}
	\caption{ Representative $4$-tiles of each class.}
	\label{fig:4-tiles}
\end{figure}

In view of Lemma \ref{prop: ref}(ii), there are $32$ different \emph{types} of $4$-tiles. Indeed, there are $2^4 = 16$ possibilities to distribute either a  form  $\diagdown$ or a  form  $\diagup$  optimal cell to the four positions of a $4$-tile. Additionally, one can do this construction for $\varsigma =1$ or $\varsigma =-1$. As we show next, the different types can be classified into six \emph{classes} which are invariant under rotation by $\pi/2$ and reflection  along the $e_1$-$e_2$-plane,  see Table~\ref{tab:full classification}. A representative of each class is shown in Figure~\ref{fig:4-tiles}. The names of the classes are inspired by their geometry: the I-tile is intermediate between the zigzag-shaped Z-tile and the diagonally rolled-up D-tile (cf.\ the example in \eqref{eq: exgraz}). Similarly, the J-tile joins the arrowhead-shaped A-tile with the E-tile,  whose periodic pattern resembles to egg cartons. 

To denote a $4$-tile we use a matrix-like notation, where the  form   of the optimal cell in the square is represented  by   $\diagdown$ or $\diagup$  in the respective position in the matrix.  The case of $\varsigma =-1$ is indicated with a  $+$-symbol in the center of the matrix, and $\varsigma =1 $ is denoted with a $-$-symbol. We use this notation since, given a $4$-tile in reference position, we have that for $i=1,\ldots,4$  the center satisfies   $(C - M_i) \cdot e_3 > 0$ if $\varsigma =-1$ (e.g.\ in Figure~\ref{fig:Z}) and $(C - M_i) \cdot e_3 < 0$ if $\varsigma =1$ (e.g.\ in Figure~\ref{fig:ex 4-tile}), see Lemma~\ref{prop: ref}(i). 

Reflection of a $4$-tile in reference position with respect to the $e_1$-$e_2$-plane interchanges the index $+$ with $-$. Moreover,  $\diagup$ and $\diagdown$ are exchanged, as observed in  Subsection~\ref{sec: opti-thm}.  Also a rotation by $\pi/2$ interchanges the forms of the optimal cells, i.e., swaps $\diagdown$ and $\diagup$, see again Subsection~\ref{sec: opti-thm}. In addition, note that, by applying a $\pi/2$ rotation, one needs to permute the entries of the matrix accordingly, e.g., 
 $$A \mapsto A^T \left( \begin{array}{cc}
0 & 1 \\ 
1 & 0
\end{array} \right)$$
for a clockwise rotation of the entries. 

\begin{table}[h]
\centering
\renewcommand{\arraystretch}{2}
\begin{tabular}{ccc}
	\hline 
	A-tile & $\mborder{\tileU{\diagdown}{\diagdown}{\diagup}{\diagup}}$, $\mborder{\tileU{\diagup}{\diagdown}{\diagup}{\diagdown}}$, $\mborder{\tileU{\diagup}{\diagup}{\diagdown}{\diagdown}}$, $\mborder{\tileU{\diagdown}{\diagup}{\diagdown}{\diagup}}$, $\mborder{\tileD{\diagdown}{\diagdown}{\diagup}{\diagup}}$, $\mborder{\tileD{\diagup}{\diagdown}{\diagup}{\diagdown}}$, $\mborder{\tileD{\diagup}{\diagup}{\diagdown}{\diagdown}}$, $\mborder{\tileD{\diagdown}{\diagup}{\diagdown}{\diagup}}$  & Figure~\ref{fig:A1}  \\ 
	\hline 
	I-tile & $\mborder{\tileU{\diagdown}{\diagdown}{\diagdown}{\diagup}}$, $\mborder{\tileU{\diagup}{\diagup}{\diagdown}{\diagup}}$, $\mborder{\tileU{\diagup}{\diagdown}{\diagdown}{\diagdown}}$, $\mborder{\tileU{\diagup}{\diagdown}{\diagup}{\diagup}}$, $\mborder{\tileD{\diagdown}{\diagup}{\diagup}{\diagup}}$, $\mborder{\tileD{\diagdown}{\diagup}{\diagdown}{\diagdown}}$, $\mborder{\tileD{\diagup}{\diagup}{\diagup}{\diagdown}}$, $\mborder{\tileD{\diagdown}{\diagdown}{\diagup}{\diagdown}}$ & Figure~\ref{fig:I1}  \\ 
	\hline 
	J-tile & $\mborder{\tileU{\diagdown}{\diagdown}{\diagup}{\diagdown}}$, $\mborder{\tileU{\diagdown}{\diagup}{\diagup}{\diagup}}$, $\mborder{\tileU{\diagdown}{\diagup}{\diagdown}{\diagdown}}$, $\mborder{\tileU{\diagup}{\diagup}{\diagup}{\diagdown}}$, $\mborder{\tileD{\diagup}{\diagup}{\diagdown}{\diagup}}$, $\mborder{\tileD{\diagup}{\diagdown}{\diagdown}{\diagdown}}$, $\mborder{\tileD{\diagup}{\diagup}{\diagdown}{\diagup}}$, $\mborder{\tileD{\diagdown}{\diagdown}{\diagdown}{\diagup}}$  & Figure~\ref{fig:J1} \\ 
	\hline 
	Z-tile & $\mborder{\tileU{\diagup}{\diagdown}{\diagdown}{\diagup}}$, $\mborder{\tileD{\diagdown}{\diagup}{\diagup}{\diagdown}}$ & Figure~\ref{fig:Z}  \\ 
	\hline 
	E-tile & $\mborder{\tileU{\diagdown}{\diagup}{\diagup}{\diagdown}}$, $\mborder{\tileD{\diagup}{\diagdown}{\diagdown}{\diagup}}$  & Figure~\ref{fig:E}  \\ 
	\hline 
	D-tile &  $\mborder{\tileU{\diagdown}{\diagdown}{\diagdown}{\diagdown}}$, $\mborder{\tileU{\diagup}{\diagup}{\diagup}{\diagup}}$, $\mborder{\tileD{\diagdown}{\diagdown}{\diagdown}{\diagdown}}$, $\mborder{\tileD{\diagup}{\diagup}{\diagup}{\diagup}}$ & Figure~\ref{fig:D1} \\ 
	\hline 
\end{tabular} 
	\renewcommand{\arraystretch}{1}
	\caption{Full classification of all possible $4$-tiles.}
	\label{tab:full classification}
\end{table}

As an example, rotation leaves the $4$-tile {\tiny $\mborder{\tileU{\DU}{\DD}{\DD}{\DU}}$ } invariant, as interchanging $\diagup$ and $\diagdown$ yields {\tiny $\mborder{\tileU{\DD}{\DU}{\DU}{\DD}}$}  and the rotation of the entries then leads to {\tiny $\mborder{\tileU{\DU}{\DD}{\DD}{\DU}}$}.  However, rotating {\tiny $\mborder{\tileU{\diagdown}{\diagdown}{\diagdown}{\diagup}}$} clockwise, i.e., first swapping $\diagdown$ and $\diagup$ to obtain {\tiny $\mborder{\tileU{\diagup}{\diagup}{\diagup}{\diagdown}}$} and then rotating the entries to {\tiny $\mborder{\tileU{\diagup}{\diagup}{\diagdown}{\diagup}}$}, yields an $4$-tile of the same class, but with different type, see Table~\ref{tab:full classification}.

\subsection{Boundary orientation and boundary angles}\label{subsec:attaching of 4-tiles}

In this subsection, we further refine the characterization of
$4$-tiles by introducing a notion of boundary orientation. To this
end, consider a $4$-tile  with notation as indicated in
Figure~\ref{fig:ex 4-tile},   placed  in reference position.  We call three points $E_{i-1}$, $M_i,$ and  $E_i$, and the two bonds in between a \df{boundary} of the $4$-tile, where the indices have again to be understood modulo $4$. We define the \emph{boundary orientation} of  $E_{i-1} \, M_i \, E_i$ by 
\begin{align}\label{eq: boundary orientation}
	\ori(E_{i-1} \, M_i \, E_i) := \left \lbrace \begin{array}{ccc}
		 \wedge   & \text{if} &    (E_i +E_{i-1})\cdot e_3 /2  <   M_i \cdot e_3 , \\ 
		 \vee  & \text{if} & (E_i +E_{i-1})\cdot e_3 /2 >  M_i \cdot e_3 ,
	\end{array} \right.
\end{align} 
and  the corresponding \emph{boundary angle} by 
\begin{align}\label{eq: boundary angle}
	 \measuredangle E_i \,  M_i   \, E_{i-1}.
\end{align}
Intuitively, the orientation describes the fact that the boundary points upwards (orientation $\wedge$) or downwards (orientation $\vee$), see Figure~\ref{fig:4-tiles} for an illustration. Boundary orientation and boundary angle are crucial for classifying admissible configurations as   they provide   compatibility conditions for neighboring $4$-tiles. To formalize this, we now introduce the notion of \emph{attached $4$-tiles}.

Given two $4$-tiles $T$ and $\tilde{T}$ with centers $C$ and $\tilde{C}$, we say that the $4$-tiles are \emph{attached to each other} if $y^{-1}(C) - y^{-1}(\tilde{C}) \in \lbrace 2 e_1, -  2 e_1,  2 e_2 ,  -2 e_2 \rbrace$. Note that $T$ and $\tilde{T}$  share exactly one of the middle points $(M_i)_{i=1}^4$ and $(\tilde{M}_i)_{i=1}^4$ (and the  adjacent   two corner points). This shared middle point is the center of the so-called  \emph{middle $4$-tile} which is formed by two optimal cells of $T$ and two optimal cells of $\tilde{T}$.

The following result will be a key tool for the classification of admissible configurations.

\begin{lemma}[Attachment of two $4$-tiles]\label{prop: attach}
If two $4$-tiles  are attached to each other, the boundary angles and the boundary orientation at the shared boundary coincide. If the boundary orientation is $\wedge$, the corresponding middle $4$-tile satisfies $\varsigma = -1$ (see Lemma~\ref{prop: ref}(i)), otherwise we have $\varsigma = 1$.
\end{lemma}

Lemma \ref{prop: attach}  will be proved in Subsection~\ref{sec: bdy
  orient}. The statement delivers necessary conditions for attaching
two $4$-tiles. In fact,  a  crucial idea for proving the main theorem,
Theorem~\ref{thm: main thm 4-tiles},  is excluding many situations by
checking that boundary angles or boundary orientations do not
match. In particular, this reasoning will allow us to prove that
admissible configurations  \emph{exclusively}  contain Z-, D-, and I-tiles. To ease the readability, from now on we include the boundary orientation in the notation, at least for the relevant tiles, i.e., the Z-, D-, and I-tiles. This allows for an easy check whether the boundary orientations match or not. 

 On  lateral boundaries, we denote boundaries with orientation $\wedge$ by $<$. Likewise, lateral boundaries with boundary orientation $\vee$  are indicated by   $>$. Table~\ref{tab:adm copl 4-tile} gives an overview of admissible $4$-tiles with the new notation.  

\begin{table}[h]
	\begin{tabular}{c|c}
		Z-tile &  $\borderOOII{\tileU{\diagup}{\diagdown}{\diagdown}{\diagup}}$, $ \borderIIOO{\tileD{\diagdown}{\diagup}{\diagup}{\diagdown}}$ \\
		\hline
		D-tile &  $\borderIIOOnwsed{\tileU{\diagdown}{\diagdown}{\diagdown}{\diagdown}}$, $\borderIIOOneswd{\tileU{\diagup}{\diagup}{\diagup}{\diagup}}$, $\borderOOIInwseu{\tileD{\diagup}{\diagup}{\diagup}{\diagup}}$, $\borderOOIIneswu{\tileD{\diagdown}{\diagdown}{\diagdown}{\diagdown}}$   \\
		\hline
		I-tile &  $\borderOOOOnwd{\tileU{\diagdown}{\diagdown}{\diagdown}{\diagup}}$, $\borderIOIOned{\tileU{\diagup}{\diagup}{\diagdown}{\diagup}}$, $\borderIIIIsed{\tileU{\diagup}{\diagdown}{\diagdown}{\diagdown}}$, $\borderOIOIswd{\tileU{\diagup}{\diagdown}{\diagup}{\diagup}}$, $\borderOOOOseu{\tileD{\diagdown}{\diagup}{\diagup}{\diagup}}$,  $\borderIOIOswu{\tileD{\diagdown}{\diagup}{\diagdown}{\diagdown}}$, $\borderIIIInwu{\tileD{\diagup}{\diagup}{\diagup}{\diagdown}}$, $\borderOIOIneu{\tileD{\diagdown}{\diagdown}{\diagup}{\diagdown}}$
	\end{tabular}
	\caption{A table of all  admissible  Z-, D-, and I-tiles with corresponding boundary orientations.}
	\label{tab:adm copl 4-tile}
\end{table}

In the notation, we  also  denote corner points pointing downwards with  $\circ$ and corner points pointing upwards with  $\bullet$ (of course, always assuming that the $4$-tile is in reference position). As an example, we refer to (\textsc{b}) and (\textsc{f}) in Figure~\ref{fig:4-tiles} for {\tiny $\borderOOOOnwd{\tileU{\diagdown}{\diagdown}{\diagdown}{\diagup}}$} and {\tiny  $\borderIIOOnwsed{\tileU{\diagdown}{\diagdown}{\diagdown}{\diagdown}}$}, respectively. Note that this notation is not part of the characterization of types, but is included only to visualize the directions along which the boundary rolls up or down, respectively.  (In fact, a $+$ in the center along with $\diagup$ pointing towards $+$ yields $\circ$ in the corresponding corner. In a similar fashion, a $-$ in the center along with $\diagup$ \emph{not} pointing towards $-$ yields $\bullet$.) This notation facilitates to determine the class of the $4$-tile as Z-tiles have no $\bullet$/$\circ$, D-tiles have two, and I-tiles have exactly one.

\begin{lemma}[Boundary orientations]\label{lemma: bo}
	The boundary orientations of the different boundaries of the Z-, D-, and I-tiles are given as indicated in Table~\ref{tab:adm copl 4-tile}.
\end{lemma}

Lemma~\ref{lemma: bo} will be proved in Subsection~\ref{sec: bdy
  orient}. We close this subsection with an example  illustrating  Lemma~\ref{prop: attach}. Let us attach the Z-tile {\tiny $\borderOOII{\tileU{\diagup}{\diagdown}{\diagdown}{\diagup}} $} and the D-tile {\tiny $\borderOOIIneswu{\tileD{\diagdown}{\diagdown}{\diagdown}{\diagdown}}$}.  From the notation we can directly see that by attaching via
\begin{align}\label{eq: exgraz}
	{\small \borderOOII{\tileU{\diagup}{\diagdown}{\diagdown}{\diagup}} \, \borderOOIIneswu{\tileD{\diagdown}{\diagdown}{\diagdown}{\diagdown}}}, 
\end{align}
the boundary orientation match at the shared boundary, i.e., the
$4$-tiles can be attached to each other provided that also the
boundary angles coincide. (This indeed holds true, as we will see
later in Lemma~\ref{prop:boundary value}.) The type of the middle
$4$-tile can be determined directly by considering the forms of the
four optimal cells in the middle, i.e., {\tiny
  $\mborder{\tileD{\diagdown}{\diagdown}{\diagup}{\diagdown}}$}.  As
the shared boundary has orientation $>$, which corresponds to  $\vee$,
Lemma~\ref{prop: attach} implies that the middle $4$-tile satisfies
$\varsigma = 1$. The latter implies a $-$-symbol in the middle of the
matrix, see the discussion below Lemma \ref{prop: ref}. Therefore,
 the middle $4$-tile is  the I-tile {\tiny
  $\borderOIOIneu{\tileD{\diagdown}{\diagdown}{\diagup}{\diagdown}}$}. Clearly,
 the  procedure applies to all combinations of $4$-tiles.

\subsection{Main  result: Characterization in terms of $4$-tiles}\label{sec: thm-4-ti} 
After having introduced the necessary notation and concepts in the previous subsections, we are ready to formulate  our main result on the characterization of admissible configurations in terms  of $4$-tiles.  To this end, we need  a variant of the  form function, the so-called \emph{type functions}:   consider an admissible deformation $y$ and let $S_1 = 0$, $S_2 = (1,0)$, $S_3=(0,1)$, and $S_4 = (1,1)$. For $i=1,\ldots,4$, we let $\sigma_i$ be the function defined on $2\Z^2$ such that $\sigma_i(k,l)$ for $k,l \in 2\Z^2$ indicates the type of the $4$-cell with center $y(S_i + (k,l))$.  The four different functions account for the fact that there are four different possibilities to tessellate $\Z^2$ with $4$-tiles. With this definition at hand, we  now  state the main result of this paper. 

\begin{theorem}[Characterization of all admissible configurations]\label{thm: main thm 4-tiles}
A deformation $y$ is admissible if and only if, possibly up to rotation of the lattice $\Z^2$ by $\pi/2$, the following holds true: 

 Only particular   types of Z-, D-, and I-tiles  are admissible, namely, for $i=1,\ldots,4$ we have 
	\begin{align}\label{eq: typi}
	 \sigma_i \colon 2\Z^2 \ra \left\{\borderOOII{\tileU{\diagup}{\diagdown}{\diagdown}{\diagup}}, \borderIIOO{\tileD{\diagdown}{\diagup}{\diagup}{\diagdown}}, \borderOOIInwseu{\tileD{\diagup}{\diagup}{\diagup}{\diagup}}, \borderIIOOnwsed{\tileU{\diagdown}{\diagdown}{\diagdown}{\diagdown}}, \borderOOOOnwd{\tileU{\diagdown}{\diagdown}{\diagdown}{\diagup}}, \borderIIIIsed{\tileU{\diagup}{\diagdown}{\diagdown}{\diagdown}}, \borderOOOOseu{\tileD{\diagdown}{\diagup}{\diagup}{\diagup}}, \borderIIIInwu{\tileD{\diagup}{\diagup}{\diagup}{\diagdown}} \right\}.
	\end{align}
Moreover, the type function is constant along $d_1$, i.e., $\sigma_i(s,t) = \sigma_i(s+2, t+2)$ for all $s,t \in 2\Z$ and the following \df{matching conditions} are satisfied:
	\begin{enumerate}
		\item[\rm (M1)] for all $s,t \in  2\Z $ we have 
	\begin{equation*}
				\begin{aligned}
					\sigma_i(s,t) \in &\left\{  \borderOOII{\tileU{\diagup}{\diagdown}{\diagdown}{\diagup}}, \borderOOIInwseu{\tileD{\diagup}{\diagup}{\diagup}{\diagup}}, \borderOOOOnwd{\tileU{\diagdown}{\diagdown}{\diagdown}{\diagup}}, \borderOOOOseu{\tileD{\diagdown}{\diagup}{\diagup}{\diagup}} \right\} \quad \iff \\				
					\sigma_i(s,t-2)   \in &\left\{  \borderOOII{\tileU{\diagup}{\diagdown}{\diagdown}{\diagup}}, \borderOOIInwseu{\tileD{\diagup}{\diagup}{\diagup}{\diagup}}, \borderIIIIsed{\tileU{\diagup}{\diagdown}{\diagdown}{\diagdown}}, \borderIIIInwu{\tileD{\diagup}{\diagup}{\diagup}{\diagdown}} \right\},					
				\end{aligned}
			\end{equation*}
		\item[\rm (M2)] for all $s,t \in 2\Z$ we have
			\begin{equation*}
				\begin{aligned}
					\sigma_i(s,t) \in &\left \lbrace \borderIIOO{\tileD{\diagdown}{\diagup}{\diagup}{\diagdown}}, \borderIIOOnwsed{\tileU{\diagdown}{\diagdown}{\diagdown}{\diagdown}},\borderIIIIsed{\tileU{\diagup}{\diagdown}{\diagdown}{\diagdown}}, \borderIIIInwu{\tileD{\diagup}{\diagup}{\diagup}{\diagdown}} \right \rbrace \quad \iff \\
					\sigma_i(s,t-2) \in &\left \lbrace \borderIIOO{\tileD{\diagdown}{\diagup}{\diagup}{\diagdown}}, \borderIIOOnwsed{\tileU{\diagdown}{\diagdown}{\diagdown}{\diagdown}},\borderOOOOnwd{\tileU{\diagdown}{\diagdown}{\diagdown}{\diagup}}, \borderOOOOseu{\tileD{\diagdown}{\diagup}{\diagup}{\diagup}} \right	\rbrace.
				\end{aligned}
			\end{equation*}	
	\end{enumerate}
\end{theorem}

The theorem gives a \emph{complete characterization} of all admissible
configurations. First, it shows that only Z-, D-, and I-tiles are
admissible. More precisely, we see that only such D-, and I-tiles from
Table \ref{tab:adm copl 4-tile} are admissible, which roll-up/down
along the same diagonal, and that the type function is constant along
the other  diagonal. In particular, no change between the direction of
rolling-up/down is admissible. This observation allows for a clear
geometric interpretation: Z-tiles correspond to flat areas and D-tiles
induce rolled-up/down areas. In order to match such $4$-tiles, the
I-tile arises naturally as a combination of the Z-tile and
D-tile. (See, e.g., Figure~\ref{fig:I1}, which is a D-tile left and a
Z-tile right.  See also  the example in \eqref{eq: exgraz}.) Clearly,
rolling-up/down exclusively along the other diagonal is admissible as
well, corresponding exactly to the other collection of D-, and I-tiles
from Table \ref{tab:adm copl 4-tile}. However, after a rotation of the
lattice $\Z^2$ by $\pi/2$, one can always  reduce to  \eqref{eq: typi}. Eventually, the matching conditions (M1) and (M2) further restrict the admissible combination of $4$-tiles, and account for the fact that the boundary orientations at shared boundaries of two attached $4$-tiles need to match, see Lemma~\ref{prop: attach}. We close this discussion by noting that the characterization cannot be simplified further, i.e., there are indeed admissible configurations $y$ which contain all eight types given in \eqref{eq: typi}. 


Let us now stress that Theorem~\ref{thm: main thm 4-tiles} implies Theorem~\ref{thm: main optimal cells}.  To see this, we observe that the type functions $\sigma_i$, $i=1,\ldots,4$, are constant along the diagonal $d_1$. This along with the fact that all types in \eqref{eq: typi} have the same form of optimal cell ($\diagdown$ or $\diagup$) along the diagonal $d_1$ (i.e., in the lower left and upper right entry)  shows  that the form function $\tau$ introduced in Subsection~\ref{sec: opti-thm} satisfies $\tau(s,t) = \tau(s+1, t+1)$ for all $s,t \in \Z$.

The fact that  all incidence angles along $d_1$ vanish and that all incidence angles along $d_2$ lie in $\lbrace 0,\gamma^*,-\gamma^*\rbrace$  (with the property that the value is constant along $d_1$)  follows by an elementary computation. We defer the exact calculation to Appendix~\ref{sec:derivation of gamma}.  At this stage, we only mention that inside Z-tiles, all incidence angles along both diagonals are equal to zero. On the other hand, for the D-tile ${\tiny  \borderIIOOnwsed{\tileU{\diagdown}{\diagdown}{\diagdown}{\diagdown}}}$ the incidence angle along $d_2$ is $\gamma^*$  and for ${\tiny  \borderOOIInwseu{\tileD{\diagup}{\diagup}{\diagup}{\diagup}}}$ it is $-\gamma^*$.
I-tiles have incidence angles  $0$  and $\pm \gamma^*$, where the sign depends on $\bullet$ or $\circ$ in the notation.


 \section{The proof of the main theorem} \label{sec:coplanar}
 This section is devoted to the proof of Theorem~\ref{thm: main thm
   4-tiles}.   This hinges on two facts, namely, that (1)
 attaching  two $4$-tiles is only possible if the boundary
 orientation at shared boundaries match  and (2) that such
 attachment  needs to lead to an admissible,  i.e., coplanar
 middle $4$-tile.  Firstly,  we use these ideas to show that actually only {Z-}, {D-}, and I-tiles are admissible, see Proposition~\ref{prop: admi}. In a second step, we further restrict the set of  admissible types by showing that D- and I-tiles necessarily need to roll-up/down along the same diagonal, see Proposition~\ref{prop:no diff diags}. This is achieved by considering four $4$-tiles arranged in a square and exploiting the aforementioned compatibility conditions. With similar techniques, we subsequently show that along one diagonal the type has to be constant, see Proposition~\ref{prop:constant type}. Eventually, we provide another auxiliary result  (Proposition \ref{prop: admi2})  stating that four $4$-tiles arranged in a square can be indeed realized by  an admissible  configuration $y$ if all compatibility conditions, including the matching conditions stated in Theorem~\ref{thm: main thm 4-tiles}, are satisfied. With these results at hand, we are then able to prove Theorem~\ref{thm: main thm 4-tiles}.

\subsection{Admissible classes of $4$-tiles}
In this subsection, we show that admissible configurations contain only   Z-, D-, and I-tiles and that pairs of such tiles can  be attached. This is achieved in two steps. We start by calculating the different boundary angles introduced in \eqref{eq: boundary angle}. Then, by  discussing the possibility of attaching two $4$-tiles along a boundary with the same boundary angle and the same boundary orientation, see \eqref{eq: boundary orientation}, we are able to show that Z-, D-, and I-tiles are admissible, while E-, A-, and J-tiles are not.

We start by observing that there are exactly three different boundary types. In view of Lemma~\ref{prop: ref}, we see that the three points forming a boundary (e.g., $E_{i-1}$, $M_i$, and $E_i$, see Figure~\ref{fig:ex 4-tile}) are completely characterized by $\varsigma \in \lbrace -1,1\rbrace$ and the form, i.e., form $\diagdown$ or form $\diagup$, of the two optimal cells adjacent to the boundary. (Strictly speaking, in Lemma~\ref{prop: ref}(ii), this was only shown once the forms of all four optimal cells are fixed, but the argument clearly localizes at each boundary.)

This leads to at most $2^3 = 8$ different boundary types, as indicated in Table~\ref{tab:classification of boundaries}. Given a $4$-tile in reference position, the boundary  type  remains invariant under reflection of the $4$-tile along the $e_1$-$e_2$-plane and the $e_2$-$e_3$-plane. This shows that the number of different boundary types  reduces to three. We indicate the corresponding boundaries as \emph{Z-, D-, and  E-boundaries}, respectively, as the corresponding $4$-tiles have exclusively such boundaries, compare also Table~\ref{tab:classification of boundaries}  with Table~\ref{tab:full classification}.   We also mention that I-tiles have both Z- and D-boundaries and that J- and A-tiles contain E-boundaries.

\begin{table}[h]
	  \centering
	  \begin{tabular}{ccc}
		  Z-boundary & D-boundary & E-boundary \\
		  \hline
		  	$\mborder{\tileU{\ph}{\ph}{\diagdown}{\diagup}}$,$\mborder{\tileD{\ph}{\ph}{\diagup}{\diagdown}}$ & $\mborder{\tileU{\ph}{\ph}{\diagup}{\diagup}}$, $\mborder{\tileU{\ph}{\ph}{\diagdown}{\diagdown}}$,$\mborder{\tileD{\ph}{\ph}{\diagup}{\diagup}}$, $\mborder{\tileD{\ph}{\ph}{\diagdown}{\diagdown}}$ & $\mborder{\tileU{\ph}{\ph}{\diagup}{\diagdown}}$,$\mborder{\tileD{\ph}{\ph}{\diagdown}{\diagup}}$   
		  \end{tabular}
	  \smallskip
	  \caption{Classification of the three types of boundaries.}
	  \label{tab:classification of boundaries}
\end{table}

\begin{lemma}[Boundary angles]\label{prop:boundary value}
	The Z-boundary angle and D-boundary angle of coplanar $4$-tiles are given by $\delta_\theta = 2\arccos\left(\sqrt{\cos\theta}\right)$. The E-boundary angle of coplanar $4$-tiles is strictly smaller than $\delta_\theta$.  
\end{lemma}
\begin{proof}
	We start by considering the Z-boundary angle. Without restriction we consider a Z-tile in reference position with notation as indicated in Figure~\ref{fig:ex 4-tile}, satisfying $M_2 = (0,s,h)$ for $s,h>0$, where $s$ and $h$ are given in Lemma~\ref{prop: ref}.    We observe that the isometry $x = (x_1,x_2,x_3) \mapsto (x_1,x_2,-x_3) +  (0,s,h) $ maps $M_1$ to $E_1$, $C$ to $M_2$, and $M_3$ to $E_2$. This yields that the Z-boundary angle coincides with $\delta_\theta$,  see   \eqref{eq: angles2} and  \eqref{eq: delta1234}.  The fact that the D-boundary angle coincides with the Z-boundary angle is postponed to Corollary~\ref{cor: D bond}, and relies on the fact that two $4$-tiles with the respective boundaries can be attached to each other, cf.\ Lemma~\ref{lem:attachment algorithm}.

 Eventually, we show that the E-boundary angle is strictly smaller. To this end, we let  $E_1=  (s,s,0)$,  $M_2= (0,s,h) $, $E_2 = (-s,s,0)  $ be again the points of the Z-tile considered above. The corresponding points of an E-tile in reference position are denoted by $\tilde{E}_1$, $\tilde{M}_2$, and $\tilde{E}_2$.  (They are obtained by changing the form of the optimal cells containing $E_1$ and $E_2$, respectively.)  By simple geometric considerations we find
	\begin{align}\label{eq: tilde}
		\tilde{E}_1 = E_1 +  (-p,-p,q),  \quad \quad  \tilde{M}_2 = M_2, \quad \quad  \tilde{E}_2 = E_2 +  (p,-p,q)  
	\end{align}
	for some $p,q>0$. One can check that $q  =  \tilde{E}_1 \cdot e_3 = \tilde{E}_2 \cdot e_3   > 2h$, see   Lemma~\ref{lem:third point of optimal cell is high}(iv)  below. Given that $|\tilde{E}_1 - \tilde{M}_2| = |  \tilde{E}_2  - \tilde{M}_2| = 1$, the  E-boundary  angle is calculated by $\arccos ((\tilde{E}_1 - \tilde{M}_2) \cdot (\tilde{E}_2 - \tilde{M}_2))$. We now compute by using \eqref{eq: tilde} and $q > 2h$ that
\begin{align*}
	 (\tilde{E}_1 - \tilde{M}_2) \cdot (\tilde{E}_2 - \tilde{M}_2)  & = ({E}_1 - {M}_2) \cdot ({E}_2 - {M}_2)   \\ & \ \ \ \ +  \begin{pmatrix} -p \\ -p \\ q  \end{pmatrix} \cdot \begin{pmatrix} -s \\ 0 \\ -h \end{pmatrix}    + \begin{pmatrix} p \\ -p \\ q \end{pmatrix} \cdot \begin{pmatrix} s \\ 0 \\ -h \end{pmatrix}  +    \begin{pmatrix} -p \\ -p \\ q \end{pmatrix}   \cdot  \begin{pmatrix} p \\ -p \\ q \end{pmatrix}  \\
	&  = ({E}_1 - {M}_2) \cdot  ({E}_2  - {M}_2) + 2ps - 2qh + q^2 \\ & > ({E}_1 - {M}_2) \cdot  ({E}_2  - {M}_2). 
\end{align*}	
As $\delta_\theta = \arccos (({E}_1 - {M}_2) \cdot (  {E}_2  - {M}_2))$ and $\arccos$ is strictly decreasing on $[-1,1]$ we find that the E-boundary angle is smaller than $\delta_\theta$. This concludes the proof. 
\end{proof}

\begin{proposition}[Nonadmissible classes of  $4$-tiles]\label{prop: admi}
	An admissible  configuration does not contain E-, A-, and J-tiles. 
\end{proposition}	

\begin{proof}
Suppose by contradiction that the configuration contains a $4$-tile of class  E, A, or J. As each E-, A-, or J-tile contains at least one E-boundary, see Table~\ref{tab:classification of boundaries} and Table~\ref{tab:full classification}, by Lemma~\ref{prop: attach} and  Lemma~\ref{prop:boundary value} we deduce that the configuration contains at least two adjacent $4$-tiles in these three classes such that the shared boundary has an E-boundary angle. For the corresponding middle $4$-tile between the two $4$-tiles we thus get that the corresponding $\delta_{13}$ or $\delta_{24}$ as defined in \eqref{eq: delta1234}  coincides  with the E-boundary angle  which is strictly smaller than $\delta_\theta$ by Lemma~\ref{prop:boundary value}. On the other hand, by \eqref{eq: angles2} we have $\delta_{13} = \delta_{24} =\delta_\theta$  for the nonplanarity angles of the middle tile, a contradiction.   
\end{proof}

\subsection{Proof of the main result}

In this subsection we give the proof of Theorem~\ref{thm: main thm 4-tiles}. The argument rests upon two propositions, showing that only certain arrangements of Z-, D-,  and  I-tiles are admissible. A third auxiliary result verifies that such arrangements are indeed admissible. We start by stating these results, whose proofs are postponed to the next subsections.   Recall the notation of the $4$-tiles in Table~\ref{tab:adm copl 4-tile}.

\begin{proposition}[Roll-up/down along one diagonal]\label{prop:no diff diags}
 Consider any four adjacent $4$-tiles of class Z, D, or I of an admissible configuration  arranged in a square. Then all D- and I-tiles  locally  roll-up/down along the same diagonal, i.e., the four $4$-tiles are either all of type 
\begin{equation}\label{eq:class of admissible 4-tile A}
	\mathcal{A} := \left \lbrace \borderOOII{\tileU{\diagup}{\diagdown}{\diagdown}{\diagup}}, \borderIIOO{\tileD{\diagdown}{\diagup}{\diagup}{\diagdown}}, \borderOOIInwseu{\tileD{\diagup}{\diagup}{\diagup}{\diagup}}, \borderIIOOnwsed{\tileU{\diagdown}{\diagdown}{\diagdown}{\diagdown}}, \borderOOOOnwd{\tileU{\diagdown}{\diagdown}{\diagdown}{\diagup}}, \borderIIIIsed{\tileU{\diagup}{\diagdown}{\diagdown}{\diagdown}}, \borderOOOOseu{\tileD{\diagdown}{\diagup}{\diagup}{\diagup}}, \borderIIIInwu{\tileD{\diagup}{\diagup}{\diagup}{\diagdown}} \right \rbrace,
\end{equation}
or all of type 
\begin{equation}\label{eq:class of admissible 4-tile B}
	\mathcal{B} := \left\{  \borderOOII{\tileU{\diagup}{\diagdown}{\diagdown}{\diagup}}, \borderIIOO{\tileD{\diagdown}{\diagup}{\diagup}{\diagdown}}, \borderOOIIneswu{\tileD{\diagdown}{\diagdown}{\diagdown}{\diagdown}}, \borderIIOOneswd{\tileU{\diagup}{\diagup}{\diagup}{\diagup}}, \borderIOIOned{\tileU{\diagup}{\diagup}{\diagdown}{\diagup}}, \borderOIOIswd{\tileU{\diagup}{\diagdown}{\diagup}{\diagup}}, \borderIOIOswu{\tileD{\diagdown}{\diagup}{\diagdown}{\diagdown}}, \borderOIOIneu{\tileD{\diagdown}{\diagdown}{\diagup}{\diagdown}}   \right\}.
\end{equation} 
\end{proposition}

Note that $\mathcal{B}$ can be obtained from $\mathcal{A}$ through a rotation of the reference lattice by $\pi/2$, and vice versa. 
The proposition shows that locally only $4$-tiles which roll along the
same diagonal  can be attached to each other.   The following result states that locally admissible configurations have the same type along one of the diagonals. 

\begin{proposition}[Arrangements along diagonals]\label{prop:constant type}
Consider four adjacent $4$-tiles of an admissible configuration with types either in $\mathcal{A}$ or in $\mathcal{B}$, see \eqref{eq:class of admissible 4-tile A}--\eqref{eq:class of admissible 4-tile B}, arranged in a square and denoted by 
 $${\begin{aligned} 
		&\nborder{\mathfrak{A}}\nborder{\mathfrak{D}}\\
		&\nborder{\mathfrak{B}}\nborder{\mathfrak{C}}.
	\end{aligned}}$$ 
If the types are in $\mathcal{A}$, we have  $\mathfrak{B} = \mathfrak{D}$, and if the  types  are in $\mathcal{B}$, we have $\mathfrak{A} = \mathfrak{C}$. 
\end{proposition}


The previous two results yield restrictions for the arrangement of $4$-tiles in admissible configurations. The next  result shows  that such arrangements are indeed admissible.
\begin{proposition}[Admissible arrangements of $4$-tiles]\label{prop: admi2}
\noindent {\rm (i)} If two coplanar $4$-tiles in $\mathcal{A}$ are attached along a boundary with matching boundary orientation, the resulting middle $4$-tile is a coplanar $4$-tile in $\mathcal{A}$. \\
\noindent {\rm (ii)} If  four adjacent coplanar $4$-tiles with types in $\mathcal{A}$  are arranged as
\begin{align}\label{eq: orideri}
{\begin{aligned} 
		&\nborder{\mathfrak{A}}\nborder{\mathfrak{D}}\\
		&\nborder{\mathfrak{B}}\nborder{\mathfrak{C}}
	\end{aligned} }
	\end{align}
	such that  $\mathfrak{B} = \mathfrak{D}$ and such that the four $4$-tiles satisfy the matching conditions {\rm (M1)}--{\rm (M2)} stated in Theorem \ref{thm: main thm 4-tiles},  there exists an admissible deformation $y\colon \lbrace 0,1,2,3,4\rbrace^2 \to \R^3$ such that the $4$-tiles of $y(\lbrace 0,1,2,3,4\rbrace^2 )$  have the types indicated in    \eqref{eq: orideri}.  
 \end{proposition}

A similar statement  holds  for $4$-tiles with types in $\mathcal{B}$ by rotation of the reference lattice by $\pi/2$. We are now in a position to prove our main result. 

\begin{proof}[Proof of Theorem \ref{thm: main thm 4-tiles}]
\noindent \emph{Step 1: $\Rightarrow$}. We recall the definition of $\sigma_i$, $i=1,\ldots,4$, before the statement of Theorem~\ref{thm: main thm 4-tiles}. Without restriction we only consider $\sigma_1$ in the following proof. By  Proposition~\ref{prop: admi}  we have that the configuration only contains Z-, D-, and I-tiles. 

We next show that all types are either in $\mathcal{A}$ or in $\mathcal{B}$, see \eqref{eq:class of admissible 4-tile A}--\eqref{eq:class of admissible 4-tile B}, i.e., rolling up/down  occurs  at most along one diagonal. Assume by contradiction that there were two $4$-tiles rolling along different diagonals, i.e., $T_1 \in \mathcal{A} \setminus \mathcal{B}$ and $T_2 \in \mathcal{B} \setminus \mathcal{A}$. Choose $s_i,t_i \in 2\mathbb{Z}$, $i=1,2$, such that $ \sigma_1(s_1,t_1)  = T_1$ and $\sigma_1(s_2,t_2) = T_2$. By  Proposition \ref{prop:no diff diags} we can apply  Proposition~\ref{prop:constant type} and thus   find $\sigma_1(s_1+r,t_1+r) = T_1$ and  $\sigma_1(s_2+r',t_2-r') = T_2$ for all $r,r' \in 2\mathbb{Z}$. For a particular choice of $r$  and $r'$  this entails  $T_1 = T_2$  or that $T_1$ is adjacent to $T_2$.  In both cases, we obtain  a contradiction  to Proposition~\ref{prop:no diff diags}.  

This shows that all types of $4$-tiles are either in $\mathcal{A}$ or
$\mathcal{B}$. Up to a rotation of the reference lattice by $\pi/2$,
we may suppose that all types of $4$-tiles lie in $\mathcal{A}$, which
 corresponds to the notation of  Theorem~\ref{thm: main thm 4-tiles}.  By Proposition \ref{prop:constant type} we get that the type function is constant along $d_1$, i.e., $\sigma_i(s,t) = \sigma_i(s+2, t+2)$ for all $s,t \in 2\Z$ and all $i=1,\ldots,4$. 

It remains to show that the \emph{matching conditions} (M1) and (M2) hold true as indicated in the statement. These properties rely on the fact that the boundary orientations of each two attaching $4$-tiles need to match,  cf.\ Lemma~\ref{prop: attach}.   

We only prove matching condition (M1) as the proof for (M2) follows along similar lines. Taking any $4$-tile in {\tiny $\left \lbrace \borderOOII{\tileU{\diagup}{\diagdown}{\diagdown}{\diagup}}, \borderOOIInwseu{\tileD{\diagup}{\diagup}{\diagup}{\diagup}}, \borderIIIIsed{\tileU{\diagup}{\diagdown}{\diagdown}{\diagdown}}, \borderIIIInwu{\tileD{\diagup}{\diagup}{\diagup}{\diagdown}} \right \rbrace$} arranged along the diagonal,  we have one  of the two possibilities 
	\[ \begin{aligned}
			&{\tiny \borderDashedSSOO{\textcolor{white}{\tileU{\diagup}{\diagdown}{\diagdown}{\diagup}}}} &&{\tiny \borderOOII{\textcolor{white}{\tileU{\diagup}{\diagdown}{\diagdown}{\diagup}}}} \qquad \quad  &&{\tiny \borderDashedSSOO{\textcolor{white}{\tileU{\diagup}{\diagdown}{\diagdown}{\diagup}}}} &&{\tiny\borderIIII{\textcolor{white}{\tileU{\diagup}{\diagdown}{\diagdown}{\diagup}}}} \\
			&{\tiny \borderOOII{\textcolor{white}{\tileU{\diagup}{\diagdown}{\diagdown}{\diagup}}}} && \ph \qquad \quad && {\tiny \borderIIII{\textcolor{white}{\tileU{\diagup}{\diagdown}{\diagdown}{\diagup}}}} &&\textcolor{white}{\tileU{\diagup}{\diagdown}{\diagdown}{\diagup}}.
	\end{aligned}\]
	 The given boundary orientations and Lemma~\ref{prop: attach} imply that only a $4$-tile from  (compare Table~\ref{tab:adm copl 4-tile})  {\tiny $\left \lbrace \borderOOII{\tileU{\diagup}{\diagdown}{\diagdown}{\diagup}}, \borderOOIInwseu{\tileD{\diagup}{\diagup}{\diagup}{\diagup}}, \borderOOIIneswu{\tileD{\diagdown}{\diagdown}{\diagdown}{\diagdown}}, \borderOOOOnwd{\tileU{\diagdown}{\diagdown}{\diagdown}{\diagup}}, \borderOOOOseu{\tileD{\diagdown}{\diagup}{\diagup}{\diagup}}  \right\rbrace$}   can be attached in the blank  position  top left  indicated by the dotted $4$-tile (where its straight boundaries represent arbitrary boundary orientations).    Within  the class of admissible $4$-tiles $\mathcal{A}$ in \eqref{eq:class of admissible 4-tile A}, exactly  {\tiny $\left\{  \borderOOII{\tileU{\diagup}{\diagdown}{\diagdown}{\diagup}}, \borderOOIInwseu{\tileD{\diagup}{\diagup}{\diagup}{\diagup}}, \borderOOOOnwd{\tileU{\diagdown}{\diagdown}{\diagdown}{\diagup}}, \borderOOOOseu{\tileD{\diagdown}{\diagup}{\diagup}{\diagup}} \right\}$} match this boundary orientation.  Conversely, a $4$-tile from {\tiny $\left\{  \borderOOII{\tileU{\diagup}{\diagdown}{\diagdown}{\diagup}}, \borderOOIInwseu{\tileD{\diagup}{\diagup}{\diagup}{\diagup}}, \borderOOOOnwd{\tileU{\diagdown}{\diagdown}{\diagdown}{\diagup}}, \borderOOOOseu{\tileD{\diagdown}{\diagup}{\diagup}{\diagup}} \right\}$} arranged along the diagonal yields one of the following possibilities
	\[ \begin{aligned}
		&\ph &&{\tiny \borderOOII{\textcolor{white}{\tileU{\diagup}{\diagdown}{\diagdown}{\diagup}}}} \qquad \quad  &&\ph &&{\tiny \borderOOOO{\textcolor{white}{\tileU{\diagup}{\diagdown}{\diagdown}{\diagup}}}} \\
		&{\tiny \borderOOII{\textcolor{white}{\tileU{\diagup}{\diagdown}{\diagdown}{\diagup}}}} && {\tiny \borderDashedIISS{\textcolor{white}{\tileU{\diagup}{\diagdown}{\diagdown}{\diagup}}}} \qquad \quad && {\tiny \borderOOOO{\textcolor{white}{\tileU{\diagup}{\diagdown}{\diagdown}{\diagup}}}} &&{\tiny \borderDashedIISS{\textcolor{white}{\tileU{\diagup}{\diagdown}{\diagdown}{\diagup}}}}.
	\end{aligned}\]	
	However, due to the given boundary orientations,   the $4$-tiles  in {\tiny $\left \lbrace \borderOOII{\tileU{\diagup}{\diagdown}{\diagdown}{\diagup}}, \borderOOIInwseu{\tileD{\diagup}{\diagup}{\diagup}{\diagup}}, \borderIIIIsed{\tileU{\diagup}{\diagdown}{\diagdown}{\diagdown}}, \borderIIIInwu{\tileD{\diagup}{\diagup}{\diagup}{\diagdown}} \right \rbrace$} are the only $4$-tiles from $\mathcal{A}$ which can be attached in the  blank position   bottom-right,  again indicated with the dotted $4$-tile.  This concludes the check of the matching conditions   (M1).

\noindent \emph{Step 2: $\Leftarrow$}.  The existence of an admissible configuration $y\colon \Z^2 \to \R^3$  follows directly from Proposition~\ref{prop: admi2}(ii) and an induction argument.  Indeed, \eqref{eq: dist} and \eqref{eq: angles} are satisfied since each cell is optimal. To see \eqref{eq: angles2}, it suffices to check that all $4$-tiles are coplanar. In fact, then \eqref{eq: angles2} follows from Lemma \ref{prop:delta bounded by 2 theta}. First,   by construction in Proposition~\ref{prop: admi2}(ii) we  get that all $4$-tiles related to the type function $\sigma_1$ are coplanar. By using  Proposition~\ref{prop: admi2}(i) we find that also the $4$-tiles related to the other type functions $\sigma_i$, $i=2,3,4$, are in $\mathcal{A}$ and are coplanar. This shows that all $4$-tiles are coplanar, as desired.  
\end{proof}

\subsection{Rolling along one diagonal} 
This subsection is devoted to the proof of Proposition~\ref{prop:no diff diags}. The proof fundamentally relies on Lemma~\ref{prop: attach}, i.e., the fact that the boundary orientations of  attached  $4$-tiles match. To this end, we will make extensive use of the matrix diagrams introduced in Table~\ref{tab:adm copl 4-tile}  in order  to exclude certain arrangements of $4$-tiles. Unfortunately, not all nonadmissible cases can be ruled out by such compatibility analysis and we also need to consider some more refined tools, based on the real three-dimensional geometry of the $4$-tiles.  For this reason, we will use the following lemma concerning the attachment of four coplanar $4$-tiles. Recall the types of $4$-tiles $\mathcal{A}$ and $\mathcal{B}$ introduced in \eqref{eq:class of admissible 4-tile A}--\eqref{eq:class of admissible 4-tile B}, as well as the different types of boundaries in Table~\ref{tab:classification of boundaries}.

\begin{lemma}[Arrangements of four $4$-tiles]\label{lemma: gap}
Consider four adjacent $4$-tiles of an admissible configuration with types either in $\mathcal{A}$ or in $\mathcal{B}$, see \eqref{eq:class of admissible 4-tile A}--\eqref{eq:class of admissible 4-tile B}, arranged in a square and denoted by 
 \begin{align}\label{eq: notat-4}
 {\begin{aligned} 
		&\nborder{\mathfrak{A}}\nborder{\mathfrak{D}}\\
		&\nborder{\mathfrak{B}}\nborder{\mathfrak{C}}.
	\end{aligned}}
\end{align}
Then: 
{\rm (i)} If three tiles are Z-tiles and one tile is an D-tile, then the D-tile is in $\lbrace \mathfrak{A}, \mathfrak{C}\rbrace$ (case $\mathcal{A}$) or in $\lbrace \mathfrak{B}, \mathfrak{D}\rbrace$ (case $\mathcal{B}$).\\
{\rm (ii)} If two tiles are Z-tiles and two tiles are D-tiles, then the Z-tiles are arranged along one diagonal and the D-tiles along the other diagonal.   \\
{\rm (iii)} If three tiles are D-tiles and one tile is a Z-tile, then the Z-tile is in $\lbrace \mathfrak{A}, \mathfrak{C}\rbrace$ (case $\mathcal{A}$) or in $\lbrace \mathfrak{B}, \mathfrak{D}\rbrace$ (case $\mathcal{B}$).\\
{\rm (iv)} The arrangement 
	\begin{equation}\label{eq: lasti}
		 {\tiny \begin{aligned}
		 &\borderIOIOned{\tileU{\diagup}{\diagup}{\diagdown}{\diagup}} && \borderOOOOseu{\tileD{\diagdown}{\diagup}{\diagup}{\diagup}} \\
		 &\borderIIIInwu{\tileD{\diagup}{\diagup}{\diagup}{\diagdown}} && \borderOIOIswd{\tileU{\diagup}{\diagdown}{\diagup}{\diagup}}.
		\end{aligned}}
	\end{equation} 
	is not admissible. 

\end{lemma}

 We postpone the proof of this lemma to  Appendix~\ref{subsection:geometric gap}   and proceed with the proof of Proposition~\ref{prop:no diff diags}.  

\begin{proof}[Proof of Proposition~\ref{prop:no diff diags}]
We proceed in two steps: in Step 1 we show that two attached $4$-tiles cannot roll-up/down along different diagonals. In Step 2 we show that in four adjacent $4$-tiles arranged in a square, the two pairs of diagonal $4$-tiles cannot roll-up/down along different diagonals.  These two steps imply the statement.

\noindent \emph{Step 1: Attached $4$-tiles.}  Up to interchanging the roles of $\bullet$ and $\circ$, and up to reflection along the $e_1$- or the $e_2$-axis, there are six different cases to address: 
$$\text{1) $\sborderneu \, \sbordernwu$, \  2) $\sborderneu \, \sbordernwd$, \  3) $\sbordernwu \, \sborderneu$, \  4) $\sbordernwu \, \sborderned$, \  5)  $\sbordernwu \, \sborderswu$ or  $\sborderneu \, \sborderseu$, \  6) $\sbordernwu \, \sborderswd$ or $\sborderneu \, \sbordersed$.}$$
Here, the symbol {\tiny $\sbordernwu$} is a placeholder both for the corresponding I-tile {\tiny $\borderIIIInwu{\tileD{\diagup}{\diagup}{\diagup}{\diagdown}}$} and the D-tile {\tiny $\borderOOIInwseu{\tileD{\diagup}{\diagup}{\diagup}{\diagup}}$ }. The meaning of the other symbols is analogous. For the proof, we refer the reader to Table~\ref{tab:adm copl 4-tile} which collects  all possible $4$-tiles.

\noindent	\emph{Case 1:} {\small $\sborderneu \, \sbordernwu$}. This case leads to a contradiction  to Proposition~\ref{prop: admi}   as necessarily the middle $4$-tile is the A-tile {\tiny $\mborder{\tileD{\diagdown}{\diagup}{\diagdown}{\diagup}}$}. As an example, among the four possibilities, we consider the case where both $4$-tiles are I-tiles. In this case, we have {\tiny $\borderOIOIneu{\tileD{\diagdown}{\diagdown}{\diagup}{\diagdown}} \, \borderIIIInwu{\tileD{\diagup}{\diagup}{\diagup}{\diagdown}}$}.
	
\noindent	\emph{Case 2:} {\small $\sborderneu \, \sbordernwd$}. This case ensues if two $4$-tiles with different boundary orientations are attached, which contradicts Lemma~\ref{prop: attach}. As an example, among the four possibilities, we consider the case where both $4$-tiles are D-tiles. In this case, we have {\tiny $\borderOOIIneswu{\tileD{\diagdown}{\diagdown}{\diagdown}{\diagdown}} \, \borderIIOOnwsed{\tileU{\diagdown}{\diagdown}{\diagdown}{\diagdown}}$.}  
	
\noindent	\emph{Case 3:} {\small $\sbordernwu \, \sborderneu$.} First, if both $4$-tiles are D-tiles, then up to a reflection along the $e_2$-axis, we are in Case 1 and obtain a contradiction as explained before. In the case that one is a D-tile and the other is an I-tile, we obtain a contradiction to Lemma~\ref{prop: attach} as then the boundary orientations do not match. In fact, these two last cases are  {\tiny $\borderOOIInwseu{\tileD{\diagup}{\diagup}{\diagup}{\diagup}} \borderOIOIneu{\tileD{\diagdown}{\diagdown}{\diagup}{\diagdown}}$} and {\tiny $\borderIIIInwu{\tileD{\diagup}{\diagup}{\diagup}{\diagdown}}\, \borderOOIIneswu{\tileD{\diagdown}{\diagdown}{\diagdown}{\diagdown}}$}.
	
We can therefore assume that both $4$-tiles are I-tiles,  i.e.,    take the form {\tiny $\borderIIIInwu{\tileD{\diagup}{\diagup}{\diagup}{\diagdown}} \, \borderOIOIneu{\tileD{\diagdown}{\diagdown}{\diagup}{\diagdown}}$}. We will now consider which $4$-tiles are admissible on top of the given $4$-tiles. Since  we have already ruled out Case 1 and the boundary orientations need to match by Lemma~\ref{prop: attach}, we see that on top of the left I-tile we can only have  ${\tiny \borderOOII{\tileU{\diagup}{\diagdown}{\diagdown}{\diagup}}}$,   ${\tiny \borderOOIInwseu{\tileD{\diagup}{\diagup}{\diagup}{\diagup}}}$, ${\tiny \borderOOOOnwd{\tileU{\diagdown}{\diagdown}{\diagdown}{\diagup}}}$, ${\tiny \borderIOIOned{\tileU{\diagup}{\diagup}{\diagdown}{\diagup}}}$,  or ${\tiny \borderOOOOseu{\tileD{\diagdown}{\diagup}{\diagup}{\diagup}}}$,  and on top of the right I-tile we can only have ${\tiny \borderOOII{\tileU{\diagup}{\diagdown}{\diagdown}{\diagup}}}$, ${\tiny \borderOOIIneswu{\tileD{\diagdown}{\diagdown}{\diagdown}{\diagdown}}}$, ${\tiny \borderOOOOnwd{\tileU{\diagdown}{\diagdown}{\diagdown}{\diagup}}}$, ${\tiny \borderIOIOned{\tileU{\diagup}{\diagup}{\diagdown}{\diagup}}}$, or ${\tiny \borderIOIOswu{\tileD{\diagdown}{\diagup}{\diagdown}{\diagdown}}}$. In any case, the $4$-tile in the middle of the four considered $4$-tiles, will be an A-tile of the form ${\tiny \mborder{\tileU{\diagup}{\diagdown}{\diagup}{\diagdown}}}$ or ${\tiny \mborder{\tileD{\diagup}{\diagdown}{\diagup}{\diagdown}}}$. This contradicts Proposition~\ref{prop: admi} and concludes the proof of Case~3.

\noindent	\emph{Case 4:} {\small $\sbordernwu \, \sborderned$}. If both $4$-tiles are D-tiles, then up to a reflection along the $e_2$-axis, we are in Case 2 and obtain a contradiction as explained before. If both $4$-tiles are I-tiles, we have  {\tiny $\borderIIIInwu{\tileD{\diagup}{\diagup}{\diagup}{\diagdown}}\borderIOIOned{\tileU{\diagup}{\diagup}{\diagdown}{\diagup}}$},   i.e., the boundary orientations are different and we obtain a contradiction to Lemma~\ref{prop: attach}.  The two remaining possibilities are ${\tiny \borderOOIInwseu{\tileD{\diagup}{\diagup}{\diagup}{\diagup}} \borderIOIOned{\tileU{\diagup}{\diagup}{\diagdown}{\diagup}}}$ and ${\tiny \borderIIIInwu{\tileD{\diagup}{\diagup}{\diagup}{\diagdown}} \borderIIOOneswd{\tileU{\diagup}{\diagup}{\diagup}{\diagup}}}$. We prove the contradiction only for the first configuration as the second configuration can be treated along similar lines. In order to do so, we proceed as in Case~3 and attach $4$-tiles at the top, yielding

\begin{equation}\label{eq:proof rolling case 4}
	\begin{aligned}
		&\TileA &&\TileB \\
 		&\borderOOIInwseu{\tileD{\diagup}{\diagup}{\diagup}{\diagup}} &&\borderIOIOned{\tileU{\diagup}{\diagup}{\diagdown}{\diagup}}.
	\end{aligned}
\end{equation} 

In \eqref{eq:proof rolling case 4}, the straight dotted lines encompass all possible boundary orientations. 
We start by noting that the I-tile in the middle of ${\tiny \borderOOIInwseu{\tileD{\diagup}{\diagup}{\diagup}{\diagup}} \borderIOIOned{\tileU{\diagup}{\diagup}{\diagdown}{\diagup}}}$ is of the form {\tiny $\borderIIIInwu{\tileD{\diagup}{\diagup}{\diagup}{\diagdown}}$}. 

Since we have already ruled out Case~1 and the boundary orientations need to match by Lemma~\ref{prop: attach}, only the $4$-tiles 
{\tiny $\borderOOII{\tileU{\diagup}{\diagdown}{\diagdown}{\diagup}}, \, \borderOOIInwseu{\tileD{\diagup}{\diagup}{\diagup}{\diagup}}, \, \borderOOOOnwd{\tileU{\diagdown}{\diagdown}{\diagdown}{\diagup}}, \, \borderIOIOned{\tileU{\diagup}{\diagup}{\diagdown}{\diagup}}, \, \borderOOOOseu{\tileD{\diagdown}{\diagup}{\diagup}{\diagup}}$} can be attached on top of the D-tile  (left),  i.e., at position $\mathfrak{A}$. Similarly,  on   top of the I-tile (right)  at position $\mathfrak{B}$   we can only attach the $4$-tiles {\tiny $\borderIIOO{\tileD{\diagdown}{\diagup}{\diagup}{\diagdown}}, \, \borderIIOOneswd{\tileU{\diagup}{\diagup}{\diagup}{\diagup}}, \, \borderOIOIswd{\tileU{\diagup}{\diagdown}{\diagup}{\diagup}}, \, \borderIIIInwu{\tileD{\diagup}{\diagup}{\diagup}{\diagdown}}, \, \borderOIOIneu{\tileD{\diagdown}{\diagdown}{\diagup}{\diagdown}}$}, see Table~\ref{tab:adm copl 4-tile}.  As the boundary orientations  between $\mathfrak{A}$ and $\mathfrak{B}$  have to match  as well,    there are only eight possibilities of the upper two $4$-tiles which are indicated in the first two columns of Table~\ref{tab:case 4}.  The two upper $4$-tiles $\mathfrak{A}$ and $\mathfrak{B}$ form middle $4$-tiles which are indicated in the third column of Table~\ref{tab:case 4}. Note that the $4$-tile attached on the bottom of this $4$-tile is exactly the middle $4$-tile between the original two $4$-tiles, i.e.,  {\tiny $\borderIIIInwu{\tileD{\diagup}{\diagup}{\diagup}{\diagdown}}$}. Therefore, in the first four cases we obtain a contradiction to Lemma~\ref{prop: attach} since the boundary orientations of the shared boundary of the two middle $4$-tiles do not match.  

For the second four cases  we need a different argument instead. To
this end, we consider also the middle $4$-tile between the D-tile and
$\mathfrak{A}$  (left middle $4$-tile) and the middle $4$-tile between
the I-tile and  $\mathfrak{B}$  (right middle $4$-tile), see the last
two columns in Table~\ref{tab:case 4}. We observe that in none of the
cases the boundary orientations of the shared boundary of the left and
right middle $4$-tiles match. This is again a contradiction to
Lemma~\ref{prop: attach}, concluding  the check of  Case~4.
	
	{\footnotesize
	\begin{table}
		\begin{tabular}{ccccc}
			At $\mathfrak{A}$ & At $\mathfrak{B}$ &  \makecell{  Middle $4$-tile\\ $\mathfrak{A}$--$\mathfrak{B}$} & \makecell{Middle $4$-tile\\ left} & \makecell{Middle $4$-tile\\ right} \\
			\hline 
			$\borderIOIOned{\tileU{\diagup}{\diagup}{\diagdown}{\diagup}}$ & $\borderIIOO{\tileD{\diagdown}{\diagup}{\diagup}{\diagdown}}$  & $\borderOIOIswd{\tileU{\diagup}{\diagdown}{\diagup}{\diagup}}$ & & \\
			$\borderIOIOned{\tileU{\diagup}{\diagup}{\diagdown}{\diagup}}$ & $\borderIIOOneswd{\tileU{\diagup}{\diagup}{\diagup}{\diagup}}$ & $\borderIIOOneswd{\tileU{\diagup}{\diagup}{\diagup}{\diagup}}$ & & \\
			$\borderIOIOned{\tileU{\diagup}{\diagup}{\diagdown}{\diagup}}$ & $\borderOIOIswd{\tileU{\diagup}{\diagdown}{\diagup}{\diagup}}$ & $\borderIIOOneswd{\tileU{\diagup}{\diagup}{\diagup}{\diagup}}$ & & \\
			$\borderIOIOned{\tileU{\diagup}{\diagup}{\diagdown}{\diagup}}$ & $\borderOIOIneu{\tileD{\diagdown}{\diagdown}{\diagup}{\diagdown}}$ & $\borderOIOIswd{\tileU{\diagup}{\diagdown}{\diagup}{\diagup}}$ & & \\

			$\borderOOII{\tileU{\diagup}{\diagdown}{\diagdown}{\diagup}}$ & $\borderIIIInwu{\tileD{\diagup}{\diagup}{\diagup}{\diagdown}}$ & $\borderOOOOseu{\tileD{\diagdown}{\diagup}{\diagup}{\diagup}}$ & $\borderOOOOseu{\tileD{\diagdown}{\diagup}{\diagup}{\diagup}}$ & $\borderOIOIswd{\tileU{\diagup}{\diagdown}{\diagup}{\diagup}}$ \\ 
			$\borderOOIInwseu{\tileD{\diagup}{\diagup}{\diagup}{\diagup}}$ & $\borderIIIInwu{\tileD{\diagup}{\diagup}{\diagup}{\diagdown}}$ & $\borderOOIInwseu{\tileD{\diagup}{\diagup}{\diagup}{\diagup}}$ & $\borderOOIInwseu{\tileD{\diagup}{\diagup}{\diagup}{\diagup}}$ & $\borderOIOIswd{\tileU{\diagup}{\diagdown}{\diagup}{\diagup}}$ \\ 
			$\borderOOOOnwd{\tileU{\diagdown}{\diagdown}{\diagdown}{\diagup}}$	& $\borderIIIInwu{\tileD{\diagup}{\diagup}{\diagup}{\diagdown}}$ & $\borderOOOOseu{\tileD{\diagdown}{\diagup}{\diagup}{\diagup}}$ &  $\borderOOOOseu{\tileD{\diagdown}{\diagup}{\diagup}{\diagup}}$  & $\borderOIOIswd{\tileU{\diagup}{\diagdown}{\diagup}{\diagup}}$\\ 
			$\borderOOOOseu{\tileD{\diagdown}{\diagup}{\diagup}{\diagup}}$ & $\borderIIIInwu{\tileD{\diagup}{\diagup}{\diagup}{\diagdown}}$ & $\borderOOIInwseu{\tileD{\diagup}{\diagup}{\diagup}{\diagup}}$ &  $\borderOOIInwseu{\tileD{\diagup}{\diagup}{\diagup}{\diagup}}$  &  $\borderOIOIswd{\tileU{\diagup}{\diagdown}{\diagup}{\diagup}}$ 

		\end{tabular} \vspace{0.2cm} 
		\caption{The eight different cases considered in Case 4.}
		\label{tab:case 4}
	\end{table}}

\noindent \emph{Case 5:} {\small $\sbordernwu \, \sborderswu$ or $\sborderneu \, \sborderseu$}: Without restriction we address only the first case as the second can be treated  analogously (and, in fact, obtained by a rotation). We have to distinguish two cases. Firstly, the $4$-tile on the left is a D-tile, i.e., {\tiny $\sbordernwseu \, \sborderswu$}.  Then up to a reflection along the $e_2$-axis, we are in Case 1 and obtain a contradiction as explained before.  Secondly, if the left $4$-tile is not a D-tile, it has to be an I-tile. We obtain the two possible configurations {\tiny $\borderIIIInwu{\tileD{\diagup}{\diagup}{\diagup}{\diagdown}} \, \borderOOIIneswu{\tileD{\diagdown}{\diagdown}{\diagdown}{\diagdown}}$}  and {\tiny $  \borderIIIInwu{\tileD{\diagup}{\diagup}{\diagup}{\diagdown}} \,  \borderIOIOswu{\tileD{\diagdown}{\diagup}{\diagdown}{\diagdown}}$} which both contradict Lemma~\ref{prop: attach} as the boundary orientations do not match.

\noindent	\emph{Case 6:} {\small $\sbordernwu \, \sborderswd$ or $\sborderneu \, \sbordersed$}. Without restriction we address only the first case as the second can be treated  analogously. If the $4$-tile on the left is a D-tile, then we have the two possibilities {\tiny $\borderOOIInwseu{\tileD{\diagup}{\diagup}{\diagup}{\diagup}}\, \borderOIOIswd{\tileU{\diagup}{\diagdown}{\diagup}{\diagup}}$} and {\tiny $\borderOOIInwseu{\tileD{\diagup}{\diagup}{\diagup}{\diagup}}\, \borderIIOOneswd{\tileU{\diagup}{\diagup}{\diagup}{\diagup}}$}. Thus, the boundary orientations do not match which contradicts Lemma~\ref{prop: attach}.  If the right $4$-tile is a D-tile,  we are in Case~4 and obtain a contradiction as explained before.

Therefore, both $4$-tiles have to be I-tiles, i.e., we have 
\begin{equation}\label{eq:proof rolling case 6}
	\borderIIIInwu{\tileD{\diagup}{\diagup}{\diagup}{\diagdown}}\borderOIOIswd{\tileU{\diagup}{\diagdown}{\diagup}{\diagup}}.
\end{equation}  
As in   Case~4, we consider two $4$-tiles attached on the top.  By using arguments similar  to the ones above, we will show that   the only possible choice how to assemble the four $4$-tiles would  be   given by
	\begin{equation}\label{eq: on top}
		 {\tiny \begin{aligned}
		 &\borderIOIOned{\tileU{\diagup}{\diagup}{\diagdown}{\diagup}} && \borderOOOOseu{\tileD{\diagdown}{\diagup}{\diagup}{\diagup}} \\
		 &\borderIIIInwu{\tileD{\diagup}{\diagup}{\diagup}{\diagdown}} && \borderOIOIswd{\tileU{\diagup}{\diagdown}{\diagup}{\diagup}}.
		\end{aligned}}
	\end{equation} 
This, however, is excluded by Lemma~\ref{lemma: gap}(iv). To see \eqref{eq: on top}, in view of the fact that we have already ruled out Cases 1--5 and the boundary orientations need to match by Lemma~\ref{prop: attach}, only the $4$-tiles 
\begin{align}\label{eq: lefti}
	\text{{\tiny $\borderOOII{\tileU{\diagup}{\diagdown}{\diagdown}{\diagup}}$, $\borderOOIInwseu{\tileD{\diagup}{\diagup}{\diagup}{\diagup}}$, $\borderOOOOnwd{\tileU{\diagdown}{\diagdown}{\diagdown}{\diagup}}$, $\borderIOIOned{\tileU{\diagup}{\diagup}{\diagdown}{\diagup}}$, $\borderOOOOseu{\tileD{\diagdown}{\diagup}{\diagup}{\diagup}}$}}
\end{align}
can be attached on top of the left I-tile in \eqref{eq:proof rolling case 6}.  Analogously,   on top of the right I-tile in \eqref{eq:proof rolling case 6} we can only attach the $4$-tiles
\begin{align}\label{eq: righti}
 	\text{{\tiny $\borderOOII{\tileU{\diagup}{\diagdown}{\diagdown}{\diagup}}$, $\borderOOIIneswu{\tileD{\diagdown}{\diagdown}{\diagdown}{\diagdown}}$, $\borderIOIOned{\tileU{\diagup}{\diagup}{\diagdown}{\diagup}}$, $\borderOOOOseu{\tileD{\diagdown}{\diagup}{\diagup}{\diagup}}$,  $\borderIOIOswu{\tileD{\diagdown}{\diagup}{\diagdown}{\diagdown}}$}},
\end{align}  
see Table~\ref{tab:adm copl 4-tile}. As in Case~4 we consider the middle $4$-tile between the left I-tile in \eqref{eq:proof rolling case 6} and the $4$-tile on top of it (left middle $4$-tile) and the middle $4$-tile between the right I-tile in \eqref{eq:proof rolling case 6} and the $4$-tile on top of it (right middle $4$-tile). In view of \eqref{eq: lefti}--\eqref{eq: righti}, there are only the cases indicated in Table~\ref{tab:case 6}. From Table~\ref{tab:case 6} we see that the boundary orientations of the shared boundary of the two middle $4$-tiles can only match if the right middle $4$-tile is of type {\tiny $\borderIIIInwu{\tileD{\diagup}{\diagup}{\diagup}{\diagdown}}$}. By \eqref{eq: righti} this shows that only {\tiny $\borderOOOOseu{\tileD{\diagdown}{\diagup}{\diagup}{\diagup}}$} can be attached on top of the right I-tile {\tiny $\borderOIOIswd{\tileU{\diagup}{\diagdown}{\diagup}{\diagup}}$}. Then, in view of \eqref{eq: lefti}, only {\tiny $\borderIOIOned{\tileU{\diagup}{\diagup}{\diagdown}{\diagup}}$} can be attached on top of the left I-tile {\tiny $\borderIIIInwu{\tileD{\diagup}{\diagup}{\diagup}{\diagdown}}$} as the other four $4$-tiles in \eqref{eq: lefti} do not math the boundary orientation of {\tiny $\borderOOOOseu{\tileD{\diagdown}{\diagup}{\diagup}{\diagup}}$}. This shows that \eqref{eq: on top} holds, and concludes the proof of Case~6.

	{\footnotesize
	\begin{table}
		\begin{tabular}{cc}
			\makecell{Middle $4$-tile\\ left} & \makecell{Middle $4$-tile\\ right} \\
			\hline 
$\borderOOOOseu{\tileD{\diagdown}{\diagup}{\diagup}{\diagup}}$ & $\borderIIOO{\tileD{\diagdown}{\diagup}{\diagup}{\diagdown}}$ \\
$\borderOOIInwseu{\tileD{\diagup}{\diagup}{\diagup}{\diagup}}$ & $\borderOIOIneu{\tileD{\diagdown}{\diagdown}{\diagup}{\diagdown}}$ \\	
				 & $\borderIIIInwu{\tileD{\diagup}{\diagup}{\diagup}{\diagdown}}$ 	
			
		\end{tabular} \vspace{0.2cm} 
		\caption{The different possible middle $4$-tiles in Case 6.}
		\label{tab:case 6}
	\end{table}
	}

\noindent \emph{Step 2: $4$-tiles on the diagonal.}  We now show  that in four adjacent $4$-tiles arranged in a square, the two pairs of diagonal $4$-tiles cannot roll-up/down along different diagonals. Up to interchanging the roles of $\bullet$ and $\circ$, and up to reflection along the $e_1$- or the $e_2$-axis, there are two cases to consider, where Case~1 represents one of the eight situations
$$	{\tiny \begin{aligned}
		&\sbordernwu \sborder \\
		&\sborder \sborderneu \quad
	\end{aligned}, \quad \begin{aligned}
		&\sborderseu \sborder \\
		&\sborder \sborderneu \quad
	\end{aligned}, \quad \begin{aligned}
		&\sbordernwu \sborder \\
		&\sborder \sborderswu \quad
	\end{aligned}, \quad \begin{aligned}
		&\sborderseu \sborder \\
		&\sborder \sborderswu \quad
	\end{aligned}, \quad \begin{aligned}
		&\sborderswu \sborder \\
		&\sborder \sborderseu \quad
	\end{aligned}, \quad \begin{aligned}
		&\sborderneu \sborder \\
		&\sborder \sborderseu \quad
	\end{aligned}, \quad \begin{aligned}
		&\sborderswu \sborder \\
		&\sborder \sbordernwu \quad
	\end{aligned}, \quad \begin{aligned}
		&\sborderneu \sborder \\
		&\sborder \sbordernwu \quad
	\end{aligned}}$$	
and Case~2 represents one of the eight situations	
$${\tiny \begin{aligned}
		&\sbordernwu \sborder \\
		&\sborder  \sborderned \quad
	\end{aligned}, \quad \begin{aligned}
		&\sborderseu \sborder \\
		&\sborder \sborderned \quad
	\end{aligned}, \quad \begin{aligned}
		&\sbordernwu \sborder \\
		&\sborder \sborderswd \quad
	\end{aligned}, \quad \begin{aligned}
		&\sborderseu \sborder \\
		&\sborder \sborderswd \quad
	\end{aligned}, \quad \begin{aligned}
		&\sborderswu \sborder \\
		&\sborder \sbordersed \quad
	\end{aligned}, \quad \begin{aligned}
		&\sborderneu \sborder \\
		&\sborder \sbordersed \quad
	\end{aligned}, \quad \begin{aligned}
		&\sborderswu \sborder \\
		&\sborder \sbordernwd \quad
	\end{aligned}, \quad \begin{aligned}
		&\sborderneu \sborder \\
		&\sborder \sbordernwd \quad
	\end{aligned}.}$$	
 Here, as in Step~1,  the symbols  $\bullet$ and $\circ$ indicate both the corresponding I-tile and D-tile. Without restriction we address only the first configuration in both cases as all other situations can be treated along similar lines.

\noindent \emph{Case 1.}  We start by introducing the labeling
$${	\begin{aligned}
			&\nbordernwu{\mathfrak{A}} \nborder{\mathfrak{D}} \\
			&\nborder{\mathfrak{B}} \nborderneu{\mathfrak{C}}.
		\end{aligned}}$$
We preliminarily note that, in view of Step~1, for $\mathfrak{B}$ and $\mathfrak{D}$  only $4$-tiles in $\mathcal{A} \cap \mathcal{B}$  are admissible,   see \eqref{eq:class of admissible 4-tile A}--\eqref{eq:class of admissible 4-tile B}, i.e., the two Z-tiles {\tiny $\borderOOII{\tileU{\diagup}{\diagdown}{\diagdown}{\diagup}}$} and {\tiny $\borderIIOO{\tileD{\diagdown}{\diagup}{\diagup}{\diagdown}}$}. 	
We distinguish three different subcases: 
	
\noindent 	\emph{Case 1.1.} If $\mathfrak{A}$ is the unique I-tile, then  Lemma~\ref{prop: attach} for the boundary between  $\mathfrak{A}$ and $\mathfrak{D}$ as well as the boundary between $\mathfrak{D}$ and $\mathfrak{C}$  implies that the $4$-tile $\mathfrak{D}$ cannot be a Z-tile.  In fact, the boundary orientation of $\mathfrak{A}$ on the right is $\wedge$ (indicated by $<$ in the notation) and the boundary orientation of $\mathfrak{C}$ on top is $\vee$.

\noindent 	\emph{Case 1.2.}   By a similar reasoning, if     $\mathfrak{A}$ is the unique D-tile  and  $\mathfrak{C}$ is the unique I-tile, Lemma~\ref{prop: attach} implies that the $4$-tile $\mathfrak{B}$ cannot be a Z-tile.

\noindent 	\emph{Case 1.3.}  If both  $\mathfrak{A}$ and $\mathfrak{C}$ are D-tiles, we again use Lemma~\ref{prop: attach} and see that  the $4$-tiles   $\mathfrak{B}$ and  $\mathfrak{D}$ can only be of type {\tiny $\borderOOII{\tileU{\diagup}{\diagdown}{\diagdown}{\diagup}}$}. Therefore, we need to  consider the configuration 
	\[ \tiny \begin{aligned}
		&\borderOOIInwseu{\tileD{\diagup}{\diagup}{\diagup}{\diagup}} && \borderOOII{\tileU{\diagup}{\diagdown}{\diagdown}{\diagup}} \\
		&\borderOOII{\tileU{\diagup}{\diagdown}{\diagdown}{\diagup}} && \borderOOIIneswu{\tileD{\diagdown}{\diagdown}{\diagdown}{\diagdown}}.
	\end{aligned} \]
The middle $4$-tile between  $\mathfrak{A}$ and $\mathfrak{B}$ is given by {\tiny $\borderIIIInwu{\tileD{\diagup}{\diagup}{\diagup}{\diagdown}}$} and the middle $4$-tile between  $\mathfrak{C}$ and $\mathfrak{D}$ is given by {\tiny $\borderIOIOswu{\tileD{\diagdown}{\diagup}{\diagdown}{\diagdown}}$}. Their shared boundary have mismatching boundary orientations, contradicting Lemma~\ref{prop: attach}.

\noindent \emph{Case~2.}  We start by introducing the labeling
	$${\begin{aligned}
			&\nbordernwu{\mathfrak{A}} \nborder{\mathfrak{D}} \\
			&\nborder{\mathfrak{B}} \nborderned{\mathfrak{C}}.
		\end{aligned}}$$		
		As in Case~1,  due to Step 1,   for $\mathfrak{B}$ and $\mathfrak{D}$  only  the two Z-tiles {\tiny $\borderOOII{\tileU{\diagup}{\diagdown}{\diagdown}{\diagup}}$} and {\tiny $\borderIIOO{\tileD{\diagdown}{\diagup}{\diagup}{\diagdown}}$} are admissible. We distinguish four different subcases:

\noindent 	\emph{Case 2.1.} If both  $\mathfrak{A}$ and $\mathfrak{C}$ are I-tiles,   Lemma~\ref{prop: attach}   implies that  the $4$-tile $\mathfrak{B}$ cannot be a Z-tile.

\noindent 	\emph{Case 2.2.} If both  $\mathfrak{A}$ and $\mathfrak{C}$ are D-tiles, then   Lemma~\ref{prop: attach}   implies that the $4$-tile $\mathfrak{B}$ cannot be a Z-tile. 

\noindent 	\emph{Case 2.3.} If $\mathfrak{A}$ is the  unique D-tile and $\mathfrak{C}$ is the unique I-tile, then   Lemma~\ref{prop: attach}   implies that the $4$-tile $\mathfrak{D}$ cannot be a Z-tile.

\noindent 	\emph{Case 2.4.} Now suppose that $\mathfrak{A}$ is the unique I-tile and $\mathfrak{C}$ is the unique D-tile. Then  $\mathfrak{B}$  and $\mathfrak{D}$ need to be of type {\tiny $\borderIIOO{\tileD{\diagdown}{\diagup}{\diagup}{\diagdown}}$}. Therefore, we need to  consider the configuration 
\begin{equation}\label{eq: 2.4}
{\tiny \begin{aligned}
		&\borderIIIInwu{\tileD{\diagup}{\diagup}{\diagup}{\diagdown}}&&\borderIIOO{\tileD{\diagdown}{\diagup}{\diagup}{\diagdown}}\\
		&\borderIIOO{\tileD{\diagdown}{\diagup}{\diagup}{\diagdown}}&&\borderIIOOneswd{\tileU{\diagup}{\diagup}{\diagup}{\diagup}},
	\end{aligned}}
	\end{equation}
	and show that it is also not admissible. The I-tile rolls up in direction top left, which has no influence in this (sub-)configuration. In other words, by replacing in \eqref{eq: 2.4} the tile $\mathfrak{A}$ with the Z-tile {\tiny $\borderIIOO{\tileD{\diagdown}{\diagup}{\diagup}{\diagdown}}$} and showing that this modified configuration is not admissible, we also find that \eqref{eq: 2.4} is not admissible.  In fact, in view of Lemma~\ref{lemma: gap}(i) and the fact that the D-tile {\tiny $\borderIIOOneswd{\tileU{\diagup}{\diagup}{\diagup}{\diagup}}$} lies in $\mathcal{B}$ (see \eqref{eq:class of admissible 4-tile B}), we see that the modified version of \eqref{eq: 2.4} is not admissible. This  concludes this step of the proof.  	 
\end{proof}

\subsection{Constant type along the diagonal}

This subsection is devoted to the proof of Proposition \ref{prop:constant type}.
 
\begin{proof}[Proof of Proposition \ref{prop:constant type}]
We assume without restriction that all four $4$-tiles lie in $\mathcal{A}$, see \eqref{eq:class of admissible 4-tile A}, as the other case is completely analogous. We consider
	\[ \small \begin{aligned} 
		&\nborder{\mathfrak{A}} \nborder{\mathfrak{D}}\\
		&\nborder{\mathfrak{B}} \nborder{\mathfrak{C}},
	\end{aligned} \]
and note that we need to show that $\mathfrak{B}$ and $\mathfrak{D}$ are of the same type.  We proceed in two steps: first, we show that $\mathfrak{B}$ and $\mathfrak{D}$ are of the same \emph{class,} i.e., both have to be either Z-, I-, or D-tiles. 	In the second step, we then conclude that they even have to be of the same \emph{type}. In the proof, we will use the following observation which directly follows from the definition of $\mathcal{A}$:	
\begin{equation}
	\begin{aligned}\label{eq: text}
		&\text{$\bullet$ Z- and D-tiles: }&&\text{All four boundary orientations  are identical}, \\
		&\text{$\bullet$ I-tiles: }&&\text{Left and upper boundary orientations  are identical,}  \\
		& &&\text{right and lower boundary orientations are identical}. 
	\end{aligned}
\end{equation}
\noindent \emph{Step 1.} In this step, we show that  $\mathfrak{B}$ and $\mathfrak{D}$ are necessarily of the same class. 
	
	\noindent \emph{Case 1.1}.  If exactly one of the two tiles  $\mathfrak{B}$ and $\mathfrak{D}$ is an I-tile, in view of \eqref{eq: text}, we obtain a contradiction to Lemma~\ref{prop: attach} as not all boundary orientations of the four shared boundaries can match.

	Thus, we can now assume that none of the tiles $\mathfrak{B}, \mathfrak{D}$ is an I-tile.  Actually, it is also not restrictive to assume that the tiles $\mathfrak{A}$ and $\mathfrak{C}$ are not of class I. Indeed, the upper left optimal cell of $\mathfrak{A}$ and the lower right optimal  $\mathfrak{C}$ have no influence on the subsequent arguments in Cases 1.2--1.4 and can readily be replaced by the other type. This allows to replace tiles of class I by types of class Z or D in $\mathcal{A}$, without affecting the following arguments. Summarizing, it suffices to consider the case that all four $4$-tiles are Z- or D-tiles.

	\noindent \emph{Case 1.2}. If three $4$-tiles are D-tiles and one tile is a Z-tile, we only have that $\mathfrak{B}$ and $\mathfrak{D}$  are not of the same class if the Z-tile lies in $\lbrace \mathfrak{B}, \mathfrak{D} \rbrace$. This contradicts Lemma~\ref{lemma: gap}(iii).

	\noindent \emph{Case 1.3}. If three $4$-tiles are Z-tiles  and one tile is a D-tile, we only have that $\mathfrak{B}$ and $\mathfrak{D}$  are not of the same class if the D-tile lies in $\lbrace \mathfrak{B}, \mathfrak{D} \rbrace$. This contradicts Lemma~\ref{lemma: gap}(i).

	\noindent \emph{Case 1.4}.  If   two $4$-tiles are of class Z and two of class D, the claim follows directly from Lemma~\ref{lemma: gap}(ii). 
	
\noindent \emph{Step 2.} In this second step we show that not only the class but also the type has to be constant along the diagonal.  First, if we had different Z-tiles or D-tiles along the diagonal,  in view of  \eqref{eq:class of admissible 4-tile A},    these two $4$-tiles would have different boundary orientations. Again by using \eqref{eq: text}, we obtain a contradiction to Lemma~\ref{prop: attach} as not all boundary orientations of the four shared boundaries can match.  
	
We now address the case that $\mathfrak{B}$ and $\mathfrak{D}$ are I-tiles. Again in view of  \eqref{eq: text} and the definition of $\mathcal{A}$, we  find 	
$$\text{either a) } \mathfrak{B}, \mathfrak{D}  \in {\tiny \left\lbrace \borderOOOOseu{\tileD{\diagdown}{\diagup}{\diagup}{\diagup}}, \borderOOOOnwd{\tileU{\diagdown}{\diagdown}{\diagdown}{\diagup}} \right\rbrace} \quad \text{ or b) } \quad \mathfrak{B}, \mathfrak{D}  \in {\tiny \left \lbrace \borderIIIInwu{\tileD{\diagup}{\diagup}{\diagup}{\diagdown}}, \borderIIIIsed{\tileU{\diagup}{\diagdown}{\diagdown}{\diagdown}} \right \rbrace}$$
since otherwise the boundary orientations do not match, contradicting Lemma~\ref{prop: attach}. Whenever the type is not constant along the diagonal, the $4$-tile in the middle of the four $4$-tiles is an A-tile which contradicts Proposition~\ref{prop: admi}. For simplicity, we show this only in case a) as case b) follows along similar lines. In fact, by Lemma~\ref{prop: attach} we find that $\mathfrak{A}$ can only be of type {\tiny $\borderIIOO{\tileD{\diagdown}{\diagup}{\diagup}{\diagdown}}$}, {\tiny $\borderIIOOnwsed{\tileU{\diagdown}{\diagdown}{\diagdown}{\diagdown}}$},  {\tiny $\borderIIIIsed{\tileU{\diagup}{\diagdown}{\diagdown}{\diagdown}}$}, or  {\tiny $\borderIIIInwu{\tileD{\diagup}{\diagup}{\diagup}{\diagdown}}$},   and $\mathfrak{C}$ can only be of type  {\tiny $\borderOOII{\tileU{\diagup}{\diagdown}{\diagdown}{\diagup}}$}, {\tiny $\borderOOIInwseu{\tileD{\diagup}{\diagup}{\diagup}{\diagup}}$},  {\tiny $\borderIIIIsed{\tileU{\diagup}{\diagdown}{\diagdown}{\diagdown}}$}, or  {\tiny $\borderIIIInwu{\tileD{\diagup}{\diagup}{\diagup}{\diagdown}}$}. Consequently, if $\mathfrak{B}$ is of type {\tiny $\borderOOOOseu{\tileD{\diagdown}{\diagup}{\diagup}{\diagup}}$},  in the middle  we find the A-tile {\tiny $\mborder{\tileU{\diagdown}{\diagdown}{\diagup}{\diagup}}$} or {\tiny $\mborder{\tileD{\diagdown}{\diagdown}{\diagup}{\diagup}}$}, and if $\mathfrak{B}$ is of type {\tiny $\borderOOOOnwd{\tileU{\diagdown}{\diagdown}{\diagdown}{\diagup}}$}, we find the A-tile
{\tiny $\mborder{\tileU{\diagdown}{\diagup}{\diagdown}{\diagup}}$} or {\tiny  $\mborder{\tileD{\diagdown}{\diagup}{\diagdown}{\diagup}}$},  see Table~\ref{tab:full classification}.  
	
\end{proof}

\subsection{Admissible arrangement of $4$-tiles}

This subsection is devoted to the proof of Proposition~\ref{prop: admi2}.

\begin{figure}[h]
  \centering
	\def\svgwidth{0.4 \columnwidth}
	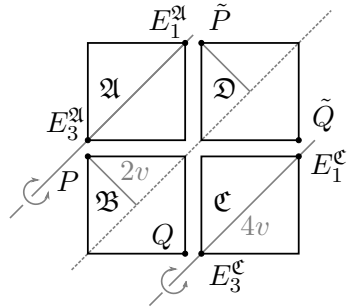

  \caption{The points and rotations indicated in the proof of Proposition~\ref{prop: admi2}.}
  \label{fig:proof sufficiency of main theorem}
\end{figure}

\begin{proof}[Proof of Proposition \ref{prop: admi2}]
Without restriction we perform the proof only for the types $\mathcal{A}$ defined in \eqref{eq:class of admissible 4-tile A}.

\noindent {(i)} We start by observing that each pair of $4$-tiles in $\mathcal{A}$  with matching boundary orientations   can be attached since all boundary angles  are either Z- or D-boundary angles, see  Table~\ref{tab:classification of boundaries} and Table~\ref{tab:full classification}, and both angles coincide with $\delta_\theta$, see Lemma~\ref{prop:boundary value}. We first show that the $4$-tile in the middle is again in $\mathcal{A}$. In a second step, we check that the middle $4$-tile is also coplanar. 

We recall that the type of the middle $4$-tile can by determined by considering the matrix notation, as exemplified in \eqref{eq: exgraz}. In view of  \eqref{eq:class of admissible 4-tile A},  we obtain the following six cases:

\noindent \emph{Case 1.} Attaching two Z-tiles, we find that the two tiles are of same type and the middle tile is the Z-tile of the other type.   

\noindent \emph{Case 2.} Attaching two D-tiles, we find that the two tiles are of same type and the middle tile is again of this type. 
   
\noindent \emph{Case 3.} Attaching two I-tiles, we can obtain all possible $4$-tiles in  $\mathcal{A}$.

\noindent \emph{Case 4.} Attaching a Z- and a D-tile, we obtain  an   I-tile in $\mathcal{A}$.  

\noindent \emph{Case 5.} Attaching a Z- and an I-tile, we obtain any Z- and  I-tile in $\mathcal{A}$. 

\noindent \emph{Case 6.} Attaching a D- and an I-tile, we obtain any D- and  I-tile in $\mathcal{A}$. 

 Note that in all cases above exactly $4$-tiles from $\mathcal{A}$ can
 occur,   and no more than those.  

It remains to show that  the resulting  middle $4$-tile  is also coplanar. As attaching two $4$-tiles does not change the optimal angle $\theta$, also the middle $4$-tile consists of four optimal cells with angle $\theta$. Therefore, relation \eqref{eq:delta eta} holds for the middle $4$-tile as well. To conclude the proof, it suffices to show that one of the nonplanarity angles $\delta_{13}$ and $\delta_{24}$ of the middle $4$-tile is equal to $\delta_\theta$. To this end, note that one of these angles coincides with the boundary angle of the shared boundary of the two $4$-tiles. By Lemma~\ref{prop:boundary value} this angle is equal  to   $\delta_\theta$.

\noindent {(ii)}  We proceed constructively to show that every configuration consisting  of four  $4$-tiles from $\mathcal{A}$ arranged in a square  satisfying the matching conditions (M1)--(M2)  is admissible, i.e., can be realized by  an admissible   deformation $y$.  By assumption, $\mathfrak{B}$ and $\mathfrak{D}$ are of the same type.  Then, one can check that, for any choice of $\mathfrak{A}, \mathfrak{C} \in \mathcal{A}$ satisfying the matching conditions (M1)--(M2), the boundary orientations of $\mathfrak{A}, \mathfrak{C}$   match with those of $\mathfrak{B}$ and $\mathfrak{D}$. 
In view of  Lemma~\ref{lem:third point of optimal cell is high}(i),   fixing $\mathfrak{B}$ in reference configuration and translating $\mathfrak{D}$ from its reference position by the vector $(2s,2s,0)$, we  see that these two $4$-tiles share exactly one corner point, and we   have  $\vert P - \tilde{P} \vert = \vert Q - \tilde{Q} \vert = \sqrt{(2s)^2 + (2s)^2} =   4v$, where $P, Q \in \mathfrak{B}$ and $\tilde{P}, \tilde{Q} \in \mathfrak{D}$ are the corner vertices  indicated   in Figure~\ref{fig:proof sufficiency of main theorem}.
By Lemma~\ref{lem:third point of optimal cell is high}(i)  the opposite corner points along the diagonal $d_1$ have distance $4v$, i.e., $\vert E_1^\mathfrak{A} - E_3^\mathfrak{A} \vert =  \vert E_1^\mathfrak{C} - E_3^\mathfrak{C} \vert = 4v$. Therefore, we can translate $\mathfrak{A}$ and $\mathfrak{C}$ from their reference positions such that their opposite corner points coincide with $P$ and $\tilde{P}$ and $Q$ and $\tilde{Q}$, respectively. Since, for every $4$-tile the distance between its center and a corner point equals $\sqrt{s^2+s^2} = 2v$, see  \eqref{eq: def of v} and  Lemma~\ref{prop: ref}(i), after  rotating  $\mathfrak{A}$ and $\mathfrak{C}$ about $(0,2s,0) + \R(1,1,0)$ and  $(2s,0,0) + \R(1,1,0)$, respectively, as indicated in Figure~\ref{fig:proof sufficiency of main theorem}, the corner points of $\mathfrak{A}$, $\mathfrak{B}$, $\mathfrak{C}$, and $\mathfrak{D}$ in the interior of the configuration coincide.
As the boundary orientations match by (M1)--(M2) and the boundary angles coincide by  Lemma \ref{prop:boundary value},  also the respective middle points coincide after rotation of $\mathfrak{A}$ and $\mathfrak{C}$.  This along with part (i) of the statement shows that  the configuration is indeed realizable by an admissible configuration $y\colon \lbrace 0, 1,2,3,4\rbrace^2 \to \R^3$. This concludes the proof.  
\end{proof}

\section*{Acknowledgements}
 This work is partially supported 
 by the FWF-DFG grant I\,4354, the FWF grants F\,65,  W\,1245, 
 I\,5149,  and P\,32788, and the OeAD-WTZ project CZ 01/2021. This work was supported by the Deutsche Forschungsgemeinschaft (DFG, German Research Foundation) under Germany's Excellence Strategy EXC 2044 -390685587, Mathematics M\"unster: Dynamics--Geometry--Structure.

\appendix

\section{Remaining proofs}

\subsection{Geometry of optimal cells and $4$-tiles}\label{subsec:proof of lemma fourthpoint}

This subsection is devoted to the proofs of the lemmas stated in Subsection~\ref{sec: 4-tiles}.

\begin{figure}[ht]
	\centering
	\def\svgwidth{0.5 \columnwidth}
	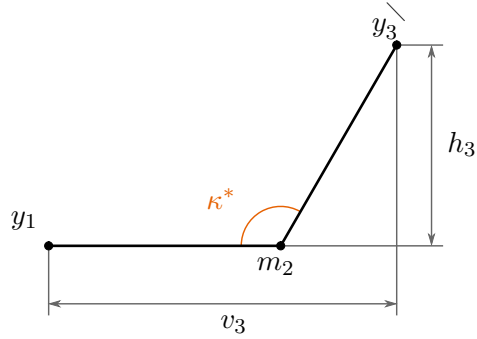

	\caption{Cross section along the plane spanned by $a$ and $n$, as defined in the proof of Lemma~\ref{lem:fourthpoint}. From this perspective the points $y_2$, $m_2$, and $y_4$   coincide.}
	\label{fig:fourth point}
\end{figure}

\begin{proof}[Proof of Lemma \ref{lem:fourthpoint}]
Recall that $m_2 :=(y_2 + y_4)/2$ is the middle point between $y_2$ and $y_4$,  cf.~Figure \ref{fig:types optimal cells}.   We define $a = m_2 - y_1$, with $\vert a \vert = d$.  Let $n$ be a normal vector to     the plane spanned by $y_1$, $y_2$, and $y_4$, in direction $(y_2 - y_1) \times (y_4 - y_1)$.  

Observe that by assumption the fourth point $y_3$ has to satisfy $\vert y_2 - y_3 \vert = \vert y_4 - y_3 \vert = 1$ and thus has to lie on the plane spanned by $a$ and $n$. 
Therefore, we can make the ansatz 
\begin{equation*}
	y_3 = y_1 + v_3 \: a \pm h_3\: n,
\end{equation*}  
where $v_3$ and $h_3$ are to be determined, see Figure~\ref{fig:fourth point}. Note that in $\pm$ we choose $+$ for form $\diagdown$ and $-$ for form $\diagup$ .  To conclude, we are left  to prove    that $v_3$ and $h_3$ can  be determined uniquely.  Since the cell is optimal, we have   $\measuredangle y_3\, m_2\, y_1 = \kappa^*$  (see \eqref{eq:alpha})   as well as  $\vert a \vert = \vert m_2 - y_1 \vert =  |m_2 - y_3  | = d$.  Consequently,   the triangle with  vertices  $ y_1$, $m_2$, and $y_3$ and thus also the values of $v_3$ and $h_3$ are uniquely determined.  
\end{proof}

For convenience, we proceed with the proof of  Lemma \ref{prop: ref} and show Lemma~\ref{prop:delta bounded by 2 theta} afterwards.

\begin{proof}[Proof of Lemma \ref{prop: ref}]
In the proof, we again use the notation indicated in Figure~\ref{fig:ex 4-tile}.  We recall the definition in \eqref{eq: delta1234} and drop for the moment the condition $\delta_{13} = \delta_{24}$ induced by \eqref{eq: angles2}.    To verify that every $4$-tile can be placed in reference position, we first rotate and translate the $4$-tile such that $C=0$ and  $M_1 = (s_1,0,h_1)$, and $M_3 = (-s_1,0,h_1)$, where    
a simple trigonometric relation yields 
\begin{align}\label{eq:M1 M3}
s_1 = \cos\left( \frac{\pi - \delta_{13}}{2} \right) = \sin(\delta_{13}/2), \quad \quad \quad h_1 =  \sin\left(\frac{\pi-\delta_{13}}{2} \right)  =  \cos(\delta_{13}/2).
 \end{align} 
 Here, we note that $s_1>0$, while $h_1$ is negative whenever $\delta_{13} > \pi$. We now show that  the coordinates of $M_2$ and $M_4$ are given by 
\begin{align}\label{eq: M24}
	M_2 = (0,s_2, h_2),  \quad \quad \quad 	M_4 = (0, -s_2, h_2),
\end{align}
where $s_2= \sin(\delta_{24}/2)$ and $h_2 =  \cos(\delta_{24}/2)$. We focus on $M_2$ since the argument for $M_4$ is analogous. For convenience, we write $M_2 = (p_1,p_2,p_3)$ and use the definition of optimal cells, i.e., $\measuredangle M_1 \, C \, M_2  = \theta= \measuredangle M_2 \, C \, M_3$ and $|M_1| = |M_2| = |M_3| = 1$, to find    
\[ \cos \theta =  M_1 \cdot M_2   = p_1 s_1 + p_3 h_1, \quad \quad  \cos\theta = M_3\cdot M_2     = - p_1 s_1 + p_3 h_1. \]
By combining the two equalities we get $p_1 = 0$.  In view of \eqref{eq:M1 M3},  $p_3$ is then given by 
\begin{equation}\label{eq: heights relation}
 p_3 =  \frac{\cos \theta}{h_1} = \frac{\cos \theta}{\cos(\delta_{13}/2)},
\end{equation}
and, since $|M_2| = 1$, we find  $p_2 = \sqrt{1-p_3^2}$.   Thus, we have $M_2 =(0,p_2,p_3)$. By a similar argument we find $M_4 = (0,-p_2,p_3)$. To conclude for \eqref{eq: M24}, we need to find the relation between $p_2$ and $p_3$. To this end, we use the fact that $\measuredangle M_2 \, C \, M_4  = \delta_{24}$ to calculate $\cos(\delta_{24}) = M_2 \cdot M_4 = p_3^2 - p_2^2$. This, together with $p_2^2+p_3^2 = 1$, verifies that  
$p_3 = \sqrt{( 1+\cos(\delta_{24})   )/2} = \cos(\delta_{24}/2)$ by using the double-angle formula. Correspondingly, we find  $p_2 = \sin(\delta_{24}/2)$. This proves \eqref{eq: M24}. Let us remark for later purposes that   \eqref{eq: heights relation} implies
\begin{equation}\label{eq:delta eta2}
	\cos(\delta_{13}/2)	\cos( \delta_{24}/2) = \cos\theta.
\end{equation}	
From the condition  $\delta_{13}=\delta_{24}$ we get that  $s=s_1=s_2$ and $h=|h_1|=|h_2| = \sqrt{1-s^2}$. We also let $\varsigma = {\rm sgn}(h_1) = {\rm sgn}(h_2)$. To conclude the proof of (i), it remains to  check that $s = \sqrt{2}v$, where $v$ is defined in \eqref{eq: def of v}, i.e., is chosen in such a way that $2v$ indicates the length of a diagonal in an optimal cell. This length can indeed be expressed as $|M_i-M_{i+1}| = \sqrt{2}s $ for $i=1,\ldots,4$, which yields the desired relation.

We proceed with the proof of (ii). By fixing $\theta$, the angle $\delta_\theta$ is also determined and, by (i), also fixing ${\rm sgn}(h_1)$ determines completely the positions of the points $(M_i)_{i=1}^4$. In view of Lemma~\ref{lem:fourthpoint}, the positions of $(E_i)_{i=1}^4$ are determined as well, as soon as the forms of the four optimal cells  are  given.    
\end{proof}

\begin{proof}[Proof of Lemma \ref{prop:delta bounded by 2 theta}]
In the proof of Lemma \ref{prop: ref} we have already verified
\eqref{eq:delta eta}, see \eqref{eq:delta eta2}. Consider    $ f_\theta\colon [0,\pi]\to \R$ defined by $ f_\theta (\delta) = 2\arccos(\cos\theta /\cos(\delta/2))$. As $\cos\theta>0$,  $ f_\theta$  is decreasing and thus has at most one fixed point. Hence, $ f_\theta$ has exactly  one fixed point given by $\delta_\theta = 2 \arccos(\sqrt{\cos\theta})$. This eventually shows that $\delta_{13}$ and $\delta_{24}$ coincide if and only if $\delta_{13} = \delta_{24} = \delta_\theta$. 
\end{proof}

We close this subsection with an elementary observation.  We again refer to the notation in Figure \ref{fig:ex 4-tile}.    

\begin{lemma}\label{lem:third point of optimal cell is high}
\noindent 
{\rm (i)} For any coplanar $4$-tile in $\mathcal{A}$ (cf.\ \eqref{eq:class of admissible 4-tile A})  in reference position,  see Lemma~\textup{\ref{prop: ref}(i)}, we have $E_1 = (s,s,0)$ and $E_3= (-s,-s,0)$. \\
\noindent {\rm (ii)}  For any Z-tile in $\mathcal{A}$ in reference position, we have $E_2 = (-s,s,0)$ and $E_4= (s,-s,0)$.   \\
\noindent {\rm (iii)}  For any D-tile in $\mathcal{A}$ in reference position,  we have $|E_2-E_4| < 4v$, where $v$ is given in  \eqref{eq: def of v}. \\
 {\rm (iv)} Assume that an optimal cell $\{y_1,\dots, y_4\}$ is
 positioned  in such a way   that $e_3 \cdot y_1 = 0$ and $e_3 \cdot y_2 = e_3 \cdot y_4 = h$. Then, depending on its form, we have $e_3 \cdot y_3 = 0$ or $e_3 \cdot y_3 > 2h$. \\
\end{lemma}
 
Similar statements as {(ii)--(iv)} hold for $\mathcal{B}$ in place of $\mathcal{A}$ by changing the roles of the diagonals.

\begin{figure}[h]
  \centering
	\def\svgwidth{0.8 \columnwidth}
	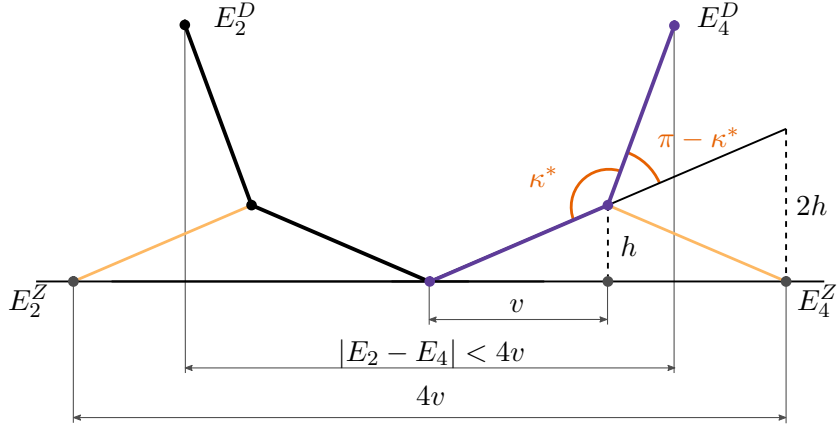

  \caption{ Cross section of a D-tile (bold, black and purple) and of a Z-tile (thin lines in light orange), positioned as in Lemma~\ref{lem:third point of optimal cell is high}. The distance between the diagonals is $4v$ for the Z-tile and smaller for the D-tile.}
  \label{fig:cross section rotated optimal cell}
\end{figure}

\begin{proof}
 Without restriction, we consider a $4$-tile in $\mathcal{A}$ in  reference position such that  $\varsigma = 1$, cf.\ Lemma~\ref{prop: ref}, as the other case only amounts to reflection along the $e_1$-$e_2$-plane. By  Lemma~\ref{prop: ref}(i) we have  that $M_1 = (s,0,h)$, $M_2 = (0,s,h)$, $M_3 = (-s,0,h)$, and $M_4 = (0,-s,h)$. The optimal cells $\{C, M_1, E_1, M_2 \}$ and $\{C, M_3, E_3, M_4\}$ are of  form $\diagup$, see Figure \ref{fig:types optimal cells} and \eqref{eq:class of admissible 4-tile A}. Thus, by Lemma~\ref{lem:fourthpoint} we get $E_1 = (s,s,0)$ and $E_3 = (-s,-s,0)$. This shows (i). We now suppose that the $4$-tile is either of  class Z or of class D, i.e., is of type {\tiny $\borderIIOO{\tileD{\diagdown}{\diagup}{\diagup}{\diagdown}}$} or  {\tiny $\borderOOIInwseu{\tileD{\diagup}{\diagup}{\diagup}{\diagup}}$}. Therefore, the two optimal cells $\{C, M_2, E_2, M_3 \}$ and $\{C, M_4, E_4, M_1\}$ are  of  form $\diagup$ (D-tile) and of form $\diagdown$ (Z-tile), which yields to a cross section  along the direction $(-1,1)$  as indicated in Figure~\ref{fig:cross section rotated optimal cell}. We now obtain
\begin{align}\label{eq:  2h}
	&E_2^D \cdot e_3 = E_4^D \cdot e_3 > 2h &&\text{for the D-tile and}\notag\\  
	&E_2^Z \cdot e_3 = E_4^Z \cdot e_3 =0 	&&\text{for the Z-tile}.
\end{align}
Indeed,  for the Z-tile this follows from Lemma~\ref{lem:fourthpoint}. For the D-tile we use Thales' intercept theorem instead, with reference to Figure~\ref{fig:cross section rotated optimal cell}. In particular, this implies (ii). Then, as in the Z-cell the distance of the diagonals  is $4v = 2\sqrt{2}s$, \eqref{eq:  2h} and Figure~\ref{fig:cross section rotated optimal cell} show that in the D-cell we have $\vert E_2^D - E_4^D \vert < 4v$. This implies  (iii). Eventually, property (iv) follows from \eqref{eq:  2h}.  
\end{proof}

\subsection{Boundary orientations and attachment of two $4$-tiles}\label{sec: bdy orient}

This subsection is devoted to the proofs of Lemma~\ref{prop: attach} and Lemma \ref{lemma: bo}.  

\begin{proof}[Proof of Lemma~\ref{prop: attach}]
	The statement for the boundary orientation and the boundary angle, defined in \eqref{eq: boundary orientation}--\eqref{eq: boundary angle}, respectively,  follows from the fact that the notions are determined uniquely by the three points which are shared by the two $4$-tiles.  More precisely, given any $4$-tile in reference position, by applying a  rotation about the $e_3$ axis composed with a further small rotation (depending on $\theta$), and a translation one can ensure that a boundary  of the $4$-tile is contained in the $e_2$-$e_3$-plane and is symmetric with respect to the $e_1$-$e_3$-plane. Provided that  $\theta$ is small, one can check that this transformation  does not change the inequality in \eqref{eq: boundary orientation}. Clearly, each two $4$-tiles with the same boundary angles can be transformed in this fashion in order to be matched along the shared boundary.

 	Consider now    two attached $4$-tiles positioned such that the middle $4$-tile is in reference position, in particular, the shared middle point of the boundary is the origin. If the boundary orientation of the shared boundary is $\wedge$, then both shared corner vertices satisfy  $E_{i-1} \cdot e_3, \, E_i \cdot e_3 < 0$, see \eqref{eq: boundary orientation},    and thus for the middle $4$-tile we have  $\varsigma = -1$.  An analogous  argument applies   if the boundary is $\vee$.  
\end{proof}

\begin{proof}[Proof of Lemma \ref{lemma: bo}]
	First, we note that, for any $4$-tile in reference position, reflection about the $e_1$-$e_2$-plane interchanges all boundary orientations since the reflection changes the sign of any $e_3$-component. Moreover, rotation around $e_3$ by  $\pi /2$   leaves the boundary orientation invariant. A rotation in the matrix notation therefore simply rotates the corresponding sides and interchanges $\vee$ with $>$ and $\wedge$ with $<$. For example, rotating {\tiny $\borderOOOOnwd{\tileU{\diagdown}{\diagdown}{\diagdown}{\diagup}}$} by $\pi/2$ counterclockwise, yields {\tiny $\borderOIOIswd{\tileU{\diagup}{\diagdown}{\diagup}{\diagup}}$}. This entails that it is enough to check the boundary orientations for one representative of any class in Table~\ref{tab:adm copl 4-tile}.

	First,  by Lemma \ref{prop: ref}  and  Lemma~\ref{lem:third point of optimal cell is high}    we get that the orientation of all boundaries of the coplanar D-tile  {\tiny $ \borderOOIInwseu{\tileD{\diagup}{\diagup}{\diagup}{\diagup}}$}  is $\vee$.  Indeed,   assume that the $4$-tile is in reference position and use the notation of  Figure \ref{fig:ex 4-tile}. Then the corner vertices $E_1 \cdot e_3 = E_3\cdot e_3 = 0 $ and $M_i \cdot e_3 = h$ for $i=1,\ldots,4$. Moreover, the optimal cells $\{C, M_2, E_2, M_3\}$ and $\{C, M_4, E_4, M_1\}$ are positioned as in Lemma~\ref{lem:third point of optimal cell is high}(iv). Thus, we can conclude that the corner vertices $E_2$ and $E_4$ have $e_3$-coordinate strictly larger than $2h$ and hence, in view of \eqref{eq: boundary orientation}, we find that the boundary orientation is $\vee$.
	
	Consider the Z-tile  {\tiny $\borderIIOO{\tileD{\diagdown}{\diagup}{\diagup}{\diagdown}}$}   in reference position. In this case, the middle points satisfy  $M_i \cdot e_3 =  h  $, $i = 1, \dots, 4$  and, in view of the forms of the four optimal cells,   the corner points satisfy $E_i \cdot e_3 = 0$, $i = 1,\dots, 4$.   Thus, all four boundaries have orientation  $\wedge$.  
	
	We observe that the above arguments actually only take into account the relative position of the two optimal cells adjacent to a boundary. Thus, one can repeat the arguments above for the I-tiles. For instance,    {\tiny $\borderIIIInwu{\tileD{\diagup}{\diagup}{\diagup}{\diagdown}}$} has two $\vee$ boundaries top and left, i.e., adjacent to $\bullet$ as in a D-tile, and two $\wedge$ boundaries right and bottom, as in a Z-tile.  
\end{proof}

\subsection{Arrangements of four $4$-tiles}\label{subsection:geometric gap}

In this subsection we prove Lemma~\ref{lemma: gap}. We start by a result about  the   mutual position of two attached $4$-tiles. To this end, recall the types of $4$-tiles $\mathcal{A}$ and $\mathcal{B}$ introduced in \eqref{eq:class of admissible 4-tile A}--\eqref{eq:class of admissible 4-tile B}, as well as   the definition of $s$ and $h$ in Lemma \ref{prop: ref}(i).

\begin{lemma}\label{lem:attachment algorithm}
Let $T$ and $\tilde{T}$  be  two attached   Z-, I-, or D-tiles of an admissible  configuration. Without restriction, up to applying an isometry,  suppose that $T$ is in reference position,  and that the shared boundary consists of the three points  $E_1$, $M_1$, $E_{4}$ and $\tilde{E}_2$, $\tilde{M}_2$, $\tilde{E}_{3}$,   respectively, referring to the notation in Figure \ref{fig:ex 4-tile}. We denote the reference position corresponding to $\tilde{T}$ by $\tilde{T}'$. We denote by $R^{\mathcal{A}}_{\alpha}$ the counterclockwise rotation around the axis $(1,1,0)$ by the angle $\alpha$, and  $R^{\mathcal{B}}_{\alpha}$ denotes the counterclockwise rotation around the axis $(-1,1,0)$ by the angle $\alpha$.

\noindent {\rm (1)} If the shared boundary is a Z-boundary of $T$, and a Z-boundary of $\tilde{T}$, then we have $\tilde{T} = (2s,0,0)+ \tilde{T}'$.

\noindent {\rm (2)} If  the shared boundary is a D-boundary of $T$ and a Z-boundary of   $\tilde{T}$, we have
\begin{align*}
\tilde{T} = \begin{cases}  R^{\mathcal{A}}_{2\varsigma_T\kappa} ( (2s,0,0)+ \tilde{T}')  & \text{ if $ T    \in \mathcal{A}$,} \\ 
  R^{\mathcal{B}}_{2\varsigma_T\kappa}( (2s,0,0) + \tilde{T}')  & \text{ if $ T     \in \mathcal{B}$,}   \end{cases}
\end{align*}
where $\kappa$ is defined in \eqref{eq:alpha}, and $\varsigma_T$ corresponding to $T$  is given   in Lemma \ref{prop: ref}. 

\noindent {\rm (3)} If  the shared boundary is a Z-boundary of  $T$ and a D-boundary of $\tilde{T}$, we have
\begin{align*}
\tilde{T} = \begin{cases}  (2s,0,0)+  R^{\mathcal{A}}_{2\varsigma_{\tilde{T}'}\kappa}   \tilde{T}'  & \text{ if $ \tilde{T}    \in \mathcal{A}$,} \\ 
 (2s,0,0)+   R^{\mathcal{B}}_{2\varsigma_{\tilde{T}'}\kappa} \tilde{T}' & \text{ if $ \tilde{T}   \in \mathcal{B}$,}   \end{cases}
\end{align*}
where $\varsigma_{\tilde{T}'}$ corresponds to $\tilde{T}'$. 
\end{lemma}

 The case of two shared D-boundaries is not addressed here as we will not need it in the sequel.   We warn the reader that,  in the applications below  without further mentioning, we will apply isometries to the tiles in order to reduce the positions to the ones indicated in the lemma.   We postpone the proof of Lemma~\ref{lem:attachment algorithm}  to the end of this subsection, and proceed with the proof of Lemma \ref{lemma: gap}. 

\begin{figure}[h]
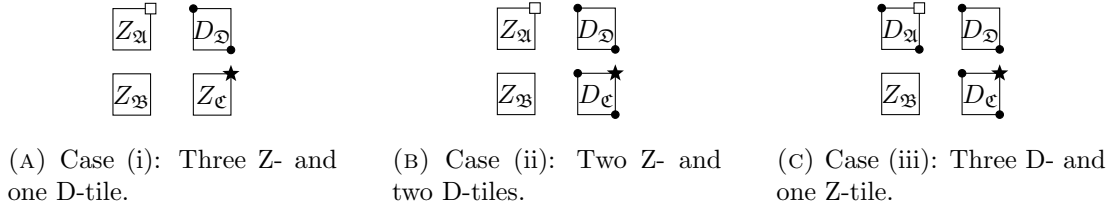

	\begin{subfigure}[b]{0.3\textwidth}
		\centering
		\[  \small \begin{aligned} 
				&\nborderne{Z_\mathfrak{A}} &&\nbordernwseu{D_\mathfrak{D}}\\
				&\nborder{Z_\mathfrak{B}} &&\nborderneStar{Z_\mathfrak{C}}
		\end{aligned} \]
		\caption{Case (i): Three Z- and one D-tile.}
		\label{fig: proof of geo gap lemma (i)}
	\end{subfigure}
	\hfill
	\begin{subfigure}[b]{0.3\textwidth}
		\centering
		\[  \small \begin{aligned} 
			&\nborderne{Z_\mathfrak{A}} &&\nbordernwseu{D_\mathfrak{D}}\\
			&\nborder{Z_\mathfrak{B}} &&\nbordernwseuneStar{D_\mathfrak{C}}
		\end{aligned} \]
		\caption{Case (ii): Two Z- and two D-tiles.}
		\label{fig: proof of geo gap lemma (ii)}
	\end{subfigure}
	\hfill
	\begin{subfigure}[b]{0.3\textwidth}
		\centering
		\[  \small \begin{aligned} 
		&\nbordernwseune{D_\mathfrak{A}} &&\nbordernwseu{D_\mathfrak{D}}\\
		&\nborder{Z_\mathfrak{B}} &&\nbordernwseuneStar{D_\mathfrak{C}}
		\end{aligned} \]
		\caption{Case (iii): Three D- and one Z-tile.}
				\label{fig: proof of geo gap lemma (iii)}
	\end{subfigure}
	
	\caption{Different cases in the proof of Lemma~\ref{lemma: gap}.}
	\label{fig: proof of geo gap lemma}
\end{figure}

\begin{proof}[Proof of Lemma \ref{lemma: gap}]
(i) Without restriction we suppose that  the tiles lie in $\mathcal{A}$ and  we suppose by contradiction   that the D-tile is given by $\mathfrak{D}$. We assume that $\mathfrak{B}$ is given in reference position. Then, by  Lemma    \ref{lem:attachment algorithm}(1) we see that ${\mathfrak{C}}$ is in reference position shifted by $(2s,0,0)$, and ${\mathfrak{A}}$ is in reference position shifted by $(0,2s,0)$. By Lemma~\ref{prop: ref} this implies that the coordinates of the points $Q$ and $P$, indicated with $\square$ and respectively $\star$ in Figure~\ref{fig: proof of geo gap lemma (i)}, are given by $Q = (s,3s,0)$ and $P = (3s,s,0)$,  respectively.   In particular, we have that $|P-Q| = 2\sqrt{2}s = 4v$, cf.\ \eqref{eq: def of v} and Lemma~\ref{prop: ref}(i), which corresponds to the length of the diagonal in $\mathfrak{D}$. For the D-tile $\mathfrak{D}$ in  $\mathcal{A}$,    however,  having   the rolling direction as given in Figure~\ref{fig: proof of geo gap lemma (i)},  cf.~\eqref{eq:class of admissible 4-tile A},   the corresponding diagonal has length smaller than $4v$ by Lemma~\ref{lem:third point of optimal cell is high}(iii), a contradiction.

(ii) Without restriction we suppose that the tiles belong to $\mathcal{A}$ and  we suppose by contradiction   that the Z-tiles are in  $\mathfrak{A}$, $\mathfrak{B}$, and that the D-tiles are in  $\mathfrak{C}$, $\mathfrak{D}$,  as in Figure~\ref{fig: proof of geo gap lemma (ii)}. We also assume that $\mathfrak{B}$ is given in reference position. By Lemma~\ref{lem:attachment algorithm}(1) we see that $\mathfrak{A}$ is in reference position shifted by $(0,2s,0)$, and thus the point $Q$, indicated by $\square$, has coordinates $Q = (s,3s,0)$.  By  Lemma~\ref{lem:attachment algorithm}(3)   the position of the tile  $\mathfrak{C}$ is obtained by taking the tile in reference configuration, rotating around the axis $(1,1,0)$ by the angle $\pm 2\kappa$, and then by a shifting by $(2s,0,0)$. As the corners where no roll-up occurs are left invariant under the rotation, we find by  Lemma~\ref{lem:third point of optimal cell is high}(i) that the point $P$, denoted by a $\star$ in Figure~\ref{fig: proof of geo gap lemma (ii)}, has coordinates $(3s,s,0)$. This implies $|P-Q| = 2\sqrt{2}s = 4v$.  As in (i), this contradicts Lemma~\ref{lem:third point of optimal cell is high}(iii) since the length of the diagonal in  the D-tile $\mathfrak{D}$   where the tile rolls-up is less than $4v$.

(iii) Again without restriction we assume that  the tiles belong to $\mathcal{A}$ and  we suppose by contradiction   that the Z-tile is in  $\mathfrak{B}$,  as in Figure~\ref{fig: proof of geo gap lemma (iii)}.  We assume that $\mathfrak{B}$ is given in reference position. By  Lemma~\ref{lem:attachment algorithm}(3)   the  position of   $\mathfrak{C}$ is obtained by taking the tile in reference configuration, rotation around the axis $(1,1,0)$ by the angle $\pm 2\kappa$, and then by a shifting by $(2s,0,0)$ (exactly in this order). As the corners where no roll-up occurs are left invariant under the rotation, we find by  Lemma~\ref{lem:third point of optimal cell is high}(i) that the point $P$,  marked with $\star$ in Figure~\ref{fig: proof of geo gap lemma (iii)},   has coordinates $P = (3s,s,0)$. In a similar fashion, the  position of   $\mathfrak{A}$ is obtained by taking the tile in reference configuration, rotating around the axis $(1,1,0)$ by the angle $\pm 2\kappa$, and then by a shifting by $(0,2s,0)$.  Lemma~\ref{lem:third point of optimal cell is high}(i)  yields   that the point  $Q$, indicated with $\square$ in Figure~\ref{fig: proof of geo gap lemma (iii)},   has coordinates $Q= (s,3s,0)$. This implies $|P-Q| = 2\sqrt{2}s = 4v$, which as in (i) and (ii) contradicts  Lemma \ref{lem:third point of optimal cell is high}(iii) since the length of the diagonal in  the D-tile $\mathfrak{D}$  where the tile roll-up is less than $4v$.

(iv) We finally show that \eqref{eq: lasti}  is not admissible. As
before, we denote the $4$-tiles by $\mathfrak{A}, \ldots,
\mathfrak{D}$, as indicated in \eqref{eq: notat-4}. Our strategy 
hinges on  (i)--(iii): we denote by $\tilde{Q}$ the right middle point of $\mathfrak{A}$ and by $\tilde{P}$ the upper middle point of $\mathfrak{C}$.  In view of Lemma~\ref{prop: ref}(i) applied on $\mathfrak{D}$, their distance necessarily needs to be $\sqrt{2}s = 2v$. We will show, however, that this is impossible.

In order to do so, we first assume that $\mathfrak{B}$ is in reference
position. In view of  Lemma~\ref{lem:attachment algorithm}(3),   the
position of the tile  $\mathfrak{C}$ is obtained by taking the tile in
reference configuration,  rotating it  around the axis $(-1,1,0)$ by the angle  $-2\kappa$, and then by a shifting by $(2s,0,0)$ (exactly in this order). In a similar fashion,    by Lemma~\ref{lem:attachment algorithm}(2)   the position of the tile  $\mathfrak{A}$ is obtained by taking the tile in reference configuration, followed by a translation by $(0,2s,0)$, and then by rotation around the axis $(1,1,0)$ by the angle  $-2\kappa$   (exactly in this order). 

We will now change the coordinate system to simplify the notational realization of the procedure: we suppose  that the common vertex of all three $4$-tiles lies in the origin and we reorient the coordinate system such that the  rotation axis $(1,1,0)$     coincides with $e_1$ and the rotation axis $(-1,1,0)$ with $e_2$, see Figure~\ref{fig:case6 rotation}. Then, the points $\tilde{Q}$ and $\tilde{P}$ are given by $\tilde{Q} = \mathcal{R}^{e_1}_{2\kappa} Q$ and   $\tilde{P} = \mathcal{R}^{-e_2}_{2\kappa} P = \mathcal{R}^{e_2}_{-2\kappa} P$,
where $Q = (v,v,-h)$ and $P = (v,-v,-h)$ are calculated by  using Lemma \ref{prop: ref},   and the rotations are given by
$$\mathcal{R}^{e_1}_{2\kappa} = \left(\begin{array}{ccc}
  1 & 0 & 0 \\
  0 & \cos(2\kappa) & -\sin(2\kappa) \\
  0 & \sin(2\kappa) & \cos(2\kappa)
  \end{array} \right), \quad \quad \quad
   \mathcal{R}^{e_2}_{-2\kappa} = \left(\begin{array}{ccc}
  \cos(2\kappa) & 0 & \sin(2\kappa) \\
  0 & 1 & 0 \\
  -\sin(2\kappa) & 0 & \cos(2\kappa)
  \end{array} \right).  $$

An elementary calculation yields 
$$  \tilde{Q} = \left( \begin{array}{c}
  v  \\
  v \\
  h 
  \end{array} \right), \quad \quad \quad   \tilde{P} =  \left( \begin{array}{c}
  0  \\
  -v \\
  0 
  \end{array} \right) +  \mathcal{R}^{e_2}_{-2\kappa} \left( \begin{array}{c}
  v  \\
  0 \\
  -h 
  \end{array} \right) = \left( \begin{array}{c}
  0  \\
  -v \\
  0 
  \end{array} \right) +  \mathcal{R}^{e_2}_{-4\kappa} \left( \begin{array}{c}
  v  \\
  0 \\
  h 
  \end{array} \right),$$
where we used the definition of $\kappa = \arctan(h/v)$,  see \eqref{eq:alpha},  and the trigonometric identities $\cos(2 \arctan(x)) = (1-x^2)/(1+x^2)$ and $\sin(2 \arctan(x)) = 2x/(1+x^2)$, as well as $(v,0,-h) = \mathcal{R}^{e_2}_{-2\kappa}(v,0,h)$. Consequently, we obtain 
\begin{align*}
  \vert \tilde{Q} - \tilde{P} \vert^2 = (2v)^2 + \left \vert    \big(\mathcal{R}^{e_2}_{0}  -   \mathcal{R}^{e_2}_{-4\kappa} \big) (v,0,h) \right \vert^2  
\end{align*}
which is strictly larger than $(2v)^2$ since $\kappa \in (0,\pi/2)$, see \eqref{eq:alpha}. This establishes a contradiction since, as stated above, the distance should be $2v$. 
\begin{figure}[h]
  \centering
	\def\svgwidth{0.3 \columnwidth}
	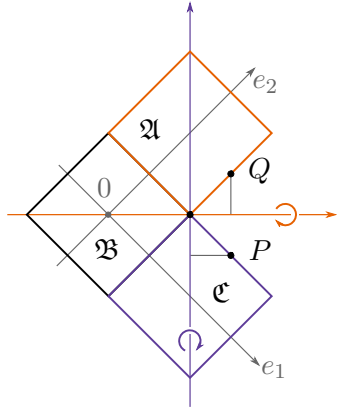

  \caption{The points $P$ and $Q$ are rotated by $2\kappa$ around the axis in the respective directions. Note that $P$ rotates clockwise.}
  \label{fig:case6 rotation}
\end{figure}
\end{proof}

\begin{proof}[Proof of Lemma \ref{lem:attachment algorithm}]

(1) Since  the shared boundary is a Z-boundary of $T$ and a Z-boundary of $\tilde{T}$, and the boundary orientations of $T$ and $\tilde{T}$ match at the shared boundary (see Lemma~\ref{prop: attach}),     Table \ref{tab:adm copl 4-tile} and Table \ref{tab:classification of boundaries} show that $\varsigma_T = \varsigma_{\tilde{T}}$ and that the middle $4$-tile between $T$ and $\tilde{T}$, denoted by $T_*$, is a Z-tile. By Lemma~\ref{prop: attach} we also find   $\varsigma_{T_*} = - \varsigma_T$. Then by  Lemma \ref{lem:third point of optimal cell is high}(ii)   it is elementary to check that $C_{\tilde{T}} - C_T = (2s,0,0)$, where $C_{\tilde{T}}$  and $C_T$ denote the centers of the $4$-tiles, respectively.

(2) We prove the result only for the particular case of the two  $4$-tiles {\tiny $\borderOOII{\tileU{\diagdown}{\diagup}{\diagup}{\diagdown}}$} and  {\tiny $\borderOOIInwseu{\tileD{\diagup}{\diagup}{\diagup}{\diagup}}$}, as depicted in Figure~\ref{fig:rotation}. In fact, the general case can be reduced to this situation by (a) replacing the optimal cells which are not adjacent to the shared boundary, as they do not affect the argument; and by (b) applying a suitable  rotation or reflection.

Suppose that $T$ is in reference position and denote the reference position of $\tilde{T}$ by $\tilde{T}'$. We define $\tilde{T}'':=\tilde{T}' + (2s,0,0)$. In view of  Lemma \ref{lem:third point of optimal cell is high}(i),(ii),  we see that $T$ and $\tilde{T}''$ share exactly one corner point $E_*  = (s,s,0)$ as depicted in Figure~\ref{fig:rotation}. 
Clearly,  the rotation around the axis $(1,1,0)$ by $2\kappa$ leaves $E_*$ invariant. We need to show that under this rotation the points $P_i$, $i=1,2$, are mapped to $\tilde{P}_i$, as depicted in Figure~\ref{fig:rotation}. We denote by $C_i$, $i=1,2$, the two points on $(1,1,0)$ which intersect the plane with normal vector $(1,1,0)$ containing $P_i$ and $\tilde{P}_{i}$.  By Lemma~\ref{prop: ref} we find that $C_1 = (s/2,s/2,0)$ and $C_2 = 0 $. We need to check that 
\begin{align}\label{eq: dist-an}
|C_i - P_i| = |C_i - \tilde{P}_i| \quad \text{ and } \quad  \measuredangle P_i \, C_i \,  \tilde{P}_i = 2\kappa \quad \text{ for $i=1,2$}.
\end{align}   
We first address $i=1$. By Lemma \ref{prop: ref}, we have $P_1 = (s,0,-h)$ and $\tilde{P}_1 = (s,0,h)$. We also note that $s = \sqrt{2}v$. This along with $C_1 = (s/2,s/2,0)$,  $\kappa = \arctan(h/v)$ (see \eqref{eq:alpha}), and the trigonometric  identity  $\cos(2 \arctan(x)) = (1-x^2)/(1+x^2)$ yields  \eqref{eq: dist-an} by an elementary computation.  

We now address $i=2$. As $C_2$ and $\tilde{P}_2$ form a diagonal of an optimal cell, see Figure~\ref{fig:rotation}, by the definition  before \eqref{eq: def of v} we get $|C_2 - \tilde{P}_2| = 2v = \sqrt{2}s$. On the other hand, by Lemma~\ref{prop: ref}, we find $P_2 = (s,-s,0)$ and therefore $|P_2 - C_2| = \sqrt{2}s$. This shows the first part of \eqref{eq: dist-an}.  To calculate  the angle, we refer to the cross section in Figure~\ref{fig:cross_sec_rot_alg}. Since this cross section is the one of an optimal cell in a rotated position, we can calculate the angle $\measuredangle P_2 \, C_2 \, \tilde{P}_2$ using the definition of $\kappa^*$ in \eqref{eq:alpha} and thus derive that $\gamma = \kappa + (\pi - \kappa^*)/2 = \kappa + \pi/2 - (\pi - 2\kappa)/2 = 2 \kappa$.  This  concludes the proof.

\begin{figure}[ht]
	\centering
	\def\svgwidth{0.8 \columnwidth}
	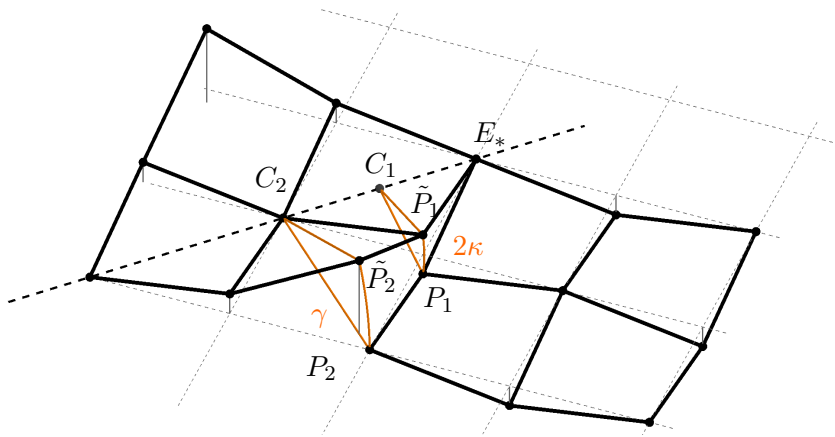

	\caption{Rotation of a Z-tile around the diagonal of a D-tile in order to match the boundaries.}
	\label{fig:rotation}
\end{figure}

\begin{figure}[ht]
	\centering
	\def\svgwidth{0.5 \columnwidth}
	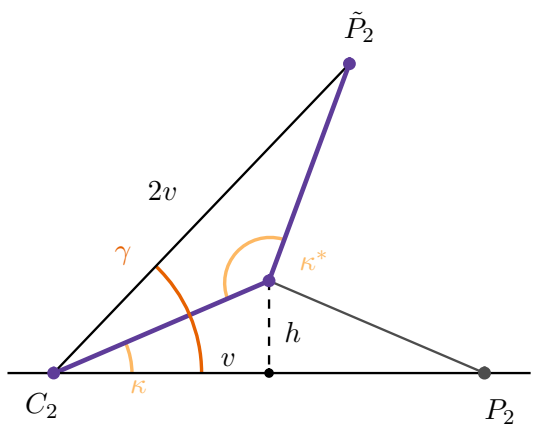

	\caption{Cross section along the plane with normal  $(1,1,0)$.}
	\label{fig:cross_sec_rot_alg}
\end{figure}

The proof of (3) is similar to (2) by interchanging the roles of the $4$-tiles. We omit the details. 
\end{proof}

We proceed with a simple consequence for boundary angles defined in  \eqref{eq: boundary angle}.

\begin{corollary}\label{cor: D bond}
The boundary angle of a D-boundary coincides the the boundary angle of a Z-boundary.
\end{corollary}

This immediately follows from Lemma~\ref{lem:attachment
  algorithm}(ii). Indeed, if the statement was not true, one could not attach the two $4$-tiles, as described in the previous proof.

\subsection{Incidence angles in coplanar $4$-tiles:  Theorem~\ref{thm: main thm 4-tiles} implies Theorem~\ref{thm: main optimal cells}}\label{sec:derivation of gamma}

In this short subsection, we   explain that Theorem~\ref{thm: main thm 4-tiles} implies Theorem~\ref{thm: main optimal cells}. In Subsection \ref{sec: thm-4-ti}, we already addressed the type function. Therefore, it remains to consider the last two items in Theorem~\ref{thm: main optimal cells}, i.e., the incidence angles defined in \eqref{eq:def of gamma hor} between optimal cells. 

 \noindent \emph{Z-tiles.} We start by showing that    the  incidence angles   between optimal cells in a   Z-tile along both diagonals are given by zero. Observe that reflection about the $e_1$-$e_2$-plane only interchanges the sign of the incidence angle. Thus, we assume  without restriction  that $\varsigma = 1$ (cf.\ Lemma~\ref{prop: ref}).  Due to symmetry, it suffices to consider one of the four bonds contained in the $4$-tile, e.g., the bond connecting $C$ and $M_1$, referring to the notation in Figure~\ref{fig:ex 4-tile}. Without restriction we only calculate the incidence angle along $d_1$ as the other one can be calculated in a similar fashion,  again exploiting symmetry.  If the $4$-tile is in reference position, Lemma \ref{prop: ref}(i) implies that $y_{\rm top}^1 = E_1 = (s,s,0)$, $y_{\rm bot}^1 =  M_4 = (0,-s,h)$, $C=0$, and $M_1 =  (s,0,h)$. Therefore,  since $n^1_{\rm top}$ is in direction $M_1 \times E_1$ and $n^1_{\rm bot}$ is in direction $-M_1 \times (M_4-M_1)$,   we find $n^1_{\rm top} = n^1_{\rm bot} = \frac{1}{\sqrt{s^2+2h^2}}(-h,h,s)$. Thus, in view of  \eqref{eq:def of gamma hor}, we get that the incidence angle is $\arccos(1) = 0$.

 \noindent \emph{D-tiles.}  We now address a D-tile in \eqref{eq: typi}, given in reference configuration.  Due to symmetry, it is again not restrictive to consider only the bond connecting $C$ and $M_1$ and to suppose that  $\varsigma = 1$. Note that, due to  Lemma \ref{prop: ref} and Lemma \ref{lem:third point of optimal cell is high}(i),  we have $E_1 = (s,s,0)$ and $E_4$ satisfies  $E_4 \cdot e_3 =  q $  and  $0 < E_4 \cdot e_1 = - E_4 \cdot e_2  =  p:=  \sqrt{s^2-q^2/2}< s$ for some $q >0$.   Since $E_1$, $M_1$, and   $M_4$  have the same position as in a Z-tile, repeating the above calculation we find that the incidence angle along $d_1$ is zero. We now consider the angle along $d_2$. To this end, we first find that $ y_{\rm top}^2  = M_2 =  (0,s,h) $ and  $ y_{\rm bot}^2  = E_4 = (p,-p,q)$,  see Lemma~\ref{prop: ref}. Therefore,  since $n^2_{\rm top}$ is in direction $M_1 \times M_2$ and $n^2_{\rm bot}$ is in direction $-M_1 \times (E_4-M_1)$, we get    $n^2_{\rm top}   = \frac{1}{\sqrt{s^2+2h^2}}(-h,-h,s)$,  and   an elementary computation yields $n^2_{\rm bot} = v/|v|$ for $v = (-h,-h,s) +   (0, q s /p,0) $. This implies $n^2_{\rm top}$ and $n^2_{\rm bot}$ are not parallel and therefore by \eqref{eq:def of gamma hor} the incidence angle,  denoted by $\gamma^*$,  is not zero.

To determine the sign of the non-zero incidence angle, we need to determine the sign of $ (y^2_{\rm top}- y_{\rm bot}^2) \cdot (n^2_{\mathrm{top}} - n^2_{\rm bot}) = (y^2_{\rm top}- y_{\rm bot}^2) \cdot n^2_{\mathrm{top}}  - (y^2_{\rm top}- y_{\rm bot}^2) \cdot  n^2_{\rm bot}$. First note that $(y^2_{\rm top}- y_{\rm bot}^2) = M_2 -  E_4  = (-p, s+p, h-q)$ which yields $(y^2_{\rm top}- y_{\rm bot}^2) \cdot n^2_{\mathrm{top}} = -\lambda  qs$, with $\lambda = 1/\sqrt{s^2 + 2h^2}$ and $(y^2_{\rm top}- y_{\rm bot}^2) \cdot  n^2_{\rm bot} =  \mu  qs^2/p  $, with $\mu = 1/\vert v \vert$. Therefore, we obtain $ (y^2_{\rm top}- y_{\rm bot}^2) \cdot (n^2_{\mathrm{top}} - n^2_{\rm bot})  = -\lambda  qs  - \mu  qs^2/p   < 0$. Hence, the incidence angle has a negative  sign, see \eqref{eq:def of gamma hor}.  Summarizing, we have shown that in  D-tiles the angles are also zero along $d_1$ and lie in $\lbrace -\gamma^*,\gamma^* \rbrace$ along $d_2$.

 \noindent \emph{I-tiles.}  It is obvious that for I-tiles, being combinations of Z- and D-tiles, we find that the incidence angles along $d_1$ are also $0$ and along $d_2$ they lie in $\lbrace -\gamma^*,0, \gamma^* \rbrace$.

 We close the proof by the observation that, due to the symmetries in Z-, D-, and I-tiles contained in $\mathcal{A}$, see \eqref{eq:class of admissible 4-tile A}, it is indeed elementary to check that $\gamma_2(s,t) = \gamma_2( s+1/2,t+1/2)$ for all $s,t\in \frac{1}{2}\Z$ with $s+t \in \Z+1/2$.

\subsection{Admissible configurations and ground states of the energy}\label{sec:energy}

This subsection is devoted to the proof of   Proposition~\ref{prop:minimizers of E}.

\begin{proof}
	\noindent \emph{Step 1.} We start by introducing a specific
        \emph{unit cell}: fix $x_0 \in \Z^2$ and denote the four
        neighbors of $x_0$ by $x_1 = x_0 + e_1$, $x_2 = x_0 + e_2$,
        $x_3 = x_0-e_1$, and $x_4 = x_0-e_2$. Given a deformation
        $y\colon \lbrace x_0,\ldots,x_4\rbrace \to \R^3$, we define
        $y_i = y(x_i)$ for $i=0,\ldots,4$, and we let $y_{5} = y_{1}$. We introduce the \textit{cell energy} by
\begin{align}
E_{\rm cell}(y) &= \frac{1}{2}\sum_{i=1}^4 v_2(|y_i-y_0|) +
  \frac{1}{2}\sum_{i=1}^4 v_2(|y_i-y_{i+1}|) \nonumber\\
  &\quad+ \sum_{i=1}^4 v_3(\theta_i) + v_3(\delta_{13}) + v_3(\delta_{24}), \label{eq:cell energy}
\end{align}
where  $\theta_i = \measuredangle y_i\,  y_{0} \,  y_{i+1}$ for $i=1,\ldots,4$, as well as $\delta_{13} = \measuredangle y_1 \, y_0\, y_3$ and $\delta_{24} = \measuredangle y_2 \, y_0 \, y_4$.  
The cell $(y_i)_{i=0}^4$ is called \textit{optimal} if it minimizes \eqref{eq:cell energy}.

 Let us start by relating  the cell energy to  the
configurational energy in  \eqref{eq: restrict energy}. To this
end, let $y\colon \Z^2 \ra \R^3$ be a deformation, and for $m \in
\mathbb{N}$ let $Q_m$ be the open  square centered at $0$ with
sidelength $2m$. For $j \in \Z^2 \cap Q_m$ we denote by $y^j = \lbrace
y_0^j,\ldots, y^j_4 \rbrace$  the  cell considered above for $x_0 = j
$. Then, in view of \eqref{eq: restrict energy},    owing to the
fact that    bonds related to  nearest-neighbors and  next-to-nearest-neighbors are contained in two cells (apart from bonds intersecting $\partial Q_m$), whereas each bond angle is contained in exactly one cell, we find for every $m \in \mathbb{N}$ that
$$E(y,Q_m) =\sum_{j\in \Z^2 \cap Q_m} E_{\rm cell}(y^j) = (2m-1)^2  \frac{1}{\# (\Z^2 \cap Q_m)}  \sum_{j\in \Z^2 \cap Q_m} E_{\rm cell}(y^j).  $$
 Then, recalling the definition in \eqref{eq: general energy} and 
 by arguing as in  \cite[Proposition 2.1]{emergence} we  have
  that $y\colon \Z^2 \to \R^3$ is a ground state if and only if
for each $x_0 \in \Z^2$ the corresponding cell $\lbrace y_0,\ldots,
y_4\rbrace$ is optimal.  Note that there exist admissible
configurations consisting of optimal cells by Theorem~\ref{thm: main
  thm 4-tiles}, e.g., a tiling with only Z-tiles.  Therefore, in
the following it suffices to minimize the cell energy and to show that
the unique minimizer is  identified by having  specific bond lengths and bond angles.  
\vspace{3mm}

	\noindent \emph{Step 2.} Let $\{y_0, y_1, y_2, y_3, y_4\}$ be an optimal cell.  We   show that $\vert y_{j} - y_{0} \vert \in (1-\eta, 1+ \eta)$ as well as $\theta_j   >  \pi/2 - \eta $ for $j=1,\ldots,4$.
	Assume first by contradiction that $\vert y_{j} - y_{0}\vert \leq 1-\eta$ for some $j = 1,\dots,4$. Then by using $v_2 \ge -1$,  $v_3 \ge 0$, the fact that $v_2$ is decreasing on $(0,1)$, and  \eqref{eq:energy a1}   we get  
	\begin{align*}
		E_{\cell}(y) &\ge  \frac{1}{2} \sum_{i=1}^4 v_2\left( \vert y_i - y_{0} \vert \right) + \frac{1}{2} \sum_{i=1}^4 v_2\left( \vert y_{i+1} - y_{i} \vert\right)   \\ 
		&\geq \frac{1}{2} v_2(1-\eta) + \frac{1}{2} \sum_{i\neq j} v_2 \left( \vert y_i - y_{0} \vert \right) + \frac{1}{2}\sum_{i=1}^4 v_2\left( \vert y_{i+1} - y_{i} \vert\right)\\
		&\geq \frac{1}{2} v_2(1-\eta) - \frac{3}{2} -2  =  \frac{1}{2} v_2(1-\eta) -\frac{7}{2} \\
		&\stackrel{\eqref{eq:energy a1}}{>} -2 + 2 v_2(\sqrt{2}) +4v_3(\pi/2) = E_{\cell}(x_0, x_1,x_2,x_3,x_4).
	\end{align*}
	In the last equation, we have also used that $v_2(1) = -1$ and
        $v_3(\pi) = 0$. 	This estimate  contradicts the   optimality of the cell. 
	
	In a similar fashion, we assume by contradiction that there
        exists some bond angle $\theta_j$, $j=1,\ldots,4$, such that
        $ \theta_j \leq \pi/2 -  \eta$.  Then,  by  $v_2 \ge -1$,
        $v_3 \ge 0$, and \eqref{eq:energy a3} we  have  
	\begin{align*}
		E_{\cell}(y) &= \frac{1}{2} \sum_{i=1}^4 v_2\left( \vert y_{i} - y_{0} \vert \right) + \frac{1}{2}\sum_{i=1}^4 v_2\left( \vert y_i - y_{i+1} \vert\right) + \sum_{i=1}^4 v_3(\theta_i) + v_3(\delta_{13}) +  v_3(\delta_{24}) \\
	&\geq  -4 + v_3(\theta_j)  \stackrel{\eqref{eq:energy a3}}{>} -2 + 2 v_2(\sqrt{2}) +4v_3(\pi/2) = E_{\cell}(x_0, x_1,x_2,x_3,x_4),
	\end{align*}
	which is again  in  contradiction  with  the optimality of $y$. 
	
	We  eventually   show that for an optimal cell the bond lengths have to be less then $1+\eta$.   Basic trigonometry  together with the least size of the bond lengths and bond angles   ensures that second-neighbor bonds have at least
length 
\begin{align}
  &2(1-\eta) \sin(\pi/4 - \eta/2) = 2(1-\eta)\frac{\sqrt{2}}{2} 
  \left(\cos(\eta/2) - \frac12
    \sin(\eta/2)\right)\nonumber\\
  &\quad > \sqrt{2} (1-\eta)^2  > 1, \label{trig}
\end{align}
where the last two  inequalities  hold for $\eta$ sufficiently small. Assume now that  $|y_j - y_{0}| \ge 1 + \eta$ for some $j=1,\dots, 4$. Then, we get by $v_2 \ge -1$,  $v_3 \ge 0$, the fact that $v_2$
increasing on $[1,\infty)$, and \eqref{eq:energy a2} that
	\begin{align*}
E_{\cell}(y) &= \frac{1}{2} \sum_{i=1}^4 v_2\left( \vert y_{i} - y_{0} \vert \right) + \frac{1}{2}\sum_{i=1}^4 v_2\left( \vert y_i - y_{i+1} \vert\right) + \sum_{i=1}^4 v_3(\theta_i) + v_3(\delta_{13}) +  v_3(\delta_{24})     \\
		&\geq -\frac{3}{2} + \frac{1}{2} v_2(1+\eta) + 2v_2(\sqrt{2}(1-\eta)^2) \\
		&\stackrel{\eqref{eq:energy a2}}{>} -2 + 2 v_2(\sqrt{2}) +4v_3(\pi/2) = E_{\cell}(x_0, x_1,x_2,x_3,x_4).
	\end{align*}
The latter inequality once again contradicts optimality and
we conclude that all first-neighbor bond lengths are at most
$1+\eta$.
		\vspace{3mm}

\noindent	\emph{Step 3.} 
To simplify notation, we denote  the collection of angles by
$\bm{\theta} \coloneqq (\theta_i)_{i=1}^4 =
(\theta_1,\dots,\theta_4)$.  We  observe that $\delta_{24}$ can be
written as a function of $\bm{\theta}$ and $\delta_{13}$, i.e.,
$\delta_{24}= f( \bm{\theta}, \delta_{13}   )$, where the function $f$
is   explicitly given  in Step 8, see \eqref{eq:f definition}.
We will not need  the exact form  of this function, but only use that it  is smooth for $\theta_i$ in a left neighborhood
of $\pi/2$ and $\delta_{13}$ in a small interval left of $\pi$,  see
Step 8 below.  Using Lemma~\ref{prop:delta bounded by 2
  theta} we find that in a cell with $\theta_1=\cdots=\theta_4 =
\theta$ it holds that $\delta_{24} =
f(\theta,\ldots,\theta,\delta_{13}) =  f_\theta(\delta_{13})$, where
$f_\theta(\delta) := 2\arccos \left(
  \cos\theta/\cos(\delta/2)\right)$.   Note  that $f_\theta$
has  a   unique fixed point $\delta_\theta:= 2 \arccos(\sqrt{\cos\theta})$.   We decompose the cell energy 
$E_{\rm cell}$   defined in \eqref{eq:cell energy} as 
\begin{align}\label{inequality}
 E_{\rm cell} (y)  &=  \sum_{i=1}^4 F(\ell_i,\ell_{i+1}, \theta_i) +  v_3(\delta_{13}) + v_3(f( \bm{\theta}, \delta_{13}   )), 
\end{align} 
 where  $\ell_i := |y_i-y_0|$ for $i=1,\ldots,4$ and 
 $$ F(\ell_i,\ell_{i+1}, \theta_i) :=    \frac{1}{4}  v_2(\ell_i) +  \frac{1}{4} 
v_2(\ell_{i  +  1}) +  \frac{1}{2}  v_2 \big((\ell_i^2 + \ell_{i  +  1}^2 - 2 \ell_i\ell_{i  +  1} \cos
\theta_i)^{1/2}\big) + v_3 (\theta_i).$$
 We have proved that, if $\{y_0,\dots,y_4\}$ is optimal,
first-neighbor bond lengths $\ell_i$ lie in 
$(1-\eta,1  +  \eta)$ and bond angles $\theta_i$  lie in  $(\pi/2
- \eta,  \pi]$. Therefore, by using the  convexity  assumption \eqref{eq:energy a4} on $F$ we find
\begin{align}\label{inequality-cvon}
 \sum_{i=1}^4 F(\ell_i,\ell_{i+1}, \theta_i)  \ge 4F(\bar{\ell},\bar{\ell},\bar{\theta}),
\end{align}
where 
\begin{align}\label{eq: mean}
 \bar{\ell}=\frac{1}{4}( \ell_1 + \dots+ \ell_4), \quad  
\bar{\theta}=\frac{1}{4}( \theta_1 + \dots+ \theta_4).
\end{align}
Note that the  inequality in \eqref{inequality-cvon} is strict whenever $\ell_i\not = \bar{\ell}$ or $\theta_i
\not = \bar{\theta}$ for some $i=1,\dots,4$.

\vspace{3mm}

\noindent	\emph{Step 4.}  We check that the map
$(\ell,\theta) \mapsto F(\ell,\ell,\theta)$ is minimized on
$(1-\eta,1+\eta) \times   (\pi/2-\eta,\pi] $  at some   $\ell^* \le 1$  and   ${\theta}^*<\pi/2$.   If  we
had  $\ell^* > 1$,  one could reduce $F$
by reducing $\ell$, noting that $v_2$ is increasing in $(1,\infty)$
and recalling \eqref{trig}. This, however, would contradict  
optimality.  We now exclude $\theta^* \ge  \pi/2$. Indeed, in this case we could decrease $\theta^*$ by $0<\tilde{\theta} \ll 1$ and by a Taylor expansion we would get that $F$ changes to first order by
$$-v_3'(\theta^*)\tilde{\theta} - v_2'\big(\sqrt{2}\ell\sqrt{1-\cos\theta^*}\big) \frac{\ell}{2\sqrt{2}} \frac{\sin\theta^* \, \tilde{\theta}}{\sqrt{1-\cos\theta^*}}. $$
By $\ell \in (1-\eta,1]$ and  \eqref{eq:energy a6} we get that the above term is negative, which contradicts minimality.

 \vspace{3mm}

\noindent	\emph{Step 5.} Next, we show that for  $\bar{\theta}$ defined in \eqref{eq: mean} it holds that  $\bar{\theta}  \le \pi/4 + \theta^*/2$. We also establish a bound from below on $\delta_{13}$ and $\delta_{24}$.  The argument is based on the observation that by \eqref{inequality},  \eqref{inequality-cvon}, and the definition of $\delta_\theta$ we find that $4F(\ell,\ell,{\theta}) + 2v_3(\delta_{\theta})$ is an upper bound for the minimal cell energy for $(\ell,\theta) \in (1-\eta,1+\eta) \times   (\pi/2-\eta,\pi]$.  By definition we have  $\delta_\theta \to \pi$ as $\theta \to \pi/2$. Thus, in view of \eqref{eq:energy a5}, the monotonicity of $v_3$,  and $\theta^*  \ge \pi/2-\eta$,   we can choose $\eta$ sufficiently small depending on $v_3$ and find $\lambda>0$ small such  that   $|v_3| \le \eps$ on $[\pi-\lambda,\pi]$, $|v_3'| \leq \eps$ on $[\pi - 2\lambda, \pi]$, and 
\begin{align}\label{eq: montoni}
v_3(\delta) >     2\epsi >  2v_3(\delta_{\theta^*}) \quad \text{ for $\delta \le \pi - \lambda$}.
\end{align}
 We  also  suppose that $\eps$ is chosen  small enough depending on $v_2$, $v_3$, and $\theta^*$ such that
\begin{align}\label{eq: montoni2}
F(\ell,\ell,\theta) >  F(\ell^*,\ell^*,\theta^*)  + 2\eps  \ \ \ \text{for} \  \theta > \frac{\pi}{4} + \frac{\theta^*}{2} \text{  and $\ell \in (1-\eta,1+\eta)$}.
\end{align}
Now,  we can suppose $\delta_{13},\delta_{24} \ge \pi-\lambda$ (recall
 that  $\delta_{24} = f(\bm{\theta},\delta_{13})$) since otherwise we get  
$${E_{\rm cell} (y) >  4F(  \bar{\ell},\bar{\ell}, \bar{\theta}) + 2v_3(\delta_{\theta^*})  \ge 4F(  \ell^*,\ell^*, {\theta}^*) + 2v_3(\delta_{\theta^*})}$$
 by  using  \eqref{inequality}, \eqref{inequality-cvon}, and  \eqref{eq: montoni}, which contradicts minimality. In a similar fashion,   we can suppose that $\bar{\theta}$ in \eqref{eq: mean} satisfies  $\bar{\theta} \le \frac{\pi}{4} + \frac{\theta^*}{2}$ as otherwise  $E_{\rm cell} (y) > 4F({\ell}^*,{\ell}^*,{\theta}^*)  + 2v_3(\delta_{\theta^*})$ follows  using  \eqref{inequality}, \eqref{inequality-cvon},   \eqref{eq: montoni}, and   \eqref{eq: montoni2}.

\vspace{3mm}

\noindent	\emph{Step 6.}   We are left with the case $\delta_{13} \ge \pi-\lambda$ and $\theta_1 + \ldots  + \theta_4 = 4 \bar{\theta}  \le \pi + 2\theta^* < 2\pi$. In this step, we  show that 
\begin{align}\label{inequality4}
 E_{\rm cell} (y)  \ge  4F(\bar{\ell},\bar{\ell},\bar{\theta})  + v_3(\delta_{13}) + v_3(f_{\bar{\theta}}(\delta_{13})) 
 \end{align}
with equality only if $\ell_i = \bar{\ell}$ and   $\theta_i
 = \bar{\theta}$ for  $i=1,\dots,4$. 
 
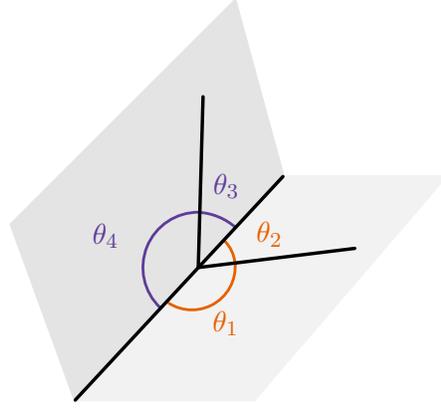
\begin{figure}
	\centering
	\def\svgwidth{0.4 \columnwidth}
\begingroup%
  \makeatletter%
  \providecommand\color[2][]{%
    \errmessage{(Inkscape) Color is used for the text in Inkscape, but the package 'color.sty' is not loaded}%
    \renewcommand\color[2][]{}%
  }%
  \providecommand\transparent[1]{%
    \errmessage{(Inkscape) Transparency is used (non-zero) for the text in Inkscape, but the package 'transparent.sty' is not loaded}%
    \renewcommand\transparent[1]{}%
  }%
  \providecommand\rotatebox[2]{#2}%
  \newcommand*\fsize{\dimexpr\f@size pt\relax}%
  \newcommand*\lineheight[1]{\fontsize{\fsize}{#1\fsize}\selectfont}%
  \ifx\svgwidth\undefined%
    \setlength{\unitlength}{206.25190891bp}%
    \ifx\svgscale\undefined%
      \relax%
    \else%
      \setlength{\unitlength}{\unitlength * \real{\svgscale}}%
    \fi%
  \else%
    \setlength{\unitlength}{\svgwidth}%
  \fi%
  \global\let\svgwidth\undefined%
  \global\let\svgscale\undefined%
  \makeatother%
  \begin{picture}(1,0.92385597)%
    \lineheight{1}%
    \setlength\tabcolsep{0pt}%
    \put(0,0){\includegraphics[width=\unitlength,page=1]{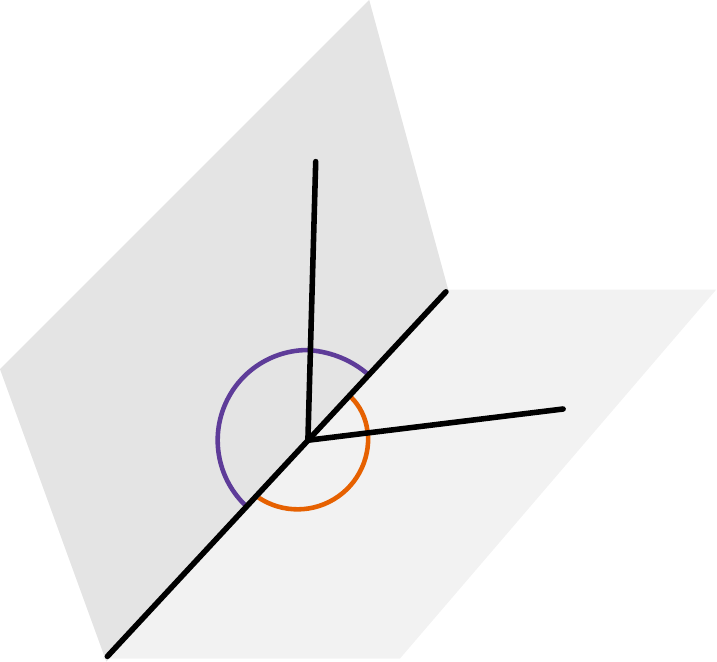}}%
    \put(0.46269435,0.16438371){\color[rgb]{0.90196078,0.38039216,0.00392157}\makebox(0,0)[lt]{\lineheight{1.25}\smash{\begin{tabular}[t]{l}$\theta_1$\end{tabular}}}}%
    \put(0.56255425,0.36658323){\color[rgb]{0.90196078,0.38039216,0.00392157}\makebox(0,0)[lt]{\lineheight{1.25}\smash{\begin{tabular}[t]{l}$\theta_2$\end{tabular}}}}%
    \put(0.46406817,0.47679235){\color[rgb]{0.36862745,0.23529412,0.6}\makebox(0,0)[lt]{\lineheight{1.25}\smash{\begin{tabular}[t]{l}$\theta_3$\end{tabular}}}}%
    \put(0.18947117,0.36283486){\color[rgb]{0.36862745,0.23529412,0.6}\makebox(0,0)[lt]{\lineheight{1.25}\smash{\begin{tabular}[t]{l}$\theta_4$\end{tabular}}}}%
  \end{picture}%
\endgroup%

	\caption{If $\delta_{13} = \pi$, then $\theta_1 + \theta_2 = \theta_3 + \theta_4 = \delta_{13} = \pi$ and thus $\theta_1 + \cdots + \theta_4 = 2\pi$. Furthermore, the angle $\delta_{24}$ can be chosen arbitrarily.}
	\label{fig:delta equals to pi}
\end{figure}

We start by noticing that $\theta_1 + \ldots  + \theta_4 <2\pi$ and $\theta_i > \pi/2 - \eta$ for $i=1,\ldots,4$  imply  $\theta_i < \pi/2 + 3\eta$ for $i=1,\ldots,4$. Therefore, the convexity estimate in \eqref{inequality-cvon} can be improved  by using the    strong  convexity assumption \eqref{eq:energy a4} on $F$, and  we find
\begin{align}\label{inequality-cvon-new}
 \sum_{i=1}^4 F(\ell_i,\ell_{i+1}, \theta_i)  \ge 4F(\bar{\ell},\bar{\ell},\bar{\theta}) + \alpha \sum_{i=1}^4 \vert \theta_i - \bar{\theta} \vert^2
\end{align}
for some $\alpha>0$. Moreover, a simple geometric  argument  shows that $\delta_{13} = \pi$ implies   $\theta_1 + \ldots  + \theta_4 = 2\pi$,  see Figure~\ref{fig:delta equals to pi}.  Therefore, by a continuity argument and    $\theta_1 + \ldots  + \theta_4 \le \pi + 2\theta^*$ we get that $\delta_{13} \le \delta^*$ for some $\delta^* < \pi$ only depending on $\theta^*$.  Consequently, we need to consider the case that $\delta_{13} \in [\pi - \lambda,\delta^*]$ and $\theta_i\in (\frac{\pi}{2}-\eta,\frac{\pi}{2}+3\eta)$. 

  If  $\alpha \sum_{i = 1}^4 |\theta_i - \bar\theta|^2 \geq 2\epsi$, by   \eqref{inequality}, \eqref{eq: montoni},  and \eqref{inequality-cvon-new} we obtain a contradiction to minimality    as  $v_3(\delta_{13}) + v_3(f(\bm{\theta},\delta_{13})) + \alpha \sum_{i = 1}^4 |\theta_i - \bar\theta|^2 \ge 2\epsi >  2v_3(\delta_{{\theta}^*})$.

If $\alpha \sum_{i = 1}^4 |\theta_i - \bar\theta|^2 < 2\epsi$, we 
now  show that $f_{\bar\theta}(\delta_{13})$ cannot be too far
away from  $f(\bm{\theta},\delta_{13})$. Eventually, this will allow
us to deduce \eqref{inequality4}.   By choosing $\eps$ small enough
and recalling that $\bar\theta<\pi/2$, we get that $\theta_i \le
\pi/2$ for $i=1,\ldots,4$.     By Taylor's Theorem there  exists
$ \bz  \in \{ t\bm{\theta} + (1-t) \bm{\bar{\theta}} \: : \: t \in [0,1] \}$,  where $\bar{\bm{\theta}} = (\bar{\theta}, \dots, \bar{\theta})$,  such that 
 \begin{align}
 f(\bm{\theta},\delta_{13}) - f_{\bar{\theta}}(\delta_{13}) &=
                                                              f(\bm{\theta},\delta_{13})
                                                              -
                                                              f(\bm{\bar{\theta}},\delta_{13})
   \nonumber\\
                                                            &=
                                                              \D_{\bm{\theta}}                                                           
                                                              f(\bm{\bar{\theta}},\delta_{13})
                                                              (\bm{\theta}
                                                              -
                                                              \bm{\bar{\theta}})
                                                              +
                                                              \dfrac{1}{2}
                                                              (\bm{\theta}
                                                              -
                                                              \bm{\bar{\theta}})
                                                              \D_{\bm{\theta}}^2
                                                              f(\bz,
                                                              \delta_{13})
                                                               (\bm{\theta} - \bm{\bar{\theta}}) \nonumber\\
 &=  \dfrac{1}{2} (\bm{\theta} - \bm{\bar{\theta}}) \D_{\bm{\theta}}^2 
                                                              f(\bz,
                                                              \delta_{13})        
   (\bm{\theta} - \bm{\bar{\theta}}) \leq \frac{\lambda_{\max}}{2} |
   \bm{\theta} - \bm{\bar{\theta}} |^2 \nonumber\\
   & =  \frac{\lambda_{\max}}{2} \sum_{i=1}^4 \vert \theta_i -\bar{\theta} \vert^2,\label{eq:eq}
 \end{align}
 where we used that $\tfrac{\partial}{\partial \theta_i} 
 f(\bm{\bar{\theta}}, \delta_{13})  = \tfrac{\partial}{\partial \theta_j}
 f(\bm{\bar{\theta}}, \delta_{13})  $ and thus  $${	
 \nabla_{\bm{\theta}} f(\bm{\bar{\theta}}, \delta_{13})  \cdot  (\bm{\theta} - \bm{\bar{\theta}})  = \tfrac{\partial}{\partial \theta_1} 
 f(\bm{\bar{\theta}}, \delta_{13})   \left( \sum_i \theta_i - \sum_i \bar{\theta} \right) = 0.}$$ 
 In \eqref{eq:eq}  we denoted the largest eigenvalue of the
Hessian  $\D_{\bm{\theta}}^2  f(\bz,\delta_{13})$    with $\lambda_{\max}$.  Using the Gershgorin circle theorem, we find $\vert \lambda_{\max} \vert \leq 4 c_f$   where we use that $f$ is smooth for $\theta_i \in I:=[\frac{\pi}{2}-\eta,\frac{\pi}{2}]$ and $\delta_{13} \in [\pi - \lambda,\delta^*]$, and define 
\begin{equation*}
	c_{f} := \max_{i,j = 1,\dots,4} \sup_{\theta_i \in I} \sup_{\delta_{13} \in [\pi - \lambda, \delta^*]} \left \vert \frac{\partial^2}{\partial \theta_i \partial \theta_j} f (\bm{\theta},\delta_{13}) \right \vert < \infty.
\end{equation*}
 The proof of the smoothness of $f$ is deferred to the end of the
 proof in Step 8.   Therefore, we  obtained 
 \begin{equation}\label{eq: est on diff of f}
 \vert f(\bm{\theta},\delta_{13}) - f_{\bar{\theta}}(\delta_{13}) \vert \leq 2 c_f \sum_{i=1}^4 \vert \theta_i -\bar{\theta} \vert^2.
 \end{equation}	
 For $\epsi$ small enough such that $ 4  c_f \epsi \leq \alpha \lambda$,  due to $\alpha \sum_{i = 1}^4 |\theta_i - \bar\theta|^2 < 2\epsi$, we have
  $$  \vert f(\bm{\theta},\delta_{13}) - f_{\bar{\theta}}(\delta_{13}) \vert \leq 2 c_f \sum_{i=1}^4 \vert \theta_i -\bar{\theta} \vert^2 \leq \frac{\alpha \lambda}{2 \epsi} \sum_{i=1}^4 \vert \theta_i -\bar{\theta} \vert^2 \leq  \lambda.$$ 
  Hence $f_{\bar\theta}(\delta_{13}) \geq \pi - 2\lambda$ as $f(\bm{\theta},\delta_{13}) \geq \pi -\lambda$  by Step 5.  Therefore,  as $|v_3'| \le \eps$ on $[\pi-2\lambda,\pi]$ and $\lambda>0$ small,  we obtain by \eqref{eq: est on diff of f}
\begin{align}\label{eq:est}
\vert v_3(f(\bm{\theta},\delta_{13})) - v_3(f_{\bar{\theta}}(\delta_{13})) \vert &\leq \vert f(\bm{\theta}, \delta_{13}) - f_{\bar{\theta}}(\delta_{13})\vert \sup_{[\pi-2\lambda, \pi]} \vert v'_3\vert\notag\\
&\leq 2 c_f \eps \sum_{i=1}^4 \vert \theta_i -\bar{\theta} \vert^2 \leq  \frac{\alpha}{2}  \sum_{i=1}^4 \vert \theta_i -\bar{\theta} \vert^2.
\end{align}
 Consequently,  \eqref{inequality4} holds by applying \eqref{inequality-cvon-new} and \eqref{eq:est} to \eqref{inequality}.

 \vspace{3mm}

\noindent	\emph{Step 7.}  We now conclude the proof by showing 
\begin{align}\label{inequality-new}
 E_{\rm cell} (y)  &   \geq 4 \left(   \frac{1}{2} 
v_2(\bar{\ell}) + \frac{1}{2}v_2 ( \sqrt{2} \bar{\ell}  (1 -\cos
\bar{\theta})^{1/2}) + v_3 (\bar{\theta}) \right)    +  2v_3(\delta_{\bar{\theta}}), 
\end{align} 
where $\delta_{\bar{\theta}} =2 \arccos(\sqrt{\cos \bar{\theta}})$, 
and that  equality  holds only if $\ell_i = \bar{\ell}$ and   $\theta_i
 = \bar{\theta}$ for  $i=1,\dots,4$.   To this end, we further estimate  \eqref{inequality4} by claiming
\begin{align}\label{inequality g}
	g(\delta_{13}) :=  v_3(\delta_{13}) + v_3(f_{\bar{\theta}}(\delta_{13}))  \ge  2v_3(\delta_{\bar{\theta}}),  
\end{align}
with equality if and only if  $\delta_{13} =
f_{\bar{\theta}}(\delta_{13})$.  Computing $
g'(\delta) = v_3'(\delta) + v_3'(f_{{\bar{\theta}}}(\delta)) \,
f'_{{\bar{\theta}}}(\delta) $ shows that
$g'(\delta_{\bar{\theta}}) = 0 $, because
$f'_{{\bar{\theta}}}(\delta_{\bar{\theta}}) = -1$ and
$\delta_{\bar{\theta}} = f_{\bar{\theta}}(\delta_{\bar{\theta}})$.
Moreover, we calculate $g''(\delta) = v_3''(\delta) +
v_3''(f_{{\bar{\theta}}}(\delta)) \, (f'_{{\bar{\theta}}}(\delta))^2 +
v_3'(f_{{\bar{\theta}}}(\delta)) \, f''_{{\bar{\theta}}}(\delta)>0$,
where we used the  monotonicity and  strict convexity of $v_3$ and the
concavity of $f_{{\bar{\theta}}}$,  which   follows from an 
elementary computation.  This indeed implies \eqref{inequality g}.

  This,  along with \eqref{inequality4},  implies that
 \eqref{inequality-new} holds, 
with equality only if all bonds of an optimal cell
have length  $\bar{\ell}$,  all angles have amplitude $\bar{\theta}$, and $\delta_{13} = \delta_{24} = \delta_{\bar{\theta}}$. Clearly, for an optimal cell, $\bar{\ell}$ and $\bar{\theta}$ are given uniquely. We finally observe that $\bar{\ell} \le 1$ and $\bar{\theta} <\pi/2$. For $\bar{\ell}$, this follows from the fact that $\ell^* \le 1$, as shown in Step 4, and $\bar{\theta}<\pi/2$ has been checked in Step~5.  
\vspace{3mm}

\noindent	\emph{Step 8.}   Let us conclude by collecting
some remarks on  
the function
$f(\bm{\theta},\delta_{13})$ used throughout the proof. If
$\delta_{13} = \pi$, we have that  $\theta_1 + \theta_2 = \theta_3 +
\theta_4 = \delta_{13} = \pi$. Therefore, $\delta_{24}$ can be
chosen arbitrarily in $[0,\pi]$, see Figure~\ref{fig:delta equals to
  pi},  and $f(\cdot,\pi)$ is hence  not defined.  For $\delta_{13} < \pi$, the definition is given by 
{\small
\begin{align}\label{eq:f definition}
&f(\bm{\theta} , \delta_{13}) = \arccos\Big\{ \cos\theta_1 \: \cos\theta_4  + (\cos\theta_2 - \cos\delta_{13} \cos\theta_1)(\cos\theta_3 - \cos\delta_{13}\cos\theta_4)/\sin^2\delta_{13}\nonumber  \\
- &\sqrt{1-\cos^2\theta_1 - \frac{(\cos\theta_2 - \cos\delta_{13} \cos\theta_1)^2}{\sin^2\delta_{13}}} \: \sqrt{1-\cos^2\theta_4 -\frac{(\cos\theta_3 - \cos\delta_{13} \cos\theta_4)^2}{\sin^2\delta_{13}}}  \Big\}, 
\end{align}}
\noindent which can be derived by elementary, yet tedious,
trigonometry. Let us now check that $f$ is smooth for all $\delta_{13}
\in[\pi-\lambda,\delta^*]$ and $\bm{\theta} \in [\pi/2-\eta,\pi]^4$
such that $f(\bm{\theta} , \delta_{13}) = \delta_{24}
\in[\pi-\lambda,\delta^*]$ which we need in Step 6 of the
proof. First, since $\delta_{24} \in[\pi-\lambda,\delta^*]$ the 
expression inside of $\arccos$ is bounded away from $-1$ and $1$. As
$\delta_{13} \in[\pi-\lambda,\delta^*]$, $\sin\delta_{13}$ is bounded
away from  0.  Thus, it suffices to check that the 
expressions   inside the square roots  are  bounded away
from  0. Indeed,   $\bm{\theta} \in [\pi/2-\eta,\pi]^4$ implies $\cos\theta_1 \: \cos\theta_4 \to 0$ as $\eta \to 0$ and $(\cos\theta_2 - \cos\delta_{13} \cos\theta_1)(\cos\theta_3 - \cos\delta_{13}\cos\theta_4)/\sin^2\delta_{13} \ge 0$. As $\cos (f(\bm{\theta} , \delta_{13}))$ lies in a neighborhood of $-1$, this is indeed only possible if the value of each of the square roots is close to $1$. 
\end{proof}

\bibliographystyle{alpha}



\end{document}